\documentclass[aps, prb, 10pt, twocolumn]{revtex4-1}
\usepackage[english]{babel}
\usepackage[colorlinks=true,citecolor=blue,urlcolor=blue,bookmarks=false]{hyperref}
\usepackage{amsmath,amssymb,amsfonts}

\newcommand{\Ht}{H_{T}}

\newcommand{\ph}{\varphi}
\newcommand{\md}{\mathrm{d}}
\newcommand{\ov}[1]{\bar{#1}}
\newcommand{\la}{\left\langle}
\newcommand{\ra}{\right\rangle}
\newcommand{\V}{\mathcal{V}}
\newcommand{\Tr}[1]{\text{Tr}\left[#1\right]}

\newcommand{\ib}{\hat{I}_B}
\newcommand{\Aij}{\mathcal{A}_{ij}(\chi)}
\newcommand{\snu}{\sqrt{\nu}}
\newcommand{\hmod}{H^{\text{mod}}_{ij}}

\usepackage{tikz}

\tikzset{
	MyPersp/.style={scale=1,x={(-0.8cm,-0.1cm)},y={(0.5cm,-0.4cm)},
    z={(0cm,1cm)}}}

\begin{document}

\title{Tunnelling current through fractional quantum Hall interferometers}

\author{O. Smits}
\email{smitso@tcd.ie}
\affiliation{Dublin Institute for Advanced Studies, School of Theoretical Physics, 10 Burlington Rd, Dublin, Ireland}
\affiliation{School of Mathematics, Trinity College, Dublin 2, Ireland}
\author{J. K. Slingerland}
\affiliation{Dublin Institute for Advanced Studies, School of Theoretical Physics, 10 Burlington Rd, Dublin, Ireland}
\affiliation{Department of Mathematical Physics, National University of Ireland, Maynooth, Ireland}

\author{S. H. Simon}
\affiliation{Rudolf Peierls Centre for Theoretical Physics, Oxford, OX1 3NP, UK}
\affiliation{Department of Mathematical Physics, National University of Ireland, Maynooth, Ireland}
\date{April 25, 2013}

\begin{abstract}
We calculate the tunnelling current through a Fabry-P\'{e}rot interferometer in the fractional quantum Hall regime. Within linear response theory (weak tunnelling but arbitrary source-drain voltage) we find a general expression for the current due to tunnelling of quasiparticles in terms of Carlson's $R$ function. Our result is valid for fractional quantum Hall states with an edge theory consisting of a charged channel and any number of neutral channels, with possibly different edge velocities and different chiralities. We analyse the case with a single neutral channel in detail, which applies for instance to the edge of the Moore-Read state. In addition we consider an asymmetric interferometer with different edge lengths between the point contacts on opposite edges, and we study the behaviour of the current as a function of varying edge length. Recent experiments attempted to measure the Aharanov-Bohm effect by changing the area inside the interferometer using a plunger gate. Theoretical analyses of these experiments have so far not taken into account the accompanying change in the edge lengths. We show that the tunnelling current exhibits multiple osculations as a function of this edge length, with frequencies proportional to the injected edge current and inversely proportional to the edge velocities. In particular the edge velocities can be measured by looking at the Fourier spectrum of the edge current. We provide a numerical scheme to calculate and plot the $R$ function, and include sample plots for a variety of edge states with parameter values which are experimentally relevant. 

\end{abstract}

\usetikzlibrary{decorations.pathmorphing,decorations.markings,trees,positioning,arrows,calc}   
\maketitle

\section{Introduction}\label{sec:introduction}
The physics of the fractional quantum Hall effect is not captured through the conventional picture of symmetry breaking and local order parameters. The effect arises in low-dimensional electronic systems and is a prime example of a topological phase of matter\citep{nayak2008,wen2004,wen2012}. A range of fractional quantum Hall phases has been discovered\citep{klitzing1980,tsui1982,willett1987,pan1999,xia2004}, each characterized by its filling fraction $\nu$, which determines the Hall conductivity through $\sigma_H = \nu \frac{e^2}{2\pi \hbar}$. Each of these phases possess a different type of order known as topological order\citep{wen1995}, referring to the presence of long-range entanglement of the ground state\citep{wen2012}. Some manifestations of topological order in the FQHE are the simultaneous formation of an energy gap in the bulk and gapless states at the edge of the system\citep{halperin1982,wen1992}, a ground-state degeneracy determined by the topology of the space-time manifold\citep{wen1990b,oshikawa2007} and exotic properties of the low-lying excitations of the system\citep{laughlin1983,mooreread1991}.

In particular, topological order predicts quasiparticle excitations known as anyons\citep{leinaas1977,wilczek1990}, which possess fractional charge\citep{laughlin1983,saminadayar1997,depicciotto1998} and obey a generalized form of exchange statistics\citep{arovas1984,mooreread1991,nayak1996}. These statistics generalize bosonic and fermionic statistics in the sense that interchange of anyons multiplies the wavefunction by a phase factor (Abelian anyons) or induces a rotation in an internal, non-local space of degenerate states (non-Abelian anyons). Non-Abelian anyons have been put forward as candidates for the realization of a topological quantum computer\citep{kitaev2003,nayak2008,preskill1998,freedman1998}.

A candidate for a non-Abelian FQH phase is the experimentally observed $\nu = \frac{5}{2}$ state\citep{willett1987,pan1999}. The Moore-Read Pfaffian state\citep{mooreread1991,greiter1991} and its particle-hole conjugate the Anti-Pfaffian\citep{levin2007,lee2007} could possible describe the corresponding topological order. These states predict the same Hall conductivity, but differ in other properties such as the type of anyons present and the effective edge theory of the system. It is both a theoretical and experimental challenge to design experiments sensitive to the topological order of the system, thereby identifying the nature of the $\frac{5}{2}$ state and other potential non-Abelian phases. 

Experiments have successfully measured the fractional charge of tunnelling quasiparticles\citep{saminadayar1997,depicciotto1998,dolev2008} for a variety of quantum Hall phases. More recent experiments aim to fully determine the topological order through use of various interferometric devices\citep{ji2003,camino2005,willett2009,willett2010,an2011,willett2012,willett2013a,willett2013b}. These experiments make use of the transport properties through these devices, which is determined by the edge where the electric current is located\citep{wen1990a,wen1992}. The electric current is chiral and flows along the edge in a single direction. Backscattering can occur through quantum point contacts, where opposite edges are forced together. This induces tunnelling of anyons between the edges\citep{wen1991,kanefisher1992}. The resulting tunnelling or backscattering current through the constrictions, $I_B$, depends on the type of anyon tunnelling.

In this work we analyse the tunnelling current through a Fabry-P\'erot interferometer in linear response theory\citep{chamon1997,fradkin1998,bishara2008,fidkowski2007,bishara2009b,levkivskyi2009,bieri2012}. An interferometer consists of multiple points contacts, see Figure~\ref{fig:interferometer}. Anyons tunnel along different trajectories, which gives rise to interference effects. In a simple picture we assign $t_1$ and $t_2$ as the complex amplitude of a quasiparticle tunnelling along the corresponding point contact. The tunnelling current follows from the absolute value
\begin{align}
I_{B} \sim |t_1 + t_2|^2 =  |t_1|^2 + |t_2|^2 + 2\text{Re}[t_1 t_2]
\end{align}
In linear response theory the form $|t_1|^2$ and $|t_2|^2$ is radically different for the case of a tunnelling anyon as compared to what would be expected for electrons. It is a non-linear function of the applied voltage, the temperature of the system, and the fractional charge and scaling dimension of the tunnelling anyon\citep{wen1991,chang2003}. The term $2\text{Re}[t_1 t_2]$ is the interference current. Interference arises due to a variety of causes, such as the Aharonov-Bohm effect, the relative phases of the tunnelling coupling constants and the dynamical interference due to the finite velocity of the anyons traversing the interferometer. Perhaps the most interesting contribution to the interference current is due to the statistics of the anyons. Anyons localized in the bulk and inside the interferometer braid with anyons tunnelling between the edges. This braiding of anyons effectively reads out the topological state of the bulk anyons, and this signature manifests itself in the interference current\citep{dassarma2006,stern2006,bonderson2006a,bonderson2006b}. Further effects arise that go beyond braiding properties which are due to coupling of bulk quasiparticles and edge degrees of freedom\citep{overbosch2007,overbosch2008,rosenow2008,rosenow2009,bishara2009a}.
\begin{figure}
\begin{tikzpicture}
\path [fill = gray!20, rounded corners] 
(0,-.15) -- ++ (1.75,0) -- ++ (.4,.7) -- ++ (.4,-.7) -- ++ (1.75,0)-- ++ 
(.4,.7) -- ++ (.4,-.7) -- ++ (1.75,0) -- ++ (0,2.6) -- ++ (-1.75,0) -- ++ 
(-.4,-.7) -- ++ (-.4,.7) -- ++ (-1.75,0) -- ++ (-.4,-.7) -- ++ 
(-.4,.7) -- ++ (-1.75,0) -- cycle;
 
\path [fill = gray!40, rounded corners] 
(0,.1) -- ++ (1.75,0) -- ++ (.4,.7) -- ++ (.4,-.7) -- ++ (1.75,0)-- ++ 
(.4,.7) -- ++ (.4,-.7) -- ++ (1.75,0) -- ++ (0,2.1) -- ++ (-1.75,0) -- ++ 
(-.4,-.7) -- ++ (-.4,.7) -- ++ (-1.75,0) -- ++ (-.4,-.7) -- ++ 
(-.4,.7) -- ++ (-1.75,0) -- cycle; 
 
\draw [rounded corners, very thick, black,style={,postaction={decorate},
        decoration={markings,mark=at position 0.54 with {\arrow[draw=black]{>}}}}]
(0,.1)    -- ++ (1.75,0) -- ++ (.4,.7) -- ++ (.4,-.7) -- ++
(1.75,0) -- ++ (.4,.7) -- ++ (.4,-.7) -- ++ (1.75,0);
\draw [rounded corners,  thick, black!65,style={,postaction={decorate},
        decoration={markings,mark=at position 0.54 with {\arrow{>}}}}]
(0,-.15) -- ++ (1.75,0) -- ++ (.4,.7) -- ++ (.4,-.7) -- ++ 
(1.75,0) -- ++ (.4,.7) -- ++ (.4,-.7) -- ++ (1.75,0);

\draw [rounded corners, very thick, black,style={,postaction={decorate},
        decoration={markings,mark=at position 0.46 with {\arrow{<}}}}]
(0,2.2) -- ++ (1.75,0) -- ++ (.4,-.7) -- ++ (.4,.7) -- ++ 
(1.75,0) -- ++ (.4,-.7) -- ++ (.4,.7)  -- ++ (1.75,0);
\draw [rounded corners, thick, black!65,style={,postaction={decorate},
        decoration={markings,mark=at position 0.46 with {\arrow{<}}}}]
(0,2.45) -- ++ (1.75,0) -- ++ (.4,-.7) -- ++ (.4, .7) -- ++ 
(1.75,0) -- ++ (.4,-.7) -- ++ (.4,.7)  -- ++ (1.75,0);

\draw [dashed, ultra thick] (2.15,.8) -- ++(0,.8);
\draw [dashed, ultra thick] (4.7,.8) -- ++(0,.8);

\draw [fill = darkgray] (2.15,.3) -- ++(.5,-.875) -- ++(-1,0)-- ++ (.5,.875);
\draw [fill = darkgray] (4.7 ,.3) -- ++(.5,-.875) -- ++(-1,0)-- ++ (.5,.875);
\draw [fill = darkgray] (2.15,2) -- ++(.5,.875) -- ++(-1,0)-- ++ (.5,-.875);
\draw [fill = darkgray] (4.7 ,2) -- ++(.5,.875) -- ++(-1,0)-- ++ (.5,-.875);
\end{tikzpicture}
\caption{Figure of an interferometer. Constrictions bring the edges together and causes tunnelling of charge from one edge to the other. Tunnelling occurs only between the inner edge states. The outer edges carry excitations of lower density Hall liquids, such as underlying fully filled Landau levels. We assume these are fully transmitted.}
\end{figure}
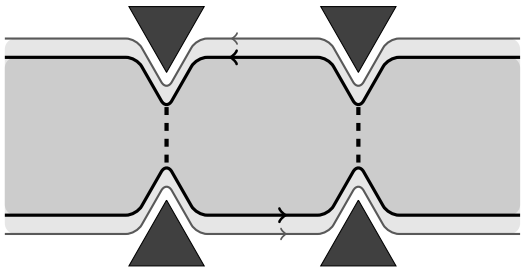\label{fig:interferometer}

We are primarily interested in the dependence of the interference current on the dynamical properties of the edge, such as the velocity of the edge channels and the applied voltage. Earlier work focused on edge states with a single characteristic velocity\citep{chamon1997} or edge states of specific quantum Hall candidates\citep{bishara2008,fidkowski2007} to obtain an expression for the interference current. We present here the more general case of an asymmetric interferometer, a generic number of edge channels with possibly different edge velocities and opposite chiralities, at both zero and finite temperatures. 

Our result is an analytic expression for the interference current in terms of a generalized hypergeometric function known as \emph{Carlson's R function} \citep{carlson1963}. This scaling function is closely related to the \emph{Lauricella hypergeometric function} \citep{lauricella1893,mathai2009}. This Lauricella function is a multivariable generalization of the Gauss hypergeometric function\citep{gradshteyn2007}, which is a function which enters the expression for the interference current for edge states described by a single velocity\citep{chamon1997}. Our expression generalizes this result for edge states consisting of an arbitrary number of decoupled channels described in the conformal limit. Each of these channels has its own corresponding velocity. We also find an expression for the interference current at zero temperature in terms of the \emph{confluent Lauricella hypergeometric function} \citep{mathai2009}, which is a multivariable generalization of the Bessel function of the first kind. Finally we obtain an expression for the two-point correlator of an anyon situated at the edge in the $(\omega, x)$-representation.

As a function of the voltage between the two edges the interference current behaves as a sum of decaying oscillations. The frequencies of these oscillations are determined by the edge lengths, edge velocities and the quasiparticle charge. For an antisymmetric interferometer this results in four frequencies appearing in the Fourier spectrum of the interference current as a function of the voltage. These four frequencies correspond to the possible combinations of one edge length and one edge velocity. Alternatively, we can fix the voltage and vary the length of one edge. This again results in oscillating behaviour with frequencies determined by the voltage, edge velocities and the quasiparticle charge. 

This behaviour of the interference current as a function of varying the edge length is relevant to experiments which measure the Aharonov-Bohm oscillations through application of a plunger gate\citep{willett2013a,willett2013b,an2011}. The plunger gate effectively deforms the area inside the interferometer through use of the Coulomb interaction. This deformation changes the Aharonov-Bohm phase of the tunnelling quasiparticles, which results in an oscillating interference current as a function of the side-gate voltage. The frequency of these oscillations, which we denote by $\phi_{AB}$, can be used to measure the charge of the tunnelling quasiparticle and the effect of quasiparticle braiding\citep{chamon1997,dassarma2006,stern2006,bonderson2006a,bonderson2006b}.

However, the change in area of the interferometer can also result in a change in the edge length, depending on the specific geometry of the interferometer. We show that for certain assumptions, such as the geometry of the device, this change in edge length results in additional oscillations in the interference current. When the change in edge length is large enough and linear with the side-gate voltage, then the interference current shows multiple oscillations characterized by the frequencies $\phi_{AB}$, $\phi_{AB} + \frac{QeV}{v_c h}$ and $\phi_{AB} + \frac{QeV}{v_n h}$. These shifted frequencies can be used to measure the edge velocity.

The paper is structured as follows. We start in Section~\ref{section:edge} with a discussion of the edge theory of a fractional quantum Hall phase. We specify the structure of the edge theory and quasiparticle operators, which is based on the decomposition in terms of a charged and neutral channel. 

In Section~\ref{sec:interferometer} we discuss the model Hamiltonian of the Fabry-P\'erot interferometer in terms of the quasiparticle operators and the corresponding linear response. This leads to an expression of the tunnelling current in terms of four-point correlators of the quasiparticle operators, as shown in Section~\ref{sec:linearresponse}. Specifically, the tunnelling current is given by evaluating the Fourier transform of these four-point correlators at the value of the Josephson frequency, see expression \eqref{tunnellingexpression}.

The four-point correlators depend on the precise nature of the edge state and they do not have a universal form. But as we show in Section~\ref{sec:correlators} the correlator has a leading dependence which does have a universal expression, which is a result of the conformal symmetry in the large system-size limit.

This leads to our main result in Section~\ref{sec:results}, which is the Fourier transform of the leading order expression of the four-point correlators at finite temperature, Eq.~\eqref{eqn:intereferencecorr}. This expression is given in terms of Carlson's $R$ function which acts as a modulating function. Since this function is somewhat obscure we summarize its properties in Appendix~\ref{app:carlsonr} and describe our method of computing the function, which is through its relation to the Lauricella function.

In Section~\ref{sec:specialcases} the main result is further explored for special cases, such as the zero temperature case. In Section~\ref{sec:plots} we plot the interference current and the $R$ function for a range of experimentally relevant parameters and analyse the result for a number of trial states for the $\nu = 5/2$, $\nu = 7/3$ and $\nu = 12/5$ plateaus. In general the $R$ function has a decaying oscillating behaviour. We show how the frequencies of these oscillations relate to the physical parameters of the system. In Section~\ref{sec:abphase} we discuss the relevance of our results to experiments involving the Aharonov-Bohm phase in the interferometer.

\section{Edge Theory} \label{section:edge}
A quantum Hall fluid is an example of a topological system \citep{wen1995,wen2004,nayak2008,wen2012}. The fluid has a mobility gap in the bulk of the system. Simultaneously, gapless states develop at the edge of the system where the confining potential crosses the Fermi level\citep{halperin1982,wen1992}. These gapless edge states are chiral and responsible for the transport properties of the fluid. 

The effective edge theory of the fractional quantum Hall effect can be seen as as a consequence of anomaly cancellation\citep{witten1989,frohlich1991a,wen1992,mooreread1991,balachandran1995,zee1995,fradkin1998,frohlich2001,georgiev2005,bieri2011}. The effective bulk theory of a quantum Hall fluid is a Chern-Simons theory; a topological gauge theory which describes the bulk of the system and develops an anomaly on the boundary where gauge invariance is broken. A dynamical edge theory forms, with the same anomaly, but opposite in sign. The combined bulk plus edge system is gauge invariant and anomaly free.

In the long-wavelength approximation the resulting edge theory is a chiral conformal field theory. The electron and quasiparticles of the theory are represented by local operators in this conformal field theory. The set of all local operators forms the chiral algebra\citep{frohlich2001}. By specifying the chiral algebra we zoom in on a candidate fractional quantum Hall state at some filling fraction $\nu$. To be a suitable candidate for a quantum Hall state, the chiral algebra needs to fulfil a number of conditions. These conditions include for instance the existence of an electron operator and the presence of a $U(1)$ symmetry. We assume such conditions are always satisfied in our discussion.

The $U(1)$ symmetry arises due to presence of the electric current. In the case of a Laughlin state the $U(1)$ symmetry is the full gauge symmetry of the bulk Chern-Simons theory. The corresponding edge theory is a chiral $\hat{u}(1)$ current algebra, also known as the chiral boson or chiral Luttinger liquid \citep{wen1990a,lee1991,floreanini1987,vondelft1998}. More complicated Abelian states involve the presence of multiple chiral bosons \citep{wen1992,blokwen1990,wen2004,boyarsky2009}. For non-Abelian quantum Hall states the $U(1)$ symmetry is also present, but only as a subgroup of a larger, more complicated gauge group\citep{mooreread1991,read1999,blokwen1990,fradkin1998,bonderson2008}. Following Ref.~\onlinecite{frohlich2001, boyarsky2009} we limit ourselves to those states described by a representation of an algebra which is formed by a direct product
\begin{align}
\mathcal{A} =  \mathcal{W}_n \otimes \hat{u}(1) \label{chiralalgebradecomposition}
\end{align}
Here $\mathcal{W}_n$ is the symmetry of the chiral algebra responsible for the non-Abelian nature of the system. Quasiparticle operators obey the same decomposition. We refer to the different terms in the product as the neutral and charged channel of the edge theory. Throughout the main text we mostly deal with a single charged and a single neutral channel, although we comment on the more general case of edge states with multiple modes.

Frequently, we deal with quantum Hall states which develop on top of one or multiple completely filled Landau levels, as is the case with candidate states for the filling fraction $\nu = 5/2$. These filled Landau levels form edge states as well, and for simplicity we assume these states completely decouple from the quantum Halls state of interest. In the presence of a point contact these filled edge states are assumed to fully transmit, meaning charge transfers only between the inner-most edges, see Figure~\ref{fig:interferometer}.

\subsection{Charged channel -- the chiral boson}
The action of the charged channel is that of the chiral boson \citep{wen1990a,lee1991,floreanini1987,levkivskyi2009,vondelft1998}. We consider a single edge with a right-moving chiral boson, held at a voltage bias $U$ in the gauge $a_x = 0$. The action is given by
\begin{align}
S_c &=  \frac{1}{4\pi} \int dt dx \left[ \partial_t\ph \partial_x\ph - v_c(\partial_x\ph)^2 \right] 	\nonumber\\
{}&+ \frac{\snu}{2\pi}  eU \int dt dx \left[\partial_x\ph \right]~.\label{gaugedaction}
\end{align}
The field is compactified by the identification \mbox{$\ph = \ph + 2\pi \nu$} and $v_c$ is the velocity of the channel. The field $\ph$ represents the charge density along the edge through the relation
\begin{align}
\rho(x) = \frac{\snu}{2\pi} \partial_x\ph~.
\end{align}
Quantization \citep{floreanini1987} results in the (non-local) equal-time commutation relations 
\begin{align}
[\ph(x), \ph(y)] &= -i \pi \text{sgn}(x-y)\nonumber \\
[\partial_x\ph(x), \ph(y)] &= -i 2\pi \delta(x-y)\label{commutationrelations}
\end{align}
with $\text{sgn}(x) = +1,0,-1$ for $x>0$, $x=0$ and $x<0$ respectively. Hamilton's equations of motion are given by
\begin{align}
(\partial_t-v_c\partial_x)\ph = -\snu eU~.
\end{align}
Differentiating with respect to $x$ shows the charge density $\rho$ is a conserved current as long as the edge is an equipotential, \mbox{$(\partial_t-v_c\partial_x)\rho = 0$}. The corresponding conserved charge is (up to normalization) identified as the total charge operator
\begin{align}
\mathcal{Q} = \int\rho(x) dx =\frac{\snu}{2\pi}\int\partial_x\ph (x) dx 
\end{align}
The Hamiltonian $K_{L,0,c}$ for a right moving edge held subject to the potential $U$ which follows from the action \eqref{gaugedaction} is
\begin{align}
K_{L,0,c} &= H_{L,0,c} - eU \mathcal{Q}~. \label{hamiltonianwithgauge}
\end{align}
The second term, $eU \mathcal{Q}$, is the coupling to the electrostatic potential. The first term corresponds to the Hamiltonian of the system in the absence of an external potential,
\begin{align}
 H_{L,0,c} &= \frac{v_c}{4\pi} \int dx  (\partial_x\ph)^2~.
\end{align}
The full Hamiltonian \eqref{hamiltonianwithgauge} is a generalization of the usual grand canonical Hamiltonian of the form \mbox{$K_0 = H_0 - \mu \hat{N}$}, with $\hat{N}$ the number operator. Instead of a number operator we use the charge operator. 

\subsection{Neutral channel and quasiparticles}\label{neutralchannel}
We do not explicitly specify the nature of the neutral channel, but only assume the decomposition \eqref{chiralalgebradecomposition}. What matters is that the full chiral algebra fixes the quasiparticle content of the theory and it comes equipped with consistent rules for fusion and braiding of these quasiparticles\citep{nayak2008,preskill1998}. Each quasiparticle is characterized by its conformal dimension and its fusion and braiding rules with respect to the remaining quasiparticles. This specifies its quantum dimension as well. 

A general quasiparticle operator factorizes as
\begin{align}
\psi^\dag(x,t) \propto \sigma(x,t) \otimes e^{i\frac{Q}{\snu}\ph(x,t)}~. \label{quasiparticleoperators} 
\end{align}
The exponentiated operator $e^{i\frac{Q}{\snu}\ph(x,t)}$ is normal ordered and corresponds to the charged channel, while $\sigma$ represents the neutral channel. The normalization factor depends on the regulators of the theory\citep{vondelft1998}. The operator is characterized by its conformal dimension, \mbox{$h_{\psi^\dag} = h_{\sigma} + h_{c}$}. The conformal dimension of the charged channel follows from the charge and the filling fraction, \mbox{$h_c = \frac{Q^2}{2\nu}$}. The commutation relations \eqref{commutationrelations} show that the operator obeys
\begin{align}
[\mathcal{Q}, \psi^\dag(x,t) ] = Q \psi^\dag(x,t) \label{commutationrelation}
\end{align}
and so the corresponding quasiparticle carries an electric charge $Qe$. 

For each quasiparticle a conjugate particle exists with opposite charge and the same conformal dimension\citep{preskill1998}. We set
\begin{align}
\psi(x,t) = \ov{\sigma}(x,t)\otimes e^{-i\frac{Q}{\snu}\ph(x,t)}~.
\end{align}
The operator $\ov\sigma$ is chosen such that the fusion product of $\sigma$ and $\ov\sigma$ contains the identity channel,
\begin{align}
\sigma \times \ov\sigma = \mathbf{1} + \ldots ~.\label{fusionsigma}
\end{align}
For non-Abelian quasiparticles we have, in general, multiple fusion channels. We assume that, for a given neutral mode, for each operator $\sigma$ there is a unique conjugate operator $\ov\sigma$ in the theory which obeys the fusion rule \eqref{fusionsigma}. This assumption is in fact a condition on the chiral algebra.

Finally, we also mention that the neutral channel traverses at some characteristic velocity $v_n$ and it is equipped with some neutral Hamiltonian, $H_n$, similar to the charged channel. However, the neutral channel does not couple to the electromagnetic field, and therefore no analogous coupling of a zero mode to the external electrostatic potential appears. Furthermore we assume the general situation in which \mbox{$v_n\neq v_c$}.

\section{Model of a Fabry-P\'erot interferometer}\label{sec:interferometer}
\subsection{Tunnelling Hamiltonian}

\begin{figure}
\begin{tikzpicture}
\draw [black,  thick] (0,0) -- ++ (6,0);
\draw [black,  thick] (0,2) -- ++ (6,0);

\draw [black, dashed, thick] (2,-.08) -- ++ (0,2.16); 
\draw [black, dashed, thick] (4.3,-.08) -- ++ (-.3,2.16); 

\draw [fill= white] (2,2) circle (.08);
\draw [fill= white] (4,2) circle (.08);
\draw [fill= white] (2,0) circle (.08);
\draw [fill= white] (4.3,0) circle (.08);

\draw [->, semithick] (1.8,.5) -- ++(0,1);
\draw [->, semithick] (4.,.5) -- ++(-.139,1);
\draw [->, dotted, very thick] (0,-.4) -- ++ (1,0);
\draw [->, dotted, very thick] (6,2.4) -- ++ (-1,0);

\node at (2,-.35) {$x_i$};
\node at (4,-.35) {$x_j$};

\node at (2,2.35) {$y_i$};
\node at (4,2.35) {$y_j$};

\node at (1.18,1) {$\V(x_i,y_i)$};
\node at (3.24,1) {$\V(x_j,y_j)$};

\end{tikzpicture}
\caption{Figure of an interferometer. Tunnelling of quasiparticles occurs at the point contacts, e.g. from $x_i$ to $y_i$ through the operator $\V_i$. The dotted arrows represent the direction of the edge currents, with a right moving current on the lower edge. In the text we set \mbox{$a=|y_j-y_i|$} and \mbox{$b=|x_j - x_i|$}.}
\end{figure}
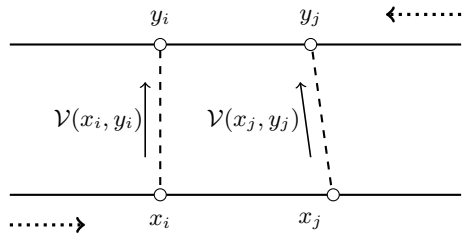

In this section we treat the basic idea behind the tunnelling formalism in a system of point contacts \citep{chamon1997,chamon1995,bishara2008,levkivskyi2009}. We consider a quantum Hall bar of infinite length at a uniform filling fraction $\nu$. The two edges, denoted as $\Sigma_{R/L}$, are disconnected and multiple constrictions are described by hopping terms allowing for the tunnelling of quasiparticles from one edge to the other. Here the subscript $L$ and $R$ denote the left (upper) and right moving (lower) edge of the system. For each edge we have an electric charge operator
\begin{align}
\mathcal{Q}_{R/L} = \int_{\Sigma_{R/L}} dx \rho_{R/L}(x)~.
\end{align}
We apply a voltage bias between the two edges, which is incorporated by fixing the electrostatic potentials $U_R$ and $U_L$ at the lower and upper edge respectively. The full Hamiltonian $K$ is given by
\begin{align}
K &= K_0 + \Ht~. \\
K_0 &= H_0 -e U_L \mathcal{Q}_L - eU_R\mathcal{Q}_R
\end{align}
Here $\Ht$ is the tunnelling Hamiltonian which is treated perturbatively with respect to $K_0$. The grand-canonical Hamiltonian $K_0$ consists of the terms coupling to the DC voltages through the charge operators and $H_0$. The Hamiltonian $H_0$ decomposes into the Hamiltonians for the decoupled left and right moving edges, $H_L$ and $H_R$. In addition $H_{L/R}$ describes both the charged and neutral channels $H_c$ and $H_n$ of each edge.

The tunnelling Hamiltonian $\Ht$ couples the edges through tunnelling of (quasi)holes and (quasi)electrons. For this we first introduce the tunnelling operators $\V$. We set $x$ and $y$ as the coordinates of the lower and upper edge respectively. A generic operator which tunnels a quasiparticle with charge \mbox{$e^*=Qe$ ($e>0$)} from the lower to the upper edge is then
\begin{align}
x \longrightarrow y ~: \qquad \V(x,y) = \psi^\dag(y)\psi(x)~.\label{tunnellingoperator}
\end{align}
The operators $\psi$ and $\psi^\dag$ are related as explained in Section~\ref{neutralchannel}. Similarly \mbox{$\V^\dag(x,y)=\psi^\dag(x)\psi(y)$} tunnels a quasi-particle from the upper to the lower edge.

We now consider a system of $N$ well-separated point contacts. Each point contact is approximated by a single tunnelling operator $\V(x_i,y_i)$ and a corresponding tunnelling coupling constant $\Gamma_i$. We have in the Schroedinger picture for the tunnelling Hamiltonian
\begin{align}
\Ht =  T + T^\dag \label{eq:tunnham}
\end{align}
where the $T$ operator is defined as
\begin{align}
T &= \sum_{i=1}^N \Gamma_i \V(x_i,y_i)~. \label{toperator}
\end{align}
Here the sum runs over the $N$ point contacts and $x_i$ and $y_i$ denote the coordinate of the $i$'th point contact on the lower and upper edge.

\subsection{Tunnelling Current}
The quantity of interest is the current running through the point contacts from one edge to the other, the so-called backscattering or tunnelling current $\langle\ib\rangle$. It is defined as the rate of change of the difference in electric charge of the edges, \mbox{$\frac{e}{2} \frac{d}{dt} \left(\mathcal{Q}_R - \mathcal{Q}_L\right)$}. Using the equations of motion for operators in the Heisenberg picture we have
\begin{align}
\ib & = -i\frac{e}{2}[\mathcal{Q}_R - \mathcal{Q}_L,T + T^\dag]
\end{align}
Here we used that the charge operators commute with the free Hamiltonian $H_0$ as the charge is conserved separately on each edge in the unperturbed system. The commutation relations \eqref{commutationrelation} imply \mbox{$\left[\mathcal{Q}_{R}, T \right]= -Q T = -\left[\mathcal{Q}_{L}, T \right]$}, and so we obtain
\begin{align}
\ib &= iQe (T - T^\dag)~.\label{app:backscatter}
\end{align}

\subsection{Linear Response} 
Initially, at some reference time $t_0$, the perturbation $\Ht$ is absent and the two edges are decoupled. At this initial time $t_0$ both edges are in thermal and (separate) chemical equilibrium with respect to the Hamiltonian $K_0$. The density matrix is given by
\begin{align}
w_0 \equiv w(t_0) &= e^{-\beta K_0}/Z~.\label{eqn:initialdensity}
\end{align}
Note that the external DC voltage is not treated perturbatively, but directly incorporated into the initial density matrix.

The perturbation $\Ht$ is adiabatically switched on at \mbox{$t>t_0$}, slowly driving the system out of equilibrium. The time evolution follows from the usual time evolution operator $U_K(t,t_0)$ which solves the Schroedinger equation with respect to $K$,
\begin{align}
i\partial_t U_{K}(t,t_0) = K U_{K}(t,t_0)\label{app:schroedinger}~.
\end{align}
In a perturbative approach\citep{rammer2007} we factorize the time evolution operator as \mbox{$U_K(t,t_0) = e^{-iK_0(t-t_0)} U_{\Ht(t')}(t,t_0)$}. Through \eqref{app:schroedinger} it follows that $U_{\Ht(t')}$ satisfies
\begin{align}
 i\partial_t U_{\Ht(t')}(t,t_0) &= \Ht(t') U_{\Ht(t')}(t,t_0)\nonumber \\
  \Ht(t) &\equiv e^{iK_0 t} \Ht e^{-iK_0 t}~.\label{eqn:tunnellingham}
\end{align}
The operator $U_{\Ht(t')}$ is expanded as a power series in the tunnelling coupling constants. At first order in $\Gamma(x,y)$ we have
\begin{align}
U_{\Ht(t')}(t,t_0) &= 1 - i \int_{t_0}^t dt' \Ht(t') + \ldots
\end{align}
The expectation value of an operator $\mathcal{O}$ is \mbox{$\la \mathcal{O}(t)\ra = \Tr{w_0 \mathcal{O}_K(t)}$} where $\mathcal{O}_K(t)$ is the Heisenberg representation of the operator
\begin{align}
\mathcal{O}_K(t) = U_{K}^\dag(t,t_0)\mathcal{O}_K(t_0)U_{K}(t,t_0)~.\label{eqn:expval}
\end{align}
At the initial time $t_0$ the perturbation $\Ht$ is absent, and so \mbox{$\mathcal{O}_K(t_0) = \mathcal{O}_{K_0}(t_0)$}. This identity together with the factorization of $U_{K}$ and expression \eqref{eqn:expval} results in
\begin{align}
\mathcal{O}_K(t) &= U_{\Ht(t')}^\dag(t,t_0) \mathcal{O}_{K_0}(t)U_{\Ht(t')}(t,t_0)
\end{align}
where we have defined
\begin{align}
\mathcal{O}_{K_0}(t) &= e^{iK_0 t} \mathcal{O}_{S}  e^{-iK_0 t} \label{eq:timedependence}
\end{align}
and $\mathcal{O}_{S}$ is the Schroedinger picture of the operator. When $U_{\Ht(t')}$ is expanded and we keep only the lowest order term we obtain for the expectation value
\begin{multline}
\la \mathcal{O}(t) \ra  = \la \mathcal{O}_{K_0}(t)\ra_0  \\
-i \int_{-\infty}^t dt'\la \left[\mathcal{O}_{K_0}(t),\Ht(t')\right]\ra_0 + \ldots ~.\label{def:expval2}
\end{multline}
Here \mbox{$\la \cdots \ra_0 \equiv \Tr{w_0\cdots}$} is the ensemble average with respect to the unperturbed thermal state of the edges, Eq.~\eqref{eqn:initialdensity}, and we have set $t_0\rightarrow -\infty$. We emphasize that this thermal state still includes the nonperturbative effect of the DC voltage. Expression \eqref{def:expval2} is the Kubo formula for the operator $\mathcal{O}$ with respect to the perturbation $\Ht$.

\subsection{Time evolution due to applied DC voltage and gauge invariance} \label{timeevolution}
In our approach a simplification is possible which elucidates some of the later manipulations. In the interaction picture the time dependence of the operators, \eqref{eq:timedependence}, follows from the edge Hamiltonian $K_0$, which includes the effect of the DC voltage bias. Since the charge operators $\mathcal{Q}_{R/L}$ commute with the Hamiltonian $H_0$ we can further factorize the time evolution operator as 
\begin{align*}
e^{-iK_0 t} = e^{-iH_0t}e^{ieU_L \mathcal{Q}_Lt}e^{i eU_R \mathcal{Q}_Rt} 
\end{align*}
The time evolution of the tunnelling operators $\V$ due to the applied bias voltage can now be made explicit. We use the commutation relations of the charge operators \eqref{commutationrelation} and the form of the tunnelling Hamiltonian \eqref{eq:tunnham}. This gives for the tunnelling operator $\V(x,y)$, 
\begin{multline}
\V_{K_0}(x,y,t) = e^{-ie(U_L \mathcal{Q}_L + U_R \mathcal{Q}_R)t } \\
\times \V_{H_0}(x,y,t) e^{ie(U_L \mathcal{Q}_L + U_R \mathcal{Q}_R)t}~.
\end{multline}
where \mbox{$\V_{H_0}(x,y,t) = e^{iH_0 t }\V_{S}(x,y,t) e^{-iH_0 t }$}. This is simplified further by using that when \mbox{$[\hat A,\hat B] = \alpha \hat B$} then \mbox{$e^{-i\beta \hat A}\hat B e^{i\beta \hat A} = \hat B e^{i\alpha\beta}$}. This gives
\begin{align}
\V_{K_0}(x,y,t)& =  e^{i\omega_Q t}\V_{H_0}(x,y,t)~.
\end{align}
Here we have defined $\omega_{Q} = Qe(U_R - U_L)/\hbar$, which is the Josephson frequency for a particle with charge $Qe$. The value of the charge $Q$ depends on the specific edge and quasiparticle under consideration. Typical experiments are carried out in the $0-100$ [$\mu$V] regime, corresponding to a Josephson frequency of $0-10^{10}$ [Hz].

We now have for the tunnelling Hamiltonian and current operator in the interaction picture
\begin{align}
\Ht(t) & =  T(t) + T^\dag(t)\label{tunnellinghamiltonian}\\
\ib(t) &\equiv e^{iK_0 t} \ib e^{-iK_0t} = iQe(T(t) - T^\dag(t))\\
T(t) &= \sum_i \Gamma_i e^{i \omega_{Q}t} \V(x_i,y_i,t) \label{eqn:tunope}\\
\V(x,y,t) &\equiv \V_{H_0}(x,y,t) = e^{iH_0 t} \V_S(x,y) e^{-iH_0 t} 
\end{align}
The effect of the DC voltage on the time evolution of the tunnelling operators $\V$ is completely captured by the phase factor $e^{i\omega_Q t}$.

The effective replacement of the tunnelling coupling constant $\Gamma$ by a time dependent one, $\Gamma\rightarrow\Gamma e^{i\omega_Q t}$, can also be obtained by performing a suitable gauge transformation; one that gauges the scalar potential of both edges $U$ to zero\citep{lee1991,chamon1997,wen2004}. Since the quasiparticle operators $\psi$ are charged, the tunnelling operators $T$ pick up a phase term $e^{i\omega_Q t}$ under this gauge transformation \citep{boyarsky2009}. 

\section{Linear response of the tunnelling current} \label{sec:linearresponse}
In the absence of the tunnelling Hamiltonian the tunnelling current vanishes, so $\langle \ib\rangle_0 = 0$. The linear response \eqref{def:expval2} for the tunnelling current \eqref{app:backscatter} is therefore
\begin{align*}
I_B(\omega_Q) \equiv \langle \ib(0)\rangle =  -i \int_{-\infty}^0 \md t \langle[\ib(0),\Ht(t)]\rangle_0~.
\end{align*}
We plug in the expressions for the tunnelling Hamiltonian \eqref{tunnellinghamiltonian} and the tunnelling current \eqref{app:backscatter} in terms of the tunnelling operators $T$. This gives
\begin{align}
I_B(\omega_Q) =Qe \int_{-\infty}^{\infty}  d t \langle[ T(t),T^\dag(0)]\rangle_0 ~.\label{CURREXPANSION}
\end{align}
The correlators of the type $\langle T T\rangle$ and $\langle T^\dag T^\dag\rangle$ vanish, as they describe overlaps of states with different electric charge. Furthermore, time translational invariance allows us to rewrite \mbox{$\la [T^\dag(t),T(0)]\ra_0 = -\la[T(-t),T^\dag(0)]\ra_0$}. A change of integration variable finally results in \eqref{CURREXPANSION}.

Next we express Eq.~\eqref{CURREXPANSION} in terms of the tunnelling operators $\V(x,y)$ by substituting Eq.~\eqref{eqn:tunope} for $T$. For that we introduce the tunnelling-tunnelling correlators between the $i$'th and $j$'th point contact
\begin{align}
G^{>}_{ij}(t)& = \langle \V(x_i,y_i,t) \V^\dag(x_j, y_j,0)\rangle_0 \label{eq:g-lesser} \nonumber \\
G^{<}_{ij}(t) &= \langle \V^\dag(x_j, y_j,0) \V(x_i,y_i,t)\rangle_0 ~.
\end{align}
This gives
\begin{align*}
\langle[ T(t),T^\dag(0)]\rangle_0 = \sum_{i,j}
\Gamma_i\Gamma_j^*  e^{i\omega_Q t}
\left[G^{>}_{ij}(t) - G^{<}_{ij}(t)\right]
\end{align*}
where $\Gamma_i$ is the tunnelling coupling constant of the $i$'th point contact. Inserting this into the expression for the tunnelling current, \eqref{CURREXPANSION}, the integration over time results in an expression in terms of the Fourier transform of the $G$-correlators
\begin{align}
I_B(\omega_Q) &=Qe \sum_{i,j} I_{ij}(\omega_Q)\label{currmaster} \\
I_{ij}(\omega_Q) &= \Gamma_i\Gamma_j^* \left[ G^{>}_{ij}(\omega_Q) - G^{<}_{ij}(\omega_Q)\right]~.  \nonumber 
\end{align}
A final simplification can be made by making use of complex conjugation, which relates \mbox{$G_{ij}^>(\omega) = \left[ G_{ji}^>(\omega)\right]^*$}, and the Kubo-Martin-Schwinger condition \citep{kadanoff1962}. The KMS condition applies to two-point equilibrium correlators and relates \mbox{$\langle \hat{A}(t)\hat{B}(0)\rangle_0 = \langle \hat{B}(0)\hat{A}(t+i\beta)\rangle_0$}. When applied to the tunnelling-tunnelling correlators $G$ we obtain
\begin{align}
G_{ij}^>(t) &= G_{ij}^<(t+i\beta)\nonumber \\
G_{ij}^>(\omega) &= e^{\beta\omega} G_{ij}^<(\omega), & T&\neq 0\label{eq:kmsrelation}
\end{align}
and so
\begin{multline}
I_{ij}(\omega_Q) +  I_{ji}(\omega_Q) 
= \\ 2|\Gamma_i\Gamma_j^{*}| ~\text{Re}\left[e^{i\tilde{\alpha}_{ij}}\left(1-e^{-\beta\omega_Q}\right)G^{>}_{ij}(\omega_Q)\right]~.
\end{multline}
Here we introduced $\tilde{\alpha}_{ij}$ as the relative phase between the coupling constants \mbox{$\Gamma_i\Gamma_j^{*} = |\Gamma_i\Gamma_j^{*}|e^{i\tilde{\alpha}_{ij}}$}. One contribution to this phase is the Aharonov-Bohm (AB) effect. Quasiparticles traversing along different point contacts enclose a different amount of flux, which causes an AB interference. This interference is independent of the applied voltage, provided the geometry is fixed as a function of this DC voltage \citep{halperin2011}; an assumption which does not always apply. We define \mbox{$\Phi_{Q} = h/(Qe)$} as the unit flux quantum for a particle with $Q$. The enclosed flux quanta between two point contacts $i$ and $j$ is then given by \mbox{$\Phi_{ij} = 2\pi(\Phi_i - \Phi_j)/\Phi_Q$}, where $\Phi_i$ is the total flux enclosed by the path of quasiparticle tunnelling along the $i$'th point contact. We have for the tunnelling current
\begin{multline}
I_B(\omega_Q) = 
Qe \Big(\sum_{i=1}^N |\Gamma_i|^2 \left(1-e^{-\beta\omega}\right)G^{>}_{ii}(\omega_Q) ~ + \\
2 \sum_{\substack{i<j}}^N 
|\Gamma_i\Gamma_j^*| ~ \text{Re}\left[e^{i\Phi_{ij}+i\alpha_{ij}}\left(1-e^{-\beta\omega_Q}\right)G^{>}_{ij}(\omega_Q)\right]\Big)\label{tunnellingexpression}
\end{multline}
where we replace \mbox{$\tilde{\alpha}_{ij} = \Phi_{ij} + \alpha_{ij}$} with $\alpha_{ij}$ the relative phase of the point contacts. The first summation is the sum of the tunnelling current through each point contact in the absence of any interference. All interference effects are encapsulated in the second summation, which we call the interference current.
\section{Correlators} \label{sec:correlators}
The tunnelling current is completely determined through the $G^>$ correlators. In terms of the quasiparticle operators~\eqref{quasiparticleoperators} these correlators are given by a product of four-point correlators, one correlator for each edge channel,
\begin{align}
G&^>_{ij}(t) =  \langle \psi^\dag(y_i, t)\psi(x_i, t) \psi^\dag(x_j,0) \psi(y_j,0)\rangle_0\nonumber \\
&= \langle e^{i\frac{Q}{\sqrt{\nu}}\ph(y_i,t)} e^{-i\frac{Q}{\sqrt{\nu}}\ph(x_i,t)}e^{i\frac{Q}{\sqrt{\nu}}\ph(x_j,0)}e^{-i\frac{Q}{\sqrt{\nu}}\ph(y_j,0)}\rangle_0\nonumber\\
&\times 
\langle \sigma(y_i,t) \ov\sigma(x_i,t)\sigma(x_j,0)\ov\sigma(y_j,0)\rangle_0~. \label{eq:correlator}
\end{align}

\subsection{The neutral mode and conformal blocks}\label{sec:neutralmodeprojection}
As it stands, the correlator for the neutral channel as stated in Eq.~\eqref{eq:correlator} is not uniquely defined. Non-Abelian quasiparticles span an internal, non-local Hilbert space. This is the realization of the non-Abelian statistical properties. In the language of conformal field theory\citep{difrancesco1995,mooreread1991} this internal space is identified as the space of conformal blocks and the correlator \eqref{eq:correlator} is a particular vector in this space. To identify this vector we first need to choose a basis in this space of conformal blocks\citep{fendley2006,fendley2007,bonderson2006a,bishara2008}

The conformal blocks in the correlator correspond to the different, possible fusion channels of the quasiparticles $\sigma$ and $\ov\sigma$. Symbolically the fusion rules of the fields $\sigma$ and $\ov\sigma$ are indicated as
\begin{align}
\sigma \times\ov\sigma = \sum_\theta N_{\sigma\ov\sigma}^\theta \theta~.
\end{align}
The sum runs over all primary states $\theta$ or quasiparticle types of the corresponding chiral algebra, including the vacuum state. The integers $N_{\sigma\ov\sigma}^\theta\geq 0$ are non-zero whenever a field $\theta$ is present in the fusion channel of $\sigma$ and $\ov\sigma$. This fusion rule signifies the possible outcomes when the two quasiparticles, $\sigma$ and $\ov\sigma$, are brought in close proximity. In this limit the quasiparticles fuse together and either form a new quasiparticle or they annihilate to the vacuum. Generally, a correlator such as Eq.~\eqref{eq:correlator} represents a superposition of possible fusion outcomes. This superposition is determined by the history of the system. 

More concrete, the correlator is a linear combination of conformal blocks, where each conformal block corresponds to an intermediate fusion channel. We write symbolically
\begin{align}
\langle \sigma\ov\sigma\sigma\ov\sigma\rangle = \sideset{}{'}\sum_{\theta} a_\theta \mathcal{E}_{\theta} ~.
\end{align}
The sum runs over those primary fields $\theta$ which appear in the fusion channel of $\sigma$ and $\ov\sigma$. With our choice of $\sigma$ and $\ov\sigma$ there is always one channel that corresponds to the identity or vacuum channel. The functions $\mathcal{E}_{\theta}$ are the conformal blocks and depend on the coordinates of the quasiparticles. The coefficients $a_\theta$ do not follow from the correlator itself but are determined by the history of the quasiparticles. 

This summation already assumes a certain order in which the quasiparticles are fused together when the correlator is evaluated. This order is essentially a choice of basis in the space of conformal blocks. A different order in which the quasiparticles are fused together corresponds to a different basis. The corresponding basis transformation that relates the two bases is determined by an object known as the $F$-matrix\citep{preskill1998}. To compute a four point correlator, such as $G^>$, we therefore need to choose a suitable basis of the space of conformal blocks for which the coefficients $a_\theta$ are known.

In the case of the $G^>$ correlators the quasiparticles are formed from the vacuum in pairs at a point contact. This means the initial fusion channel is the vacuum channel with respect to this basis. Put differently, the tunnelling operator $\V(x_i,y_i)$ creates a quasiparticle-anti-quasiparticle pair from the vacuum at the $i$'th point contact. It is therefore natural to use this basis, as the correlator is a single conformal block with respect to it,
\begin{align}
\langle \sigma(y_i,t) \ov\sigma(x_i,t)\sigma(x_j,0)\ov\sigma(y_j,0)\rangle = \mathcal{E}_{\text{vac}}~.
\end{align}
Pictorially we have\citep{fendley2007,bishara2008}
\begin{center}
\begin{picture}(130,30)
\put(-35,0){$\mathcal{E}_{\text{vac}} = $}
\put(-2,-2){$\sigma(y_i,t)$}
\put(43,20){$\ov\sigma(x_i,t)$}
\put(93,20){$\sigma(x_j,0)$}
\put(75,3){$\text{vac}$}
\put(30,0){\line(1,0){100}}
\put(55,0){\line(0,1){15}}
\put(105,0){\line(0,1){15}}
\put(135,-2){$\ov\sigma(y_j,0)$~.}
\end{picture}
\end{center}
\ \\
We now identified the vector in the space of conformal blocks corresponding to the $G^>$ correlator. However, a problem with this basis is that it makes use of fusing quasiparticles on different edges. The conformal block $\mathcal{G}_{\text{vac}}$ has components which corresponds to overlaps between the two edges. We need to project out these overlaps, before explicitly calculating the correlator\citep{fendley2007,bishara2008}. 

To perform this projection, we switch to a basis in which we first fuse together the quasiparticles on the same edge, followed by fusion of the these fusion products. We have
\begin{multline}
\langle \sigma(y_i,t) \ov\sigma(x_i,t)\sigma(x_j,0)\ov\sigma(y_j,0)\rangle = \\ a_{\text{vac}} \mathcal{F}_{\text{vac}} + \sideset{}{'}\sum_\theta a_\theta \mathcal{F}_{\theta}\label{conformalblock}
\end{multline}
where the basis is now given by
\begin{center}
\begin{picture}(130,30)
\put(-30,0){$\mathcal{F}_\theta = $}
\put(-2,-2){$\sigma(y_i,t)$}
\put(43,20){$\ov\sigma(y_j,0)$}
\put(93,20){$\sigma(x_i,t)$}
\put(80,3){$\theta$}
\put(30,0){\line(1,0){100}}
\put(55,0){\line(0,1){15}}
\put(105,0){\line(0,1){15}}
\put(135,-2){$\ov\sigma(x_j,0)$~.}
\end{picture}
\end{center}
\ \\
Note that the quasiparticles of each edge are paired together and in particular the vacuum channel is always present. The coefficients $a_{\text{vac}}$ and $a_\theta$ follow from the basis transformation which relates the blocks $\mathcal{F}$ and $\mathcal{E}$, and they are determined by the components of the $F$-matrix \citep{difrancesco1995}. In particular,
\begin{align}
a_{\text{vac}} = F\left[\begin{matrix}
\sigma &\ov\sigma \\
\ov\sigma & \sigma
\end{matrix}\right]_{\text{vac},\text{vac}}
\end{align}
All conformal blocks $\mathcal{F}_{\theta}$ as appearing in Eq.~\eqref{conformalblock} with a fusion channel different from the vacuum  ($\theta\neq \text{vac}$) vanish in the large system-size limit. This is the limit in which the size of each edge is taken to infinity, but where the distance between the point contacts is held fixed. The conformal block that remains corresponds to the vacuum channel, and it factorizes into a product of two-point correlators. We have \mbox{$\mathcal{E}_{\text{vac}} = a_{\text{vac}}~ \mathcal{F}_{\text{vac}} + \ldots$} and so
\begin{multline}
\langle \sigma(y_i,t) \ov\sigma(x_i,t)\sigma(x_j,0)\ov\sigma(y_j,0)\rangle_0 \\
=a_{\text{vac}} 
\langle \sigma(y_i,t) \ov\sigma(y_j,0)\rangle_0 \langle \ov\sigma(x_i,t)\sigma(x_j,0)\rangle_0 + \cdots~. \label{neutralmodecorrelator}
\end{multline}
The dots represent finite-size corrections which will be ignored. The two-point correlators are non-zero only when $\sigma$ and $\ov\sigma$ fuse to the identity, which is why we started with this assumption. What we have accomplished here is a disentangling of the edges. In this basis the projection onto well-separated edges can be performed. 

\subsection{Two-point correlator of a conformal field theory}\label{twopointcorrelator}
Two-point correlators in a conformal field theory are strongly constrained due to symmetries of the CFT\citep{belavin1984,difrancesco1995}. Following Ref.~\onlinecite{difrancesco1995} we first consider the two-point correlator of some quasiparticle (primary) operator $\mathcal{O}$,
\begin{align}
\langle  \mathcal{O}(z_1) \ov{\mathcal{O}}(z_2)\rangle &=
\frac{1 }{(z_1-z_2)^{g}}~.  \label{eq:corrcomplex}
\end{align}
Here the $z_i$ are complex coordinates of the plane, the parameter $g$ is called the algebraic decay and it is related to the scaling or conformal dimension $h$ of the field $\mathcal{O}$ and $\ov{\mathcal{O}}$ through $g = \frac{1}{2}h$. The fields $\mathcal{O}$ and $\ov{\mathcal{O}}$ must have the same conformal dimension or else the correlator vanishes identically.

A temperature is introduced through the conformal mapping of the plane to the cylinder, given by $z=\exp(2\pi i T w/v)$ where $T$ is the temperature of the system, $v$ is the velocity of the channel and we work in units where $k_B = \hbar =1$. The fields transform covariantly \citep{belavin1984,difrancesco1995} according to $\mathcal{O}(w) = \left(\frac{dz}{dw}\right)^{h}\mathcal{O}(z)$, which leads to
\begin{align}
\la \mathcal{O}(w_1) \ov{\mathcal{O}}(w_2)\ra 
& =  \frac{\left(\pi T/v\right)^{g}}{\sin(\pi T(w_1-w_2)/v)^{g}}~.
\end{align}
This transformation introduces a compactification of the coordinates, which is a geometric realization of the temperature. The Euclidean-time expression is obtained through the relation $w= v\tau \pm i x$. The sign choice determines the chirality of the CFT, and a minus sign $(-)$ results in a right moving channel. The real-time expression is obtained by performing a Wick rotation. The rotation introduces the infinitesimal regulator\citep{vondelft1998}, which we call $\delta$. We have $w_1-w_2 = \delta + i(vt_{12} - x_{12})$, where $t_{12} = t_1-t_2$. This results in
\begin{align}
\la \mathcal{O}(x,t) \ov{\mathcal{O}}(0,0)\ra &= \frac{\left(\pi T/v\right)^{g}}{\sin(\pi T(\delta + i(t+x/v)))^{g}}
\end{align}
This correlator is sometimes referred to as the greater Green's function. In the end the propagator is neatly summarized as
\begin{align}
\langle \mathcal{O}(x,t)& \ov{\mathcal{O}}(0,0)\rangle = v^{-g}P_{g}(t-x/v)\nonumber\\
P_{g}(t) &= \begin{cases}
\dfrac{1}{(\delta + it)^{g}} & T=0\\
\dfrac{\left(\pi T\right)^{g}}{  \sin(\pi T(\delta + it))^{g}} & T>0
\end{cases}\label{eqn:propagatorcase}
\end{align}
For completeness, we have included the zero-temperature limit. Putting everything together we obtain for the correlator of the neutral mode
\begin{multline}
\langle \sigma(y_i,t) \ov\sigma(x_i,t)\sigma(x_j,0)\ov\sigma(y_j,0)\rangle_0 = \\
a_{\text{vac}} 
\langle \sigma(y_i,t) \ov\sigma(y_j,0)\rangle_0 \langle \ov\sigma(x_i,t)\sigma(x_j,0)\rangle_0 + \cdots \approx 
\\
\frac{a_{\text{vac}} }{v^{2g_n}} P_{g_n}(t + \eta a/v_n) P_{g_n}(t - \eta b/v_n)\label{two-point-neutralmode}
\end{multline}
Here $a = |y_i - y_j|$, $b=|x_j-x_i|$, and $v_n$ and $g_n = \frac{1}{2} h_{\sigma}$ are the velocity and algebraic decay of the neutral channel. The parameter $\eta = \pm 1$ denotes the chirality of the neutral channel relative to the charged mode, with $(\eta = +)$ representing the same chirality. 

\subsection{Charged mode}
The charged mode is Abelian, meaning all fusion channels are unique and the projection onto disentangled edges can be done without having to perform a change of basis in the space of conformal blocks. The factorization is
\begin{align}
\langle e^{i\frac{Q}{\sqrt{\nu}}\ph(y_i,t)}& e^{-i\frac{Q}{\sqrt{\nu}}\ph(x_i,t)}e^{i\frac{Q}{\sqrt{\nu}}\ph(x_j,0)}e^{-i\frac{Q}{\sqrt{\nu}}\ph(y_j,0)}\rangle_0\nonumber \\
&= \langle  e^{i\frac{Q}{\sqrt{\nu}}\ph(y_i,t)} e^{-i\frac{Q}{\sqrt{\nu}}\ph(y_j,0)}\rangle_0 \nonumber\\
{}&\times \langle  e^{-i\frac{Q}{\sqrt{\nu}}\ph(x_i,t)}e^{i\frac{Q}{\sqrt{\nu}}\ph(x_j,0)}\rangle_0 + \cdots \label{abelianexpression}
\end{align}
The dots represent finite-size correction which we ignore. It is tempting to apply the results for the two-point correlator as motivated in Section~\ref{twopointcorrelator}. Although this leads to the correct result, it glosses over the fact that the correlator for the charged mode is taken with respect $K_0$ instead of the usual conformal Hamiltonian $H_0$.

At this stage we recall that we have already taken into account the effect of the coupling terms $-e U_L \mathcal{Q}_L - eU_R\mathcal{Q}_R$ on the time evolution of the quasiparticle operators $e^{i\frac{Q}{\sqrt{\nu}}\ph(y_i)}$. This was done in Section~\ref{timeevolution} and leads to the phase factor $e^{i\omega_Q t}$. However the coupling terms also appear in the initial density matrix, $e^{-\beta K_0}$, which can potentially give rise to extra contributions. We now show that these contributions are attributed to finite-size correction which we ignore. We do this by explicitly calculating the propagator with respect to $K_0$ using a mode expansion.

We assume a finite system length $L$ with periodic boundary conditions\citep{wen1990a,vondelft1998} on $\partial\ph$ and switch to a Fourier decomposition for $\partial_x\ph$,
\begin{align}
&\sqrt{\frac{L}{2\pi}}\partial_x\ph(x) = \rho_0 - i\sum_{k> 0} \sqrt{k} (e^{ikx} b_k- e^{-ikx} b_k^\dag)   \\
&\sqrt{\frac{L}{2\pi}}\ph(x) = \ph_0 + x \rho_0 - \sum_{k> 0} \frac{1}{\sqrt{k}} (e^{ikx} b_k+ e^{-ikx} b_k^\dag)~.\nonumber
\end{align}
%and the inverse relation
%\begin{align*}
%\rho_0 &=  \sqrt{\frac{2\pi}{L}} \int_{-L/2}^{L/2} dx \partial_x\ph  \\
%\ph_0 &=  \sqrt{\frac{2\pi}{L}} \int_{-L/2}^{L/2} dx \ph  \\
%b_k  &=  i\sqrt{\frac{2\pi}{L}} \int_{-L/2}^{L/2} dx \partial_x\ph e^{-ikx} \\
%b_k^\dag  &= -i \sqrt{\frac{2\pi}{L}} \int_{-L/2}^{L/2} dx \partial_x\ph e^{ikx}
%\end{align*}
Here $k ={2\pi n}/{L}$ with $n$ integer $>0$. The zero mode $\rho_0$ is proportional to charge operator $\rho_0 = \sqrt{{2\pi}/{\nu L}}\mathcal{Q}$. From the commutation relations \eqref{commutationrelations} we obtain for the modes
\begin{align}
[\ph_0, \rho_0] &= i & [b^\dag_k, b_k] =   1
\end{align}
and the remaining commutation relations vanish. Up to a constant term the normal ordered Hamiltonian is given by
\begin{align}
K_0 &=\frac{v_c}{2}\rho_0^2 -\snu e U \sqrt{\frac{L}{2\pi}}\rho_0 + v_c \sum_{k>0} k b^\dag_k b_k
\end{align}
From the Hamiltonian we derive the time evolution of $\ph(x,t)$ with respect to $K_0$. This gives
\begin{multline}
\ph(x,t) = -\snu e Ut + \sqrt{\frac{2\pi}{L}}\ph_0 + \sqrt{\frac{2\pi}{L}}  (x+v_c t) \rho_0 \\
- \sqrt{\frac{2\pi}{L}}\sum_{k> 0} \frac{1}{\sqrt{k}} (e^{ik(x+v_ct)} b_k+ e^{-ik(x+v_ct)} b_k^\dag)~.
\end{multline}
The Hilbert space is constructed in the usual way \citep{vondelft1998}, meaning we have a vacuum state $|0\rangle$ which satisfies $\rho_0|0\rangle = b_k|0\rangle = 0$. The modes $b_k^\dag$ with $k<0$ act as creation operators of momentum modes on this state. Since $\ph_0$ does not enter the Hamiltonian, the operator $\rho_0$ is conserved and can be diagonalized simultaneously with the Hamiltonian. The operator $e^{i\alpha \ph_0}$ creates a charged eigenstate, as it raises the eigenvalue of $\rho_0$ by $\alpha$, i.e. $\rho_0(e^{i\alpha \ph_0}|0\rangle) = \alpha (e^{i\alpha \ph_0}|0\rangle)$. In the large system-size limit the overlap between states with different charge $\alpha$ vanishes\citep{vondelft1998}. The allowed values for $\alpha$ which construct a state with non-zero norm depends on the chiral algebra. 

The normal ordered exponential operator is defined as
\begin{multline}
:e^{i\alpha \ph(x,t)} : = 
e^{-i\alpha \snu eU t} 
e^{i\alpha \sqrt{\frac{2\pi}{L}}\ph_0}e^{i\alpha\sqrt{\frac{2\pi}{L}}  (x+v_c t) \rho_0} \\
\times\prod_{k>0} e^{-i\alpha\sqrt{\frac{2\pi}{Lk}}e^{-ik(x+v_ct)} b_k^\dag} e^{-i\alpha\sqrt{\frac{2\pi}{Lk}}e^{ik(x+v_ct)} b_k}
\end{multline}
with the time evolution again due to $K_0$. We are interested in the two-point correlator,
\begin{align}
\frac{1}{Z}\Tr{e^{-\beta K_0} e^{i\alpha \ph(x,t)} e^{-i\alpha \ph(0,0)} }
\end{align}
Note again that the time evolution is with respect to $K_0$. The correlator is computed for each mode separately, since the different modes commute and so $Z = Z_0 \prod_k Z_k$. The contributions of the non-zero modes to this correlator is the same as in the zero-bias case, see e.g. Ref.~\onlinecite{vondelft1998} for details on this computation or Ref.~\onlinecite{ilieva2001} for an alternative derivation. We have
\begin{align}
\frac{1}{Z_k}\text{Tr}\big[&e^{-\beta v_c b^\dag_k b_k} 
e^{-i\alpha\sqrt{\frac{2\pi}{Lk}}e^{-ik(x+v_ct)} b_k^\dag} \nonumber\\
&e^{-i\alpha\sqrt{\frac{2\pi}{Lk}}e^{ik(x+v_ct)} b_k}  
e^{i\alpha\sqrt{\frac{2\pi}{Lk}} b_k^\dag} e^{i\alpha\sqrt{\frac{2\pi}{Lk}}b_k}\bigr] = \nonumber \\
&\qquad v^{-g_c} P_{g_c}(t+x/v_c) + \ldots
\end{align}

The normalization of the two-point correlator is unity in this limit. For the zero mode we first note that compactification of the boson $\ph = \ph + 2\pi\nu$ restricts the spectrum (eigenvalues) of $\rho_0$ to $\frac{k}{\sqrt{2\pi L}\nu}$ with $k$ integer. This gives for the correlator, 
\begin{align*}
\frac{1}{Z_0}&\text{Tr}[e^{-\beta (\frac{v_c}{2}\rho_0^2 -\snu e U \sqrt{\frac{L}{2\pi}}\rho_0)} e^{-i\alpha \snu eU t} 
e^{i\alpha \sqrt{\frac{2\pi}{L}}\ph_0}\\
&\times e^{i\alpha\sqrt{\frac{2\pi}{L}}  (x+v_c t) \rho_0}e^{-i\alpha \sqrt{\frac{2\pi}{L}}\ph_0}] = e^{-i\alpha \snu eU t}  + \ldots
% \nonumber\\
 %\frac{1}{Z_0}\text{Tr}[&e^{-\beta (\frac{v_c}{2}\rho_0^2 -\snu e U \sqrt{\frac{L}{2\pi}}\rho_0)} 
%e^{i\alpha \sqrt{\frac{2\pi}{L}}\ph_0}e^{-i\alpha \sqrt{\frac{2\pi}{L}}\ph_0}\nonumber\\
%&e^{i\alpha\sqrt{\frac{2\pi}{L}}  (x+v_c t) \rho_0}
%e^{-i\alpha \sqrt{\frac{2\pi}{L}} i\alpha\sqrt{\frac{2\pi}{L}}  (x+v_c t) [\rho_0, \ph_0]} ] =\nonumber\\
%\frac{1}{Z_0}\text{Tr}[&e^{-\beta (\frac{v_c}{2}\rho_0^2 -\snu e U \sqrt{\frac{L}{2\pi}}\rho_0) + i\alpha\sqrt{\frac{2\pi}{L}}  (x+v_c t) \rho_0}]\nonumber\\
% &\times e^{-i\alpha^2 (x+v_c t)\frac{2\pi}{L}} = e^{-i\alpha^2 (x+v_c t)\frac{2\pi}{L}} \times\nonumber\\
% &\frac{ \vartheta\left(-\frac{i \beta }{2\pi \nu }\frac{\snu e U}{2\pi}
%+\frac{\alpha  (x+v_c t)  }{2\pi L \nu}
%  ;  \frac{i \beta }{2\pi \nu }  \frac{ v_c}{\pi\nu L }\right) }
%{  \vartheta\left(-\frac{i \beta }{2\pi \nu }\frac{\snu e U}{2\pi}
%  ;  \frac{i \beta }{2\pi \nu }  \frac{ v_c}{\pi\nu L }\right)
% }=
% \nonumber \\
 \end{align*}
%In the last step we used,
%\begin{multline}
%\text{Tr}[e^{-A\rho_0^2 - B\rho_0}] =\\
% \sum_{n=-\infty}^\infty e^{-\frac{A}{2\pi L \nu^2}k^2 + \frac{B}{\sqrt{2\pi L}\nu} k}  =\vartheta(z; \tau)
%\end{multline} 
%where $\tau = \frac{i A}{(2\pi\nu)^2L }$ and $z = -\frac{i B}{2\pi\sqrt{2\pi L} \nu}$ and $\vartheta$ is the Jacobi-Theta function. 
%
The dots represent finite-size corrections. The effect of the zero mode coupling to the external potential on the two-point correlator is the phase factor $e^{-i\alpha \snu eU t}$. We obtain for the two-point correlator with the time evolution due to $K_0$
\begin{multline}
\frac{1}{Z}\Tr{e^{-\beta K_0} e^{i\alpha \ph(x,t)} e^{-i\alpha \ph(0,0)} } = \\ e^{-i\alpha \snu eU t} v^{-g_c} P_{g_c}(t+x/v_c)~.
\end{multline}
Finally, for the expression for the desired correlator appearing in Eq.~\eqref{abelianexpression} we strip off the phase factor $e^{-i\alpha \snu eU t}$, since we already extracted this through the manipulation performed in Section~\ref{timeevolution} -- it leads to the phase factor $e^{i\omega_Q t}$, which is already taken into consideration in the expression for the tunnelling current \eqref{tunnellingexpression}. We obtain
\begin{multline}
\langle e^{i\frac{Q}{\sqrt{\nu}}\ph(y_i,t)} e^{-i\frac{Q}{\sqrt{\nu}}\ph(x_i,t)}e^{i\frac{Q}{\sqrt{\nu}}\ph(x_j,0)}e^{-i\frac{Q}{\sqrt{\nu}}\ph(y_j,0)}\rangle_0 =\\
v^{-2g_c} P_{g_c}(t+a/v_c)P_{g_c}(t-b/v_c) + \ldots \label{two-point-chargedmode}
\end{multline}
This form matches with what we obtain by simply replacing the two-point correlators \eqref{abelianexpression} by the propagators $P_g$.

\subsection{Quasiparticle braiding and bulk-edge coupling} \label{sec:quasiparticlebraiding}
\begin{figure}
\begin{tikzpicture}

\draw [|->, semithick, black] (-4.3,.2) -- ++(0,1.8);
\node at (-4.6,1.9) {$t$};

\draw [thick, rotate=25, black!85] (-1.4,1.8) ellipse (1 and .5);
\fill [white] (-2.2,1.3) rectangle (-1.9,1.8);
\draw [thick, black!85,style={,postaction={decorate},
        decoration={markings,mark=at position 0.5 with {\arrow{>}}}}] (-2.05,.66) -- ++ (0,1.4);
\draw [thick, black!85] (-2.05,.3) -- ++ (0,-.3);
\draw [black!85, fill = black!85] (-2.85,.52) circle (.03);
\node at (-2.3,1.9) {$\chi$};

\draw [black!85, fill = black!85] (-1.18,1.54) circle (.03);

\draw [black!50, fill= gray!95,MyPersp] (3.18,1) circle (.1);

\draw [black!50,  thick,MyPersp] (0,0) -- ++ (6,0);
\draw [black!50,  thick,MyPersp] (0,2) -- ++ (6,0);

\draw [black!50, dotted, thick,MyPersp] (2,0) -- ++ (0,2.16); 
\draw [black!50, dotted, thick,MyPersp] (4,0) -- ++ (0,2.16); 
\draw [black!50, thick,MyPersp] (4,.08) -- ++ (-2,0); 
\draw [black!50, thick,MyPersp] (4,1.9) -- ++ (-2,0); 

\draw [black!85, fill = black!85,MyPersp] (4,1) circle (.04);
\draw [black!85, fill = black!85,MyPersp] (2,1) circle (.04);

\draw [fill= white,MyPersp] (2,1.93) circle (.1);
\draw [fill= white,MyPersp] (4,1.93) circle (.1);
\draw [fill= white,MyPersp] (2,0.07) circle (.1);
\draw [fill= white,MyPersp] (4,0.07) circle (.1);

\end{tikzpicture}
\caption{Quasiparticles inside the interferometer braid with quasiparticles tunnelling along the point contacts. At lowest order the effect of braiding is captured by the corresponding braiding diagram, which is determined from the topological quantum field theory.}\label{fig:quasiparticlebraiding}
\end{figure}
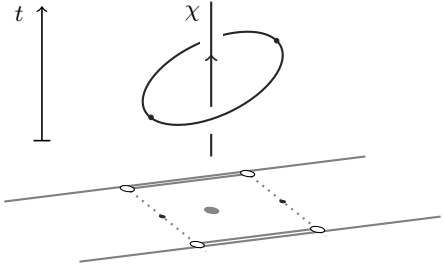

The correlators of the neutral and charged modes, equations~\eqref{two-point-neutralmode} and \eqref{two-point-chargedmode}, encapture part of the dynamical effects of quasiparticles traversing along the edge. The other dynamical contribution is due to the AB phase. In addition, there is also a topological contribution to the tunnelling current due to braiding of bulk and edge quasiparticles \citep{fradkin1998,dassarma2006,stern2006,bonderson2006a,bonderson2006b,bishara2008}. The correlator $G^>_{ij}$ is interpreted as the amplitude of the process in which a pair of quasiparticles $\psi$ and $\psi^\dag$ are created from the vacuum at the $j$'th point contact and annihilate to the vacuum at the $i$'th point contact. If one or multiple quasiparticles is present between these point contacts, the resulting amplitude contains a contribution coming from the quasiparticle braiding. This so-called matrix element is depicted in Figure~\ref{fig:quasiparticlebraiding}.

More generally, Figure~\ref{fig:quasiparticlebraiding} represents the expectation value of Wilson lines computed with respect to the full topological quantum field theory and it is fully determined in terms of the $S$-matrix\citep{bonderson2006b}. To fully determine this expectation value we require to specify the exact TQFT and the configuration and state of the bulk quasiparticles. In general the outcome is some complex valued function $\Aij$, bounded by $|\Aij|\leq 1$, which depends on the topological quantum number $\chi$ associated with the bulk anyons inside the interferometer. For the $G^>$ correlators we have
\begin{multline}
G^>_{ij} = a_{\text{vac}}~\Aij \times (\text{dynamical contributions}) \\
+ \text{finite-size effects}~.
\end{multline}
The effect of quasiparticle braiding is a topological effect, due to the statistical properties of the anyons. In the case of the Moore-Read state the effect leads to what is known as the even-odd effect\citep{fradkin1998,dassarma2006,stern2006,bonderson2006a,bonderson2006b,bishara2008}. When there are bulk quasiparticles present inside the interferometer and these quasiparticles are located far from the edge then the interference current due to tunnelling of the $e/4$ quasiparticle vanishes when the number of bulk quasiparticles is odd. When the number is even the interference current re-emerges.

The situation is more complicated when the bulk quasiparticles are close enough to the edge of the system. In that case the coupling between the bulk quasiparticles and edge degrees of freedom needs to be taken into account\citep{overbosch2007,overbosch2008,rosenow2008,rosenow2009,bishara2009a}. This coupling can induce tunnelling of the neutral degrees of freedom associated with the non-Abelian statistics from the bulk quasiparticles to the edge theory. One result is that even in the case of an even number of bulk quasiparticles located inside the interferometer this bulk-edge coupling can effectively flush out the interference current. Averaged over time the tunnelling of neutral degrees of freedom can greatly reduce the strength of the interference current. We do not take into account the effect of bulk-edge coupling, but we do note that this effect can be relevant to recent experiments\citep{an2011,willett2013a,willett2013b}

\subsection{\texorpdfstring{$G^>$}{G} correlators and its Fourier transform}
The expression for the $G^>$ correlator \eqref{eq:correlator} follows straightforwardly from combining the correlators for the neutral and charged mode, \eqref{two-point-neutralmode} and \eqref{two-point-chargedmode}.
\begin{align}
G_{ij}^>(t) = {}& a_{\text{vac}}\langle  \psi^\dag(y_i,t) \psi(y_j,0)\rangle  \langle \psi(x_i,t) \psi^\dag(x_j,0)\rangle +\ldots \nonumber\\
= {}& \Aij v_n^{-2g_n} v_c^{-2g_c} P_{g_c}(t+\tfrac{a}{v_c}) P_{g_n}(t+\eta\tfrac{a}{v_n})\nonumber\\
&\times P_{g_c}(t-\tfrac{b}{v_c}) P_{g_n}(t-\eta\tfrac{b}{v_n})~. \label{correlator}
\end{align}
Here we have defined $a= |y_{i}-y_j|$ and $b= |x_{i}-x_j|$ as the distance between the $i$'th and $j$'th point contact along the upper and lower edge respectively. Recall furthermore that $\eta=\pm$ represents the chirality of the neutral channel relative to the charged channel and $\Aij$ is due to braiding of quasiparticles.

For the tunnelling current we need the Fourier transform of the $G^>$ correlator. In Appendix~\ref{sec:integrals} we show how this Fourier transform is obtained. We first treat the contribution due to tunnelling along a single point contact, $G_{ii}^>$. The correlator for $G_{ii}^>(t)$ is independent of position, since $a=b=0$ in \eqref{correlator}. We have
\begin{align}
G_{ii}^>(t) = v_n^{-2g_n} v_c^{-2g_c} P_{2g}(t)~.
\end{align}
with $g = g_n+g_c$ twice the total scaling dimension of the quasiparticle. Using the result of \eqref{app:fourierprop} gives
\begin{align}
G_{ii}^>(\omega) = 
e^{\frac{\omega}{2T}} \frac{\left(2\pi T\right)^{2g-1} }{v_n^{2g_n} v_c^{2g_c}} B\left(g+i\frac{\omega}{2\pi T},g-i\frac{\omega}{2\pi T}\right) \label{eqn:ftcorr}
\end{align}
Here $B(x,y)$ is the Euler beta function and we have set the integral regulator $\delta$ to zero. We treat the zero temperature case later on. 

The expression for the more general case $(i\neq j)$ is more complicated. We write the Fourier transform of $G_{ij}^>(\omega)$ $(i\neq j)$ as the integral definition of Carlson's $R$ function\citep{carlson1963}. This is a multivariable generalization of the Gauss hypergeometric function. An alternative way of representing Carlson's $R$ function is through the fourth Lauricella hypergeometric function\citep{lauricella1893,carlson1963,mathai2009}, see also the appendix. We have cf. Eq.~\eqref{app:fouriercarlson} the following expression,
\begin{align}
G_{ij}^>(\omega_Q) = \Aij H_{ij}^{\text{mod}}(\omega_Q) G_{ii}^>(\omega_Q)\label{eqn:intereferencecorr}
\end{align}
where all (trajectory-dependent) interference effects are hidden away in the modulating function $H^{\text{mod}}$ given by
\begin{align}
H_{ij}^{\text{mod}}&(\omega_Q) = e^{\pi T (b-a)\left(\frac{g_c}{v_c} + \eta \frac{g_n}{v_n}\right)} \nonumber\\
\times R\Big( &g-i\frac{\omega_Q}{2\pi T}; \lbrace g_c,g_c,g_n,g_n\rbrace ;  \nonumber\\
&e^{-2\pi T\frac{a}{v_c}},e^{2\pi T\frac{b}{v_c}},e^{-\eta 2\pi T\frac{a}{v_n}},e^{\eta 2\pi T\frac{b}{v_n}} \Big)~. \label{modulatingfunction}
\end{align}
The $R$ function is treated extensively in Ref.~\onlinecite{carlson1963} and we have summarized some of its properties in Appendix~\ref{app:rfunction}. In particular, the order in which the parameters appear in \eqref{modulatingfunction} is relevant for its evaluation. Furthermore the $R$ function allows for certain transformations of the arguments, see also the appendix. Computation of the $R$ function is explained in Appendix~\ref{app:computingr} using results of Ref.~\onlinecite{laarhoven1988}. We mention one transformation in particular which is equation~\eqref{app:intrep}. Through this transformation we have the equivalent expression of the modulating function \eqref{modulatingfunction}. This transformation effectively switches \mbox{$a\leftrightarrow b$} in the expression of $H_{ij}^{\text{mod}}$ and simultaneously changes the sign of $\omega_Q$, 
\begin{align}
H_{ij}^{\text{mod}}&(\omega_Q) = e^{\pi T (a-b)\left(\frac{g_c}{v_c} + \eta \frac{g_n}{v_n}\right)} \nonumber\\
\times R\Big( &g+i\frac{\omega_Q}{2\pi T}; \lbrace g_c,g_c,g_n,g_n\rbrace ;  \nonumber\\
&e^{2\pi T\frac{a}{v_c}},e^{-2\pi T\frac{b}{v_c}},e^{\eta 2\pi T\frac{a}{v_n}},e^{-\eta 2\pi T\frac{b}{v_n}} \Big)~.
\end{align}
The function $\Aij$ describes the effect of possible quasiparticle braiding entering the correlator $G_{ij}$. Finally, in the expression for $G_{ij}^>(\omega)$ we recover the expression for the single point contact case, Eq.~\eqref{eqn:ftcorr}. The effect of the spatial separation of the point contacts, and thus all interference effects, is completely captured by the modulating function $H^{\text{mod}}$. 

Since the $R$ function is so closely related to the Lauricella function we also mention the form of the Fourier transform in terms of this function. The exact relation is explained in the appendix. Here we assume for simplicity a symmetric interferometer $b=a$. Assuming $\frac{a}{v_n} > \frac{a}{v_c}$ and the expression reduces to
\begin{multline}
H_{ij}^{\text{mod}} (\omega_Q)= e^{-2\pi T \frac{a}{v_n} g}
e^{i\omega_Q\frac{a}{v_n}} \\
\times F_D^{(3)}\big(g-i\frac{\omega_Q}{2\pi T};\lbrace g_c, g_c, g_n\rbrace ;2g;
1-e^{-2\pi T a(\frac{1}{v_n}+\frac{1}{v_c})},\\
1-e^{-2\pi T a(\frac{1}{v_n}-\frac{1}{v_c})},
1-e^{-4\pi T\frac{a}{v_n}}\big)~.
\end{multline}
This expression no longer depends on the chirality parameter $\eta = \pm$. The symmetric interferometer does not distinguish between chiral and anti-chiral edge states.

\section{Expression for the tunnelling current}\label{sec:results} 

We combine the expression for the tunnelling current \eqref{tunnellingexpression} with the expression for the Fourier transform of the correlators \eqref{eqn:ftcorr} and \eqref{eqn:intereferencecorr} and obtain the following expression
\begin{multline}
I_B\left(\omega_Q\right) = 2Qe \frac{\left(2\pi T\right)^{2g-1}}{v_n^{2g_n} v_c^{2g_c}} a_{\text{vac}}|\Gamma_{\text{eff}}(\omega_Q)|^2 \\
\times\sinh\left(\frac{\omega_Q}{2T}\right)
  B\left(g+i\frac{\omega_Q}{2\pi T},g-i\frac{\omega_Q}{2\pi T}\right)~.\label{eqn:tunn-curr}
\end{multline}
In the spirit of Ref.~\onlinecite{chamon1997} we have combined the effects due to interference into an effective tunnelling coupling amplitude,
\begin{multline}
|\Gamma_{\text{eff}}(\omega_Q)|^2 =\sum_{i=1}^N |\Gamma_i|^2 +\\
2 \sum_{\substack{i<j}}^N 
|\Gamma_i\Gamma_j| ~\text{Re}[\Aij e^{i\Phi_{ij} +i \alpha_{ij}}
H_{ij}^{\text{mod}}(\omega_Q)]~.
 \label{eqn:effectivecoupling}
\end{multline}
The function $H_{ij}^{\text{mod}}(\omega_Q)$ is given by \eqref{modulatingfunction}, which we call the modulating function. The $\Gamma_i$'s are the tunnelling coupling constant of the $i$'th point contact and $\alpha_{ij}$ is the relative phase between $\Gamma_i$ and $\Gamma_j$. We also introduced the Aharonov-Bohm phase $\Phi_{ij}$ and contributions due to quasiparticle braiding are attributed to $\Aij$. The disentangling of the conformal blocks results in the factor $a_{\text{vac}}$. Within our setup only $H_{ij}^{\text{mod}}$ depends explicitly on the external voltage bias. The tunnelling constants $\Gamma_i$ depend on the exact geometry of the interferometric device, and so the normalization of the current is not universal.

Expression~\eqref{eqn:effectivecoupling} for the tunnelling current is of the form,
\begin{align}
I_B(\omega_Q) &\sim \Bigl(\sum_i|\Gamma_i|^2  + \sum_{i<j}^N F^{\text{mod}}_{ij}\Bigr)I_{\text{single pc.}}(\omega) \label{generalformcurrent} \\
&\equiv I_0 + I_{\text{osc}} \nonumber\\
F^{\text{mod}}_{ij} &= 2 |\Gamma_i\Gamma_j|~ \text{Re}[\Aij e^{i\Phi_{ij} +i \alpha_{ij}}
H_{ij}^{\text{mod}}(\omega_Q)]\nonumber
\end{align}
All interference effects are contained in the function $F^{\text{mod}}_{ij}$, which  we call the interference term. We only deal with interference between pairs of point contacts; there are no interference effects involving tunnelling along three or more point contacts. This is due to the linear response approximation, which only takes into account effects up to order $|\Gamma_i\Gamma_j|$.

The modulating function $H_{ij}^{\text{mod}}$ is a function of the different energy scales, which are set by the temperature and voltage bias, and the scales associated with the velocity and distance between the point contacts,
\begin{align}
\left\{\frac{v_c}{a}, \frac{v_c}{b}, \frac{v_n}{a}, \frac{v_n}{b}, \frac{k_B T}{\hbar}, \omega_Q \right\}~.
\end{align}
These parameters enter the expression for the function $H_{ij}^{\text{mod}}$ through dimensionless combinations, and the function depends on the relative scales. The modulating function is, up to an exponential factor, determined by Carlson's $R$ function which we treat in the appendix. The $R$ function is a scaling function, which manifests itself through the homogeneous scaling transformation \eqref{app:homog}. It is computed through its relation to the Lauricella function and the corresponding Taylor series as described in Appendix~\ref{app:computingr}. 

The expression for the interference current is very general, and the price we pay for this is a limited intuition when it comes to the behaviour of the corresponding modulating function, $H^{\text{mod}}$. We can still summarize the general behaviour of the function as a function of the physical parameters. As a function of increasing voltage $\omega_Q$ the modulating function is the sum of multiple, decaying oscillations. The frequencies of the oscillations are determined by the edge lengths and edge velocities. The temperature and algebraic decay determines the relative amplitudes of the oscillations. In addition, for large temperatures $H^{\text{mod}}$ decays exponentially. Some of these features are proven analytically, while others follow empirically from numerical analyses.
\section{Special cases and generalizations}\label{sec:specialcases}
The main result of our work is the expression for the interference term \eqref{eqn:effectivecoupling} for the tunnelling current \eqref{eqn:tunn-curr} in terms of the $R$ function \eqref{modulatingfunction}. Here we consider several limits and generalizations, such as the zero temperature limit and other cases in which the expression for the modulating function $H^{\text{mod}}$ simplifies. This relates our results to earlier work \citep{chamon1997,bishara2008,fidkowski2007,bieri2012,bishara2009b}. We consider the generalization to more than two modes and discuss a relation to the two-point quasiparticle propagator.

Recall that we use $g_c$ and $g_n$ to denote the algebraic decay of the charged and neutral channel, and $v_c$ and $v_n$ the corresponding edge velocities and $\eta=\pm$ as the chirality of the neutral mode. In the case of three or more point contacts we obtain a modulating function for each unique pair of point contacts, $H_{ij}^{\text{mod}}$. We use $a$ and $b$ to denote the length between the $i$'th and $j$'th point contact along the upper and lower edge respectively. In principle, these lengths depends on $i$ and $j$, so $a = a_{ij}$ and $b = b_{ij}$. However, we omit these subscripts for the sake of breviety.

 Finally, we set \mbox{$g=g_c+g_n$} as the total algebraic decay and work in units where \mbox{$k_B = \hbar = 1$}.

\subsection{Zero temperature limit}
The zero temperature limit can be obtained in two ways. The first is to start with the expression for the propagator at zero temperature, \eqref{eqn:propagatorcase}, and follow the same steps as in the finite-temperature case by computing the Fourier transform of $G^>$ and $G^<$. Alternatively, we can start with the expression for the tunnelling current at finite temperature, and from here take the zero temperature limit. Both routes should produces the same result. 

However, the first route leads to an obstruction. When we attempt to determine the interference current we encounter the following integral (see also Appendix~\ref{app:zeroT})
\begin{multline}
G_{ij}^>(\omega)- G_{ji}^<(\omega) \sim  \int_{-\infty}^\infty dt e^{i\omega t} \times\\
\Big[ P_{g_c}(t+ \frac{a}{v_c})P_{g_n}(t+\eta \frac{a}{v_n}) P_{g_c}(t-\frac{b}{v_c})P_{g_n}(t-\eta \frac{b}{v_n}) \\
-
\bigl(t \longleftrightarrow -t\bigr)\Big]\label{zerot-integral}
\end{multline}
We do not know how to solve this integral with these general parameters and we are not aware of a reference in which it is treated. Therefore we proceed with the other route, in which we start with the finite temperature expression, Eq.~\eqref{eqn:tunn-curr}, and take the zero temperature limit. For the current we find the usual power-law behaviour times an effective coupling amplitude
\begin{multline}
I_B\left(\omega_Q\right) = \\
 2Qe\frac{2\pi }{v_n^{2g_n} v_c^{2g_c}}a_{\text{vac}}|\Gamma_{\text{eff}}(\omega_Q)|^2  |\omega_Q|^{2g-1} \text{sgn}(\omega_Q)~.
\end{multline}
The expression for $|\Gamma_{\text{eff}}(\omega_Q)|^2$ is the same as in the finite temperature case, Eq.~\eqref{eqn:effectivecoupling}, but with a different expression for the modulating function $H_{ij}^{\text{mod}}$. We have worked out the zero temperature limit of $H_{ij}^{\text{mod}}$ in Appendix~\ref{app:zeroT}. The result is
\begin{multline}
H_{ij}^{\text{mod}}(\omega) = e^{i\omega \frac{a}{v_n}} \Phi_2^{(3)}\Big(\lbrace g_c, g_c, g_n\rbrace; 2g; \\
-i\omega( \frac{a}{v_n} + \eta \frac{a}{v_c}), -i\omega (\frac{a}{v_n} - \eta \frac{b}{v_c}), -i\omega(\frac{a}{v_n} - \frac{b}{v_n})\Big)\label{modulatingzerot}
\end{multline}
The function $\Phi_2^{(3)}$ is the \emph{confluent} \emph{Lauricella} \emph{hypergeometric} \emph{function} of 3 variables \citep{mathai2009} and its series representation is given by Eq.~\eqref{app:confllaur}. It can be extended to include more than two modes per edge. This expression for $H_{ij}^{\text{mod}}$ should also be obtained by direct computation of the integral \eqref{zerot-integral}.

In the symmetric case where $a=b$ the modulating function reduces to
\begin{multline*}
H_{ij}^{\text{mod}}(\omega) = e^{i\omega \frac{a}{v_n}}  \times\\
 \Phi_2^{(2)}\Bigl(\lbrace g_c,  g_c\rbrace; 2g; 
-i\omega a ( \frac{1}{v_n} + \frac{1 }{v_c}), -i\omega a(\frac{1}{v_n} - \frac{1 }{v_n})\Bigr)~.
\end{multline*}
The function $\Phi_2^{(2)}$ is known as a Humbert confluent hypergeometric function of two variables \citep{erdelyi1953,gradshteyn2007}. For the symmetric interferometer the chirality of the neutral mode has no effect on the current.

\subsection{Equal velocities and chiralities}
For equal velocities and equal chiralities between the two channels we set $v=v_n=v_c$ and $\eta = +1$. This is effectively an edge with a single channel. The modulating function $H^{\text{mod}}_{ij}(\omega)$ reduces to the Gauss hypergeometric function.
\begin{align}
H^{\text{mod}}_{ij}&(\omega) \longrightarrow \nonumber \\
& e^{\pi Tg\frac{(b-a)}{v}} R\left( g- i\tfrac{\omega}{2\pi T}; g,g; e^{-2\pi T \frac{a}{v}}, e^{2\pi T \frac{b}{v}}\right)\nonumber \\
={}& e^{-\pi T \frac{(a+b)}{v}}e^{ i\omega\frac{b}{v}}\times  \nonumber \\
{}&
{}_2F_1 \left(g - i\tfrac{\omega}{2\pi T}; g; 2g; 1-e^{-2\pi T \frac{(a+b)}{v}}\right) \label{equalvelo}
\end{align}
The function ${}_2F_1$ is the Gauss hypergeometric function. For equal distances between the tunnelling points on both edges ($a=b$) the expression coincides with that found in Ref.~\onlinecite{chamon1997}, although to arrive at this expression we require some manipulations of the Gauss function. These can be found at e.g. [p. 1009] in Ref.~\onlinecite{gradshteyn2007}.
\begin{multline}
\text{eq. \eqref{equalvelo}~} = 2\pi\frac{\Gamma[2g]}{\Gamma[g]}\frac{e^{-2\pi gT\frac{a}{v}}}{\sinh(\tfrac{\omega}{2T})} \times\\
\text{Im}\left[\frac{e^{i\omega\frac{a}{v}}{}_2F_1 \left(g; g - i\tfrac{\omega}{2\pi T};1 - i\tfrac{\omega}{2\pi T}; e^{-4\pi T \frac{a}{v}}\right)}{\Gamma[g+i\tfrac{\omega}{2\pi T}]\Gamma[1-i\tfrac{\omega}{2\pi T}]}\right]~.\label{eq:othergauss}
\end{multline}
Expression \eqref{eq:othergauss} seems obscure and overly complicated in comparison with \eqref{equalvelo}. However, the representation \eqref{eq:othergauss} is an expansion in terms of the parameter $e^{-4\pi T \frac{a}{v}}$, which tends to zero for large temperature. In contrast, the expansion appearing in \eqref{equalvelo} is in terms of $1-e^{-2\pi T \frac{(a+b)}{v}}$, meaning the argument of the Gauss function tends to one for high temperatures. The exact behaviour of the Gauss function around unit argument is problematic, and leads to slow convergence of its Taylor series or even singular behaviour. In fact, the standard way of analysing the behaviour of ${}_2F_1(a,b;c;1-z)$ for $z \rightarrow 0$ is by first transforming it into a function of the form ${}_2F_1(a',b';c';z)$.
 
The zero temperature limit can again be obtained in two ways: by directly computing the Fourier transform or by taking the zero temperature limit of the finite temperature expression. In this case it is possible to determine the Fourier transform directly, which we have done in Appendix~\ref{zerot-singlemode}. We also show that this Fourier transform matches with the zero-temperature limit, demonstrating the equivalence of both routes. We find
\begin{multline}
H_{ij}^{\text{mod}}(\omega) = \Gamma\bigl[g+\frac{1}{2}\bigr]\Bigl(\frac{a+b}{4v}|\omega|\Bigr)^{\frac{1}{2}-g} \\
\times e^{-i\frac{a-b}{2v}\omega}  J_{g-\frac{1}{2}}\Bigl(\frac{a+b}{2v}|\omega|\Bigr) ~.\label{besselcurrent}
\end{multline}
Here $J_{g-\frac{1}{2}}(x)$ is the Bessel function of the first kind and this expression matches with what was found in Ref.~\onlinecite{chamon1997} when we set $a=b$. 
 
\subsection{Fast charged channel}
We consider the limit where the energy scales associated with the charged mode are far greater than the remaining energy scales,
\begin{align}
\frac{v_c}{a}, \frac{v_c}{b} \gg \frac{v_n}{a}, \frac{v_n}{b}, \frac{k_BT}{\hbar}, \omega_Q~.
\end{align}
The scales on the right hand side are that of the neutral mode, the temperature scale and the applied voltage bias.  In this limit the modulating function is
\begin{align}
H^{\text{mod}}_{ij}(\omega) &=
e^{\pi T (b-a) \frac{g_n}{v_n}} \times \nonumber\\
 R\Big( g-i\frac{\omega}{2\pi T};& \lbrace 2g_c,g_n,g_n\rbrace ;  1,e^{- 2\pi T\frac{a}{v_n}},e^{ 2\pi T\frac{b}{v_n}} \Big)~. \nonumber\\
&=e^{-\pi T (a+b) \frac{g_n}{v_n}}   
e^{-2\pi T\frac{b}{v_n}g_c}
e^{i\omega\frac{ b}{v_n}}
\times\nonumber\\
F_1\big(g-i\frac{\omega}{2\pi T};& \lbrace 2g_c,g_n\rbrace; 2g; 1-e^{-2\pi T\frac{b}{v_n}}, 1- e^{- 2\pi T\frac{a+b}{v_n}}\big)
\end{align}
On the final line we obtain the first Appell hypergeometric function of two variables\citep{gradshteyn2007}, $F_1(\alpha;\beta,\gamma;z_1,z_2)$. When $g_c = 0$ this function reduces to the case of a single edge mode Eq.~\eqref{equalvelo}, as expected.

\subsection{Large interferometer and high temperature limit}\label{sec:hightemperature}
For well separated contacts we consider large $a+b$. In Appendix~\ref{app:asymptotic} we show how this behaviour can be extracted from the integral. This limit suppresses the interference current exponentially according to
\begin{align}
H_{ij}^{\text{mod}} \longrightarrow \exp\Bigl(-\pi T (a+b)\sum_i \frac{g_i}{v_i}\Bigr) \label{suppression}
\end{align}
This is interpreted as an effective dephasing length
\begin{align}
L_T = \frac{\hbar}{\pi k_B T} \left[\sum_i \frac{g_i}{v_i}\right]^{-1}~.
\end{align}
Beyond this scale the interference current is suppressed as $I \propto e^{-(a+b) / L_T}$ with $a+b$ the total circumference of the interferometer. A similar analysis applies for high temperatures. Setting
\begin{align}
k_BT_L = \frac{1}{\pi(a+b)}\left[\sum_i \frac{g_i}{v_i}\right]^{-1}
\end{align}
and the interference signal vanishes as $I \propto e^{-T/ T_L}$. In general the decoherence effects are reduced by decreasing the temperature. See Ref.~\onlinecite{bishara2009b} for further discussion on energy scales and visibility of the interference signal.

\subsection{Asymmetric interferometer}
We now consider the limit where the length of one edge approaches zero. We set $a = 0$ which effectively merges the point contacts on one edge. We obtain
\begin{align}
H_{ij}^{\text{mod}}(\omega) &= e^{-\pi T a\left(\frac{g_c}{v_c} + \eta \frac{g_n}{v_n}\right)}\times \nonumber\\
 R\Big( g-i\frac{\omega}{2\pi T}&; \lbrace g,g_c,g_n\rbrace ; 1,e^{-2\pi T\frac{a}{v_c}},e^{-\eta 2\pi T\frac{a}{v_n}} \Big)
\end{align}
The reduction of this expression to the corresponding hypergeometric form depends on the sign of $\eta$. For \mbox{$\eta = +1$} we have
\begin{multline}
H_{ij}^{\text{mod}}(\omega) = e^{-\pi T a\left(\frac{g_c}{v_c} + \frac{g_n}{v_n}\right)} \times \\
 F_1\bigl(g-i\frac{\omega}{2\pi T}; g_c ,g_n;2g; 
 1-e^{- 2\pi T\frac{a}{v_c}}, 1-e^{-2\pi T\frac{a}{v_n}} \bigr) \nonumber
\end{multline}
while for $\eta = -1$ we obtain
\begin{multline}
H_{ij}^{\text{mod}}(\omega) =  e^{-\pi T a\left(\frac{g_c}{v_c} + \frac{g_n}{v_n}\right)}  e^{-2\pi T\frac{a}{v_n}g_c}
  e^{i\omega \frac{a}{v_n}}\times \\
 F_1\bigl(g-i\frac{\omega}{2\pi T}; g ,g_c;2g; 
 1-e^{- 2\pi T\frac{a}{v_n}}, 1-e^{-2\pi Ta(\frac{1}{v_c}+\frac{a}{v_n})} \bigr)~.\nonumber
\end{multline}
Here $F_1$ is the Appell hypergeometric function of two variables\citep{gradshteyn2007}. Using transformation properties of the Appell function, which can be found in e.g. [p. 1020] in Ref.~\onlinecite{gradshteyn2007}, for the case of $\eta=-1$ we can obtain a single expression for the $H_{ij}^{\text{mod}}(\omega)$ function given by
\begin{multline}
H_{ij}^{\text{mod}}(\omega) = e^{-\pi T a\left(\frac{g_c}{v_c} + \eta \frac{g_n}{v_n}\right)} \times \\
 F_1\bigl(g-i\frac{\omega}{2\pi T}; g_c ,g_n;2g; 
 1-e^{- 2\pi T\frac{a}{v_c}}, 1-e^{-2\pi T \eta \frac{a}{v_n}} \bigr) ~.
\end{multline}

\subsection{More than two channels}\label{sec:generalcase}
Our result for the interference current generalizes to edges which consists of more than one mode, all with different velocities. We can also include the possibility of different edge velocities for each edge. The edge velocity is not a topologically protected property of the edge mode, and its value(s) can depend on the exact geometric details of the corresponding device. 

The generalized result is obtained if we assume the modes decouple in a similar fashion as in the two-channel case or that the propagator factorizes along the lines of \eqref{correlator}. In these cases the correlator $G^>$ generalizes to
\begin{multline}
G^>_{ij}(t) = a_{\text{vac}}\Aij \prod_{i=1}^m v_{i,L}^{-g_i}v_{i,R}^{-g_i} P_{g_i}(t+\eta_i a/v_{i,L})\\
\times P_{g_i}(t-\eta_i b/v_{i,R})
\end{multline}
Here $g_i$ and $\eta_i$ are the algebraic decay and chirality of the $i$'th edge channel and $v_{i,R}$ and $v_{i,L}$ the velocity of the $i$'th edge mode on the lower and upper edge. The function $\Aij$ accounts for possible braiding of quasiparticles and $a_{\text{vac}}$ arises due to disentangling of the edges. The current is still determined by the Fourier transform, and the only change arises in the modulating function and the normalization of the tunnelling current which now involves all of the velocities, see \eqref{eqn:tunn-curr}. We have 
\begin{multline}
 H_{ij}^{\text{mod}}(\omega) = \exp\Bigl(\pi T\sum_{i=1}^m \eta_i g_i \bigl(\frac{b}{v_{i,R}} - \frac{a}{v_{i,L}}\bigr)\Bigr) \times\\
R\Big( g-i\frac{\omega}{2\pi T};
  \lbrace g_i,g_i\rbrace_{i=1}^m ; \lbrace e^{- 2\pi T \eta_i\frac{a}{v_{i,L}}},e^{ 2\pi T\eta_i\frac{b}{v_{i,R}}}\rbrace_{i=1}^m \Big)
\end{multline}
Here the arguments are ordered sets consisting of the algebraic decay and energy scales,
\begin{align*}
& \lbrace g_i,g_i\rbrace_{i=1}^m = \lbrace g_1,g_1,g_2,g_2,\cdots, g_m, g_m\rbrace \\
&\lbrace e^{-\eta_i 2\pi T \frac{a}{v_{i,L}}},e^{\eta_i 2\pi T\frac{b}{v_{i,R}}}\rbrace_{i=1}^m = \\
& ~\lbrace e^{-\eta_i 2\pi T \frac{a}{v_{1,L}}}, e^{\eta_i 2\pi T \frac{b}{v_{n,R}}},
\cdots, e^{-\eta_i 2\pi T \frac{a}{v_{n,L}}}, e^{\eta_i 2\pi T \frac{b}{v_{n,R}}}\rbrace ~.
\end{align*}
and $g = \sum_i g_i$. Computation of this function is similar to the two-channel case and covered in Appendix~\ref{app:computingr}.

\subsection{Two-point correlators and the \texorpdfstring{$R$}{R} function}
The tunnelling correlator $G^>_{ij}$ is constructed through projection onto decoupled edges, which results in a decomposition in terms of a product of two-point correlators. A simpler expression arises when we consider the the two-point propagator of a non-Abelian anyon on a single edge. We have in the conformal limit
\begin{multline}
X^>_{(m)}(t,x) \equiv \langle \psi^\dag(x,t) \psi(0,0)\rangle = \\ \prod_{i=1}^{m} |v_i|^{-g_i} P_{g_i}(t- x/v_i)~.
\end{multline}
Here we absorb the chirality of each mode into the velocity $v_i$, which can therefore take on negative values. The corresponding Fourier transform with respect to time is
\begin{multline}
X^>_{(m)}(\omega, x) =(2\pi T)^{g-1} H_{(m)}(\omega,x) \left[ \prod_{i=1}^{m} |v_i|^{-g_i}\right] \\
\times e^{\frac{\omega}{2T} - \omega\delta} B\left(\frac{g}{2} + i\frac{\omega}{2\pi T},\frac{g}{2} - i\frac{\omega}{2\pi T}\right)
\end{multline}
where $g=\sum_i g_i$ and all spatial dependence is captured by the function
\begin{multline}
H_{(m)}(\omega,x) = e^{-\pi T x\sum_i \frac{g_i}{v_i}} \\ \times R\left(\frac{g}{2} - i\frac{\omega}{2\pi T}; \lbrace g_i\rbrace_{i=1}^m ; 
\lbrace e^{-2\pi T \frac{x}{v_i}}\rbrace_{i=1}^m  \right)~.
\end{multline}
This is the equilibrium two-point quasiparticle propagator in a frequency-coordinate representation. 
% This result can be used to formulate a convolution integral of the modulating function, since the expression for $G^>_{ij}(\omega_Q)$ is a product of quasiparticle propagators, see equation \eqref{correlator}. We have
%\begin{align}
%H^{\text{mod}}_{ij}&(\omega_Q) = \frac{}{B\Bigl(g+i\frac{\omega_Q}{2\pi T}, g-i\frac{\omega_Q}{2\pi T}\Bigr)}\nonumber\\
%\times\int \frac{d\omega}{2\pi} 
%&B\Bigl(\frac{g}{2}+i\frac{\omega+\omega_Q}{2\pi T}, \frac{g}{2}-i\frac{\omega+\omega_Q}{2\pi T}\Bigr)  \nonumber\\
%\times &B\Bigl(\frac{g}{2}+i\frac{\omega-\omega_Q}{2\pi T}, \frac{g}{2}-i\frac{\omega-\omega_Q}{2\pi T}\Bigr)e^{\frac{\omega}{T} - 2\omega\delta} \nonumber\\
%\times &H_{(n=2)}\Bigl(\omega + \frac{\omega_Q}{2}, a\Bigr) H_{(n=2)}\Bigl(\omega - \frac{\omega_Q}{2}, -b\Bigr)~.
%\end{align}
\section{Plots of the modulating function and interference current}\label{sec:plots}
In this section we plot the modulating function and the corresponding interference current. Based on experiments  \citep{willett2009,willett2010,willett2012,willett2013a,willett2013b} we take the distance between two point contacts to be around $2$ [$\mu$m]. For the velocity no experimental data is available, but numerics \citep{hu2009} suggests a much faster velocity for the charged mode compared to the neutral mode on the order of $v_c/v_n \sim 10$ and $v_c \sim 10^{4}$ [m/s]. The applied voltage bias lies typically in the range of 10 to 50 [$\mu$V] and temperature ranges in the order of $10-25$ [mK]. We assume a lower temperature of $\sim 1$ [mK] as this significantly improves the rate of convergence of the series used to compute the expression for the tunnelling current, see Appendix~\ref{app:computingr}. 

In this section we are mainly interested in the behaviour of the $R$ function. The factor $\Aij e^{i\Phi_{ij}+i\alpha_{ij}}$ is due to quasiparticle braiding, the AB phase and the relative phase between the tunnelling amplitudes of the point contacts. They are assumed to be independent of the applied voltage bias and we set the total factor to unity. We comment on the AB effect in the next section.

The final parameters that need to be fixed are model-dependent, and correspond to the filling fraction $\nu$, the algebraic decay of the quasiparticle propagators $g_n$ and $g_c$, and the quasiparticle charge $Qe$. For a given edge state a renormalization group analysis predicts the quasiparticle with the lowest algebraic decay, $g_n+g_c$, to be the most relevant perturbation \citep{kanefisher1992,moon1993,fendley2007}. Quasiparticles with a larger algebraic decay are less relevant in the language of the renormalization group and we ignore their contributions in the plots. 

A second effect is that the effective magnetic length, $l_B^2 = \hbar /(QeB)$, is larger for quasiparticles with a smaller charge. The bare tunnelling matrix element depends on this length scale, and it is expected that a smaller charge correspond to larger matrix elements. Some trial states predict multiple quasiparticles with the same algebraic decay. In these cases the contributions to the tunnelling current is expected to arise from the quasiparticles with the smaller charge.

Computation of Carlson's $R$ function is not completely straightforward. The function admits a multivariable Taylor expansion or one can resort to numerical integration of the Fourier transform $G^>$. Using combinatoric results of Ref.~\onlinecite{laarhoven1988} the Taylor expansion is cast into a single summation, which we explain in Appendix~\ref{app:computingr}. We use this expansion for computing the $R$ function. 

For physically relevant values of the input parameters both the series expansion and numerical integration schemes converge very slowly. In particular a higher temperature scale reduces the convergence rate significantly. We apply a series acceleration using the CNCT method \citep{ulrich1999} to partially remedy this problem, see also the appendix. However, even the CNCT method is not practical for high temperatures and to our knowledge an efficient numerical scheme is still lacking.

Due to these convergence problems we are not able to compute the $R$ function for all ranges of the physical parameters. For instance, we mostly assume temperatures of 1 or even 0 [mK] for the sake of convergence of the modulating function. We also plot the modulating function over a range of the source-drain voltage which lies outside of what is reached in experiments. We have chosen for this range as we want to demonstrate the nontrivial behaviour of the $R$ function over a greater voltage range. Finally, we point out that currently the edge velocities have not been measured and it is possible that the values used in the plots are inaccurate. 

\subsection{The tunnelling current without interference}
Before we provide some plots of the modulating function and the interference current we first discuss the general behaviour of the tunnelling current in the absence of interference\citep{wen1990a}. This expression also enters the result for the total interference current. It is given by Eq.~\eqref{eqn:tunn-curr} in terms of the Euler beta and the hyperbolic sine function, with the tunnelling amplitude held constant. The tunnelling current is characterized by the total algebraic decay, and we discuss two particular values of $g$ for which the function simplifies. These follow from the properties of the gamma function\citep{erdelyi1953}
\begin{align}
g &= \frac{1}{2}~: & I_B(\omega_Q) &\propto \tanh\Bigl(\frac{\omega_Q}{2 T}\Bigr) \nonumber\\
g &= 1~: & I_B(\omega_Q) &\propto   T\omega_Q  
\end{align}
In the limit of $\omega_Q\rightarrow \infty$ and $g=\frac{1}{2}$ the expression for the tunnelling current approaches a constant value, while $g=1$ grows linearly with $\omega_Q$. For the remaining cases the current decays to zero for $g<\frac{1}{2}$, grows sublinearly for $\frac{1}{2} < g < 1$, and grows superlinearly for $g> 1$. Finally, at zero temperature the expression for the current follows the power law behaviour 
\begin{align}
I_B\propto |\omega_Q |^{2g-1} \text{sgn}(\omega_Q)
\end{align}
while for high temperatures the function follows
\begin{align}
I_B\propto \omega_Q T^{2g-2}~.
\end{align}

\subsection{The tunnelling current with interference}
 
\begin{figure}[!tb]
\includegraphics[width=.5\textwidth]{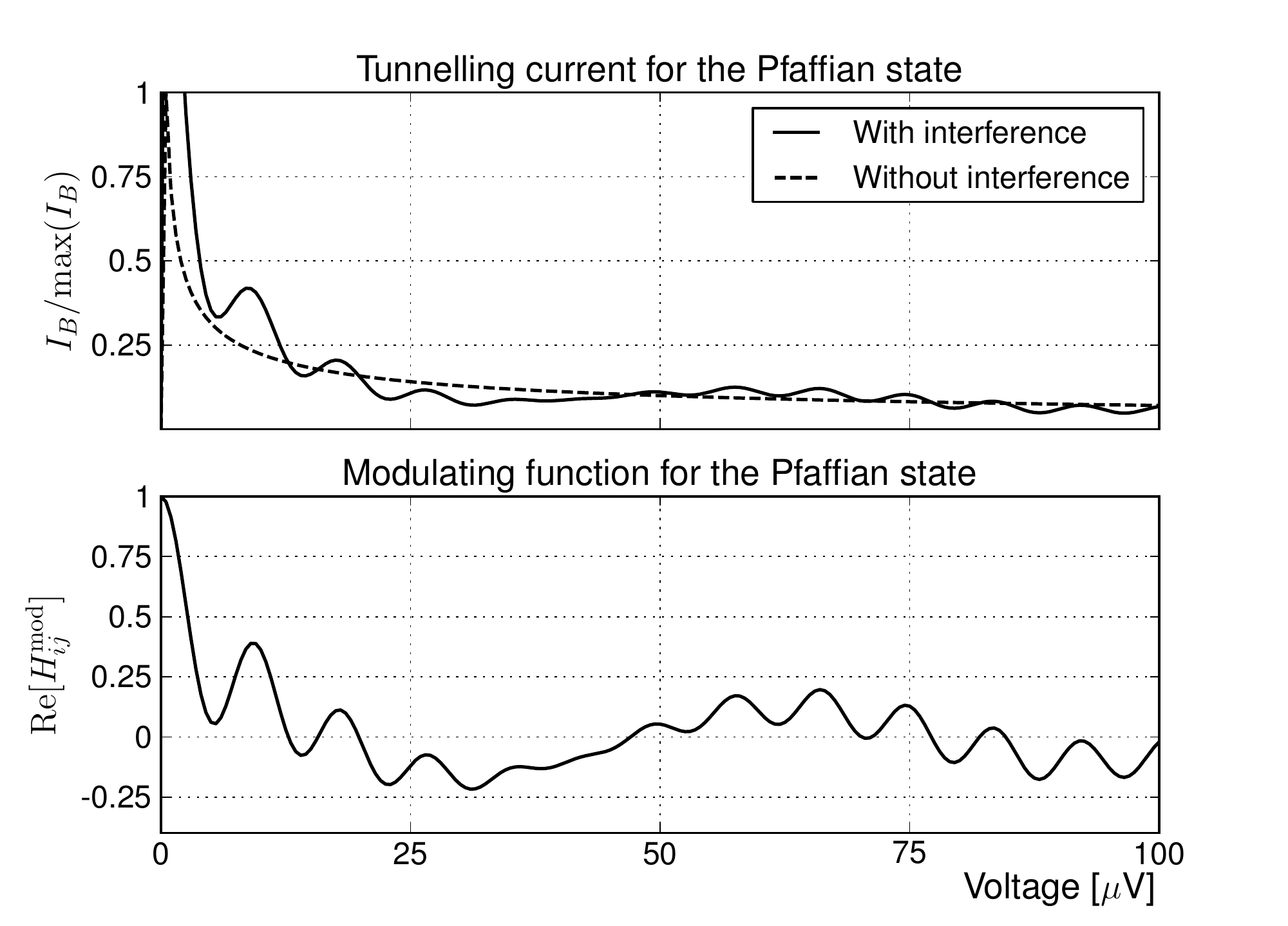}
\caption{The tunnelling current through an interferometer with and without the modulating function. The current is normalized by the maximal value of the tunnelling current without interference ($\text{max}(I_B(H^{\text{mod}} = 0))$). The quasiparticle is the $Qe = e/4$ QP of the Pfaffian state which has $g_n  = g_c = \frac{1}{8}$. The remaining parameters are $v_c =7\cdot 10^3$ [m/s], $v_n = 1\cdot 10^3$ [m/s], $T = 1$ [mK], $a = 2.0$ [$\mu$m] and $b = 1.8$ [$\mu$m]. All coupling constants are equal. Interference effects due to braiding with bulk quasiparticles is absent, i.e. we set $\Aij = 1$.}\label{fig:rfunctiongeneral}
\end{figure}

The upper panel of figure \ref{fig:rfunctiongeneral} is a plot of the total tunnelling current (eq. \eqref{eqn:tunn-curr}) with and without interference for the case of the Moore-Read / Pfaffian quantum Hall trial state\citep{mooreread1991,greiter1991} for the $\nu=5/2$ plateau. The lower panel of figure \ref{fig:rfunctiongeneral} is a plot of the corresponding modulating function $\text{Re}[H^{\text{mod}}_{ij}]$, given by equations~\eqref{eqn:effectivecoupling} and \eqref{modulatingfunction}. The parameters for the  set $g_c=g_n=1/8$ and $Qe = e/4$. This result is also analysed in Ref.~\onlinecite{bishara2008}. See the figure caption for the exact values of all parameters. 

The normalization of the current, which is the prefactor appearing in expression \eqref{eqn:tunn-curr}, contains the tunnelling coupling constants $\Gamma_i$. These factors are non-universal, meaning the normalization of the current is non-universal as well. In Figure~\ref{fig:rfunctiongeneral} the current without interference is normalized by its maximum value, \mbox{$I_B(\text{no interference})/ \text{max}(I_B(\text{no interference}))$}. The normalization of the current with interference is chosen such that when the two currents cross in Figure~\ref{fig:rfunctiongeneral} the modulating function vanishes $H^{\text{mod}} = 0$.

\subsection{Voltage and geometry dependent oscillations and corresponding frequencies} \label{sec:frequencyanalysis}
\begin{figure}[tb]
\includegraphics[width=.5\textwidth]{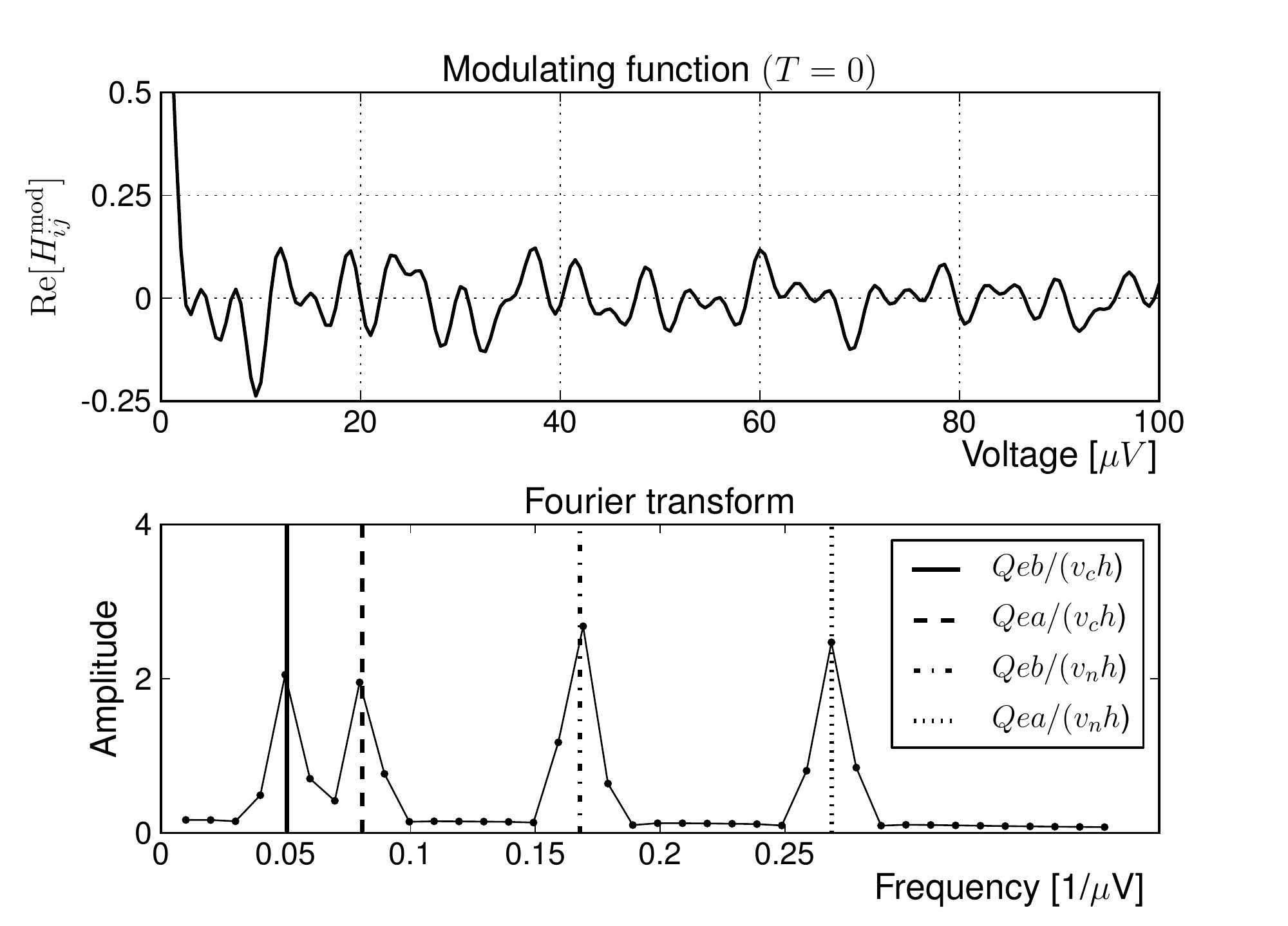}
\caption{Upper panel: the modulating function for a fictitious quasiparticle with $g_c = 1/8$ and $g_n = 1/6$ at $T=0$ [K] as a function of the voltage. Furthermore, 
 $Qe = e/4$, $v_c = 3\cdot 10^3$ [m/s], $v_n = 9\cdot 10^2$ [m/s], $a=4.0$ [$\mu$m] and $b=2.5$ [$\mu$m]. Lower panel: the corresponding (windowed) Fourier transform. The vertical lines represent the predicted frequency components given by $Qex_i/(v_i h)$.}\label{fig:ft_analysis_voltage}
\end{figure}
The modulating function $\text{Re}[H^{\text{mod}}_{ij}]$ shows multiple oscillations and decays when $V\rightarrow\infty$, see the lower panel of figure \ref{fig:rfunctiongeneral} and the upper panel of figure \ref{fig:ft_analysis_voltage}. A numerical analysis (figure \ref{fig:ft_analysis_voltage}) shows that for an asymmetric interferometer ($a\neq b$) and two different edge velocities the modulating function consists of four oscillating signals with frequencies
\begin{align}
f_{x_j, v_i} = \frac{Qex_j}{v_i h}\label{frequenciesforvaryingvoltage}
\end{align}
where $x_j = a,b$ and $v_j = v_c,v_n$. These frequencies can be extracted from the $(x,t)$ representation of the tunnelling-tunnelling correlators $G^>_{ij}(t)$, see Eq.~\eqref{correlator}. The peak values appearing in this correlator correspond to the frequencies \eqref{frequenciesforvaryingvoltage}. We also find that the frequencies are independent of the temperature and algebraic decay -- these parameters only influence the total and relative amplitudes of the oscillations. In the limit of a symmetric interferometer ($a\approx b$) the number of contributing oscillating frequencies drops from four to two, since $f_{a,v_i} \approx f_{b,v_i}$. In this regime the two oscillations form a modulating signal with 'fast' and 'slow' frequencies $\frac{Qea}{2h}(\frac{1}{v_n} \pm \frac{1}{v_c})$, which was also found in Ref.~\onlinecite{bishara2008}. It is also possible that the edge velocity for each channel is different on opposite edges. In that case we still have four different frequencies in the Fourier spectrum, even in the case of a symmetric interferometer.

\begin{figure}[bt]
\includegraphics[width=.5\textwidth]{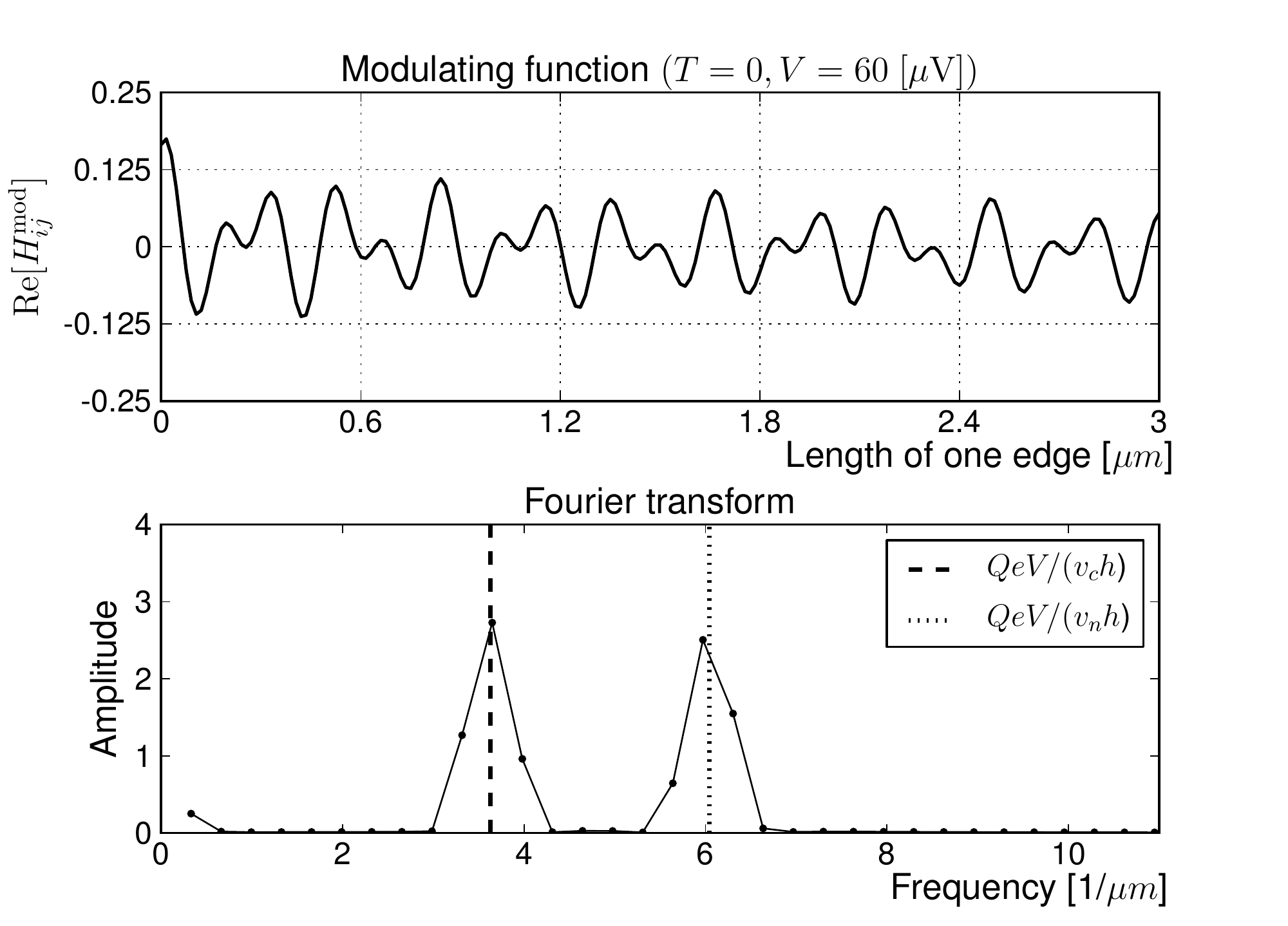}
\caption{Upper panel: the modulating function for a quasiparticle with $g_c = g_n = 1/8$ at $T=0$ [K] while varying the distance of one edge ($0\leq a \leq 3.0$ [$\mu$m]). The applied voltage is kept constant at $V=60$ [$\mu$V]. Furthermore $Qe=e/4$, $v_c = 1\cdot 10^3$ [m/s], $v_n=6\cdot 10^2$ [m/s] and $b=2.5$ [$\mu$m]. Lower panel: the corresponding windowed Fourier transform with peaks at the predicted frequencies $QeV/(v_c h)$ and $QeV/(v_n h)$.}\label{fig:ft_analysis_edge}
\end{figure}

The second analysis we perform looks at the oscillating behavior of the modulating function as a function of the length of one edge, while keeping all other parameters fixed. These are the oscillations in $H^{\text{mod}}_{ij}$ when $a$ is varied. The frequencies of these oscillations are obtained through a numerical Fourier transform, see figure \ref{fig:ft_analysis_edge}. The modulating function shows a similar decaying, oscillating behaviour as in the case of varying the voltage, with frequencies given by
\begin{align}
f_{v_i} = \frac{QeV}{v_i h}~.\label{frequenciesforvaryingedgelength}
\end{align}
Since the other edge length $b$ is kept constant we observe only two contributing frequencies. For the case of a single edge velocity these frequencies can be extracted from the expression of the current at zero temperature, \eqref{besselcurrent}, by making use of properties of the Bessel function. However, we are not able to extract the frequencies in expression \eqref{frequenciesforvaryingedgelength} analytically for the more general case. We suspect that such a result can be obtained from the $R$ function through an asymptotic expansion, which we leave as an open problem. We expect that these results carry over to the more general case of several edge channels and different velocities, see Section~\ref{sec:generalcase}.

\subsection{Effect of temperature}
\begin{figure}[tb]
\includegraphics[width=.5\textwidth]{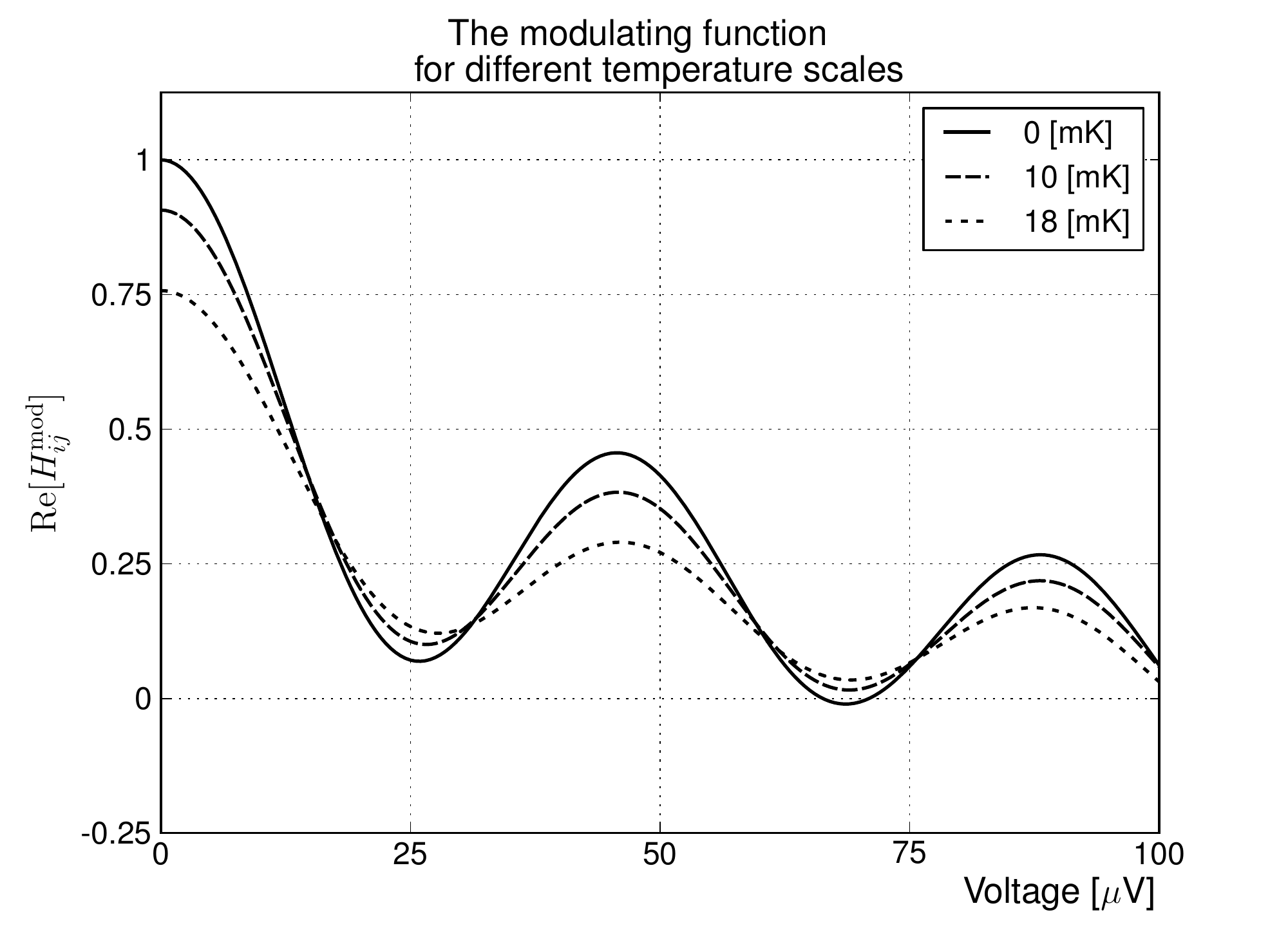}
\caption{The modulating function at three different temperatures, keeping all other parameters fixed. The tunnelling quasiparticle is the $e/4$ quasiparticle of the Pfaffian state with $g_c = g_n = \frac{1}{8}$. The remaining parameters are $v_c = 9\cdot 10^4$ [m/s], $v_n = 9\cdot 10^3$ [m/s], $a = 3.5$ [$\mu$m] and $b = 3.5$ [$\mu$m].}\label{fig:plots-dependencies}
\end{figure}

In figure  \ref{fig:plots-dependencies} the modulating function is plotted for the temperatures 0 [mK], 10 [mK] and 18 [mK]. The $T = 0$ case is computed using the confluent Lauricella hypergeometric function, as explained in Appendix~\ref{app:zeroT}. Computation of the confluent Lauricella function is very similar to the finite temperature case. 

The convergence of the series representation used to compute $R$ function becomes progressively worse for temperature scales larger compared to the remaining energy scales. Computing the $R$ function using the series expansion in this regime becomes impracticable, even when we employ a series acceleration. This type of slow convergence is similar to that exhibited by the Gauss function ${}_2F_1(a,b;z)$ when $|z|\rightarrow 1$. For the Gauss function a set of linear transformations exist which allow one to avoid this $|z| = 1$ singularity \citep{gradshteyn2007}, see also Eq.~\eqref{eq:othergauss} and the corresponding discussion. We are not aware of a generalized type of transformations applicable to the $R$ function. Due to this slow converge for high temperatures we frequently put $T=0$ or $T=1$ [mK] throughout this work.

From figure \ref{fig:plots-dependencies} we observe that the oscillations are independent of the temperature. Other numerical analyses suggest that this remains valid for other physical parameters as well. Instead the temperature appears to be responsible for the relative and absolute amplitudes of the oscillations which were studied in the previous section. In particular, higher temperatures cause an exponential suppression of the function as was found in Section~\ref{sec:hightemperature}. Lower temperature increase the visibility of the interference signal. 

The experiments are typically performed at temperatures of $T = 25$ [mK] or lower. Numerically, we have not been able to reach temperatures higher than $T=20$ [mK]. We expect that the behaviour of the modulating function as predicted by our results remains valid in this regime. In particular we expect that the frequencies of the oscillations \eqref{frequenciesforvaryingvoltage} and \eqref{frequenciesforvaryingedgelength} are independent of the temperature, although we have not proven this analytically.

\subsection{\texorpdfstring{$\nu = 5/2$}{nu = 5/2} state}
\begin{table}[!hbt]
\begin{center}
\begin{tabular}{l||c|c|c|c|c}
Model ($\nu= 5/2$)& $Q$ & $g_c$ & $g_n$ & $g$ & $\eta$\\
\hline\hline
Moore-Read  & $1/4$ & $1/8$ & $1/8$ & $1/4$ & +\\
\hline
Anti-Pfaffian  & $1/4$ & $1/8$ & $3/8$& $1/2$ & --\\
\hline
(3,3,1)  & $1/4$ & $1/8$ & $1/4$ & $3/8$ & +\\
\hline
Laughlin$_{1/2}$ & $1/2$ & $1/2$ & $0$ & $1/2$ & 
\end{tabular}
\caption{Parameters for quasiparticles for different edge models of the $\nu=\frac{5}{2}$ state. Listed are the quasiparticle charge $e^* = Qe$, the algebraic decay of the quasiparticle's neutral $g_n$ and charged $g_c$ channel, the total algebraic decay $g=g_c+g_n$ and the chirality. The $e/2$ Laughlin quasiparticle is present in all three states. Data obtained from Ref.~\onlinecite{bishara2009b}.}\label{tab:quasiparticles}
\end{center}
\vspace{-0.6cm}
\end{table}

\begin{figure}[hbt]
\includegraphics[width=.5\textwidth]{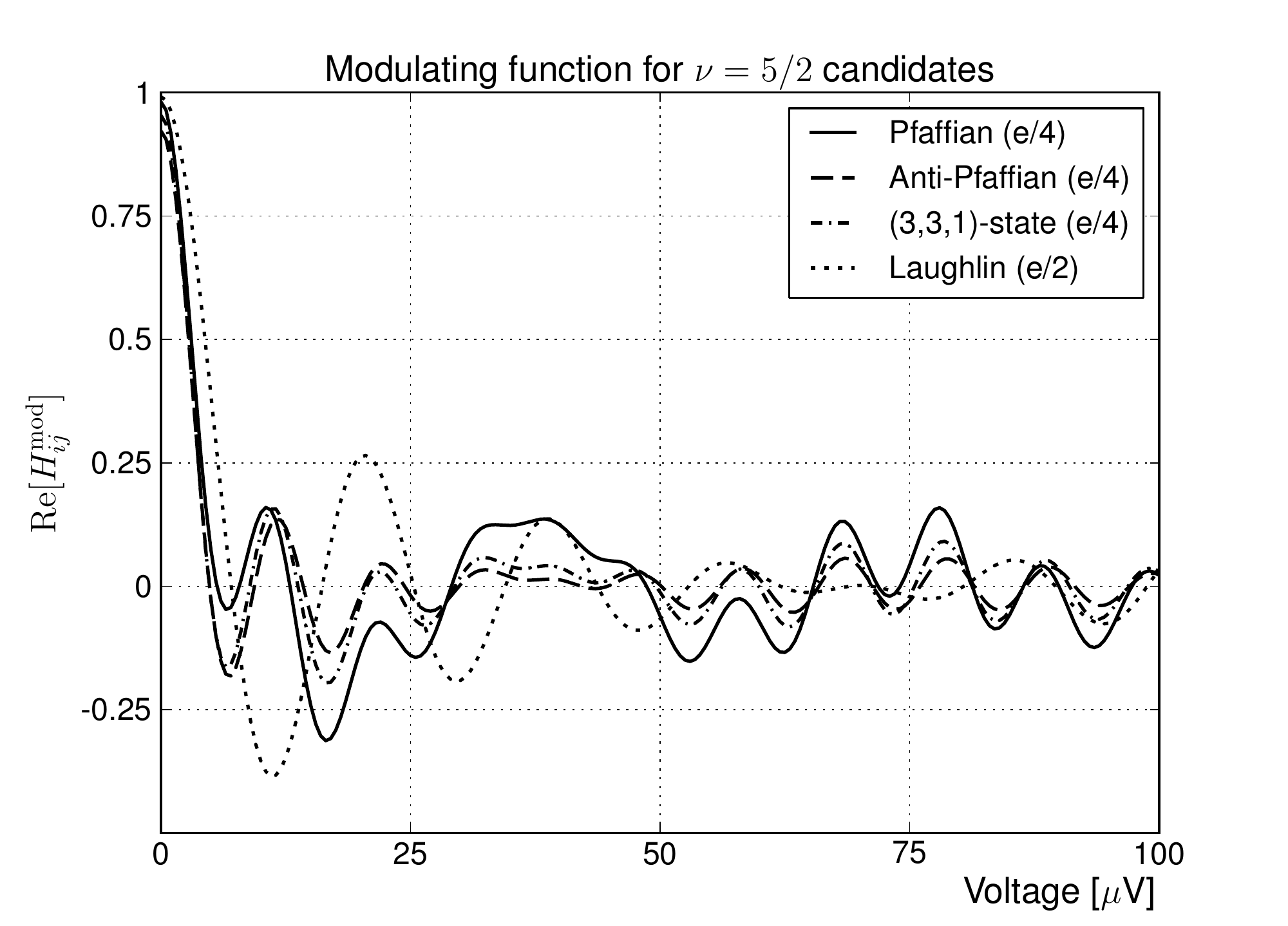}
\caption{The modulating function of four candidate states for the $\nu = 5/2$ state. The proposed states and corresponding quasiparticles are listed in table \ref{tab:quasiparticles}. The parameters used for this plot are $v_c = 5\cdot 10^3$ [m/s], $v_n = 1.4\cdot 10^3$ [m/s], $T = 1$ [mK], $a = 2.4$ [$\mu$m] and $b = 2.1$ [$\mu$m].}\label{fig:particles_5_2}
\end{figure}

The most prominent state for which the corresponding topological phase is conjectured to be non-Abelian is the $\nu = 5/2$ state\citep{willett1987,pan1999}. In table \ref{tab:quasiparticles} we list some of the proposed edge states for the $\nu = \frac{5}{2}$ state and their quasiparticle properties. The edge states we consider are the Moore-Read state \citep{mooreread1991,greiter1991} also known as the Pfaffian, its particle-hole conjugate the Anti-Pfaffian \citep{lee2007,levin2007} and the (331)-state \citep{halperin1983}. See also Ref. \onlinecite{bishara2009b}. Of these the (331)-state is an Abelian theory. The proposed edge theories consist in all cases of a decoupled neutral and charged channel as described in Section~\ref{section:edge}. In the case of the Anti-Pfaffian the neutral and charged channels have opposite chiralities. All of these edge theories predict a charge of $Qe = e/4$ associated with the quasiparticle with the lowest algebraic decay. Furthermore, the quasiparticle with second-smallest algebraic decay is for all cases a Laughlin-type anyon with a charge of $e^* = e/2$ and algebraic decays of $g_c = 1/2$ and $g_n = 0$. Figure \ref{fig:particles_5_2} is a plot of the corresponding modulating functions for the different edge theories, including the $e/2$ quasiparticles.

As we mentioned before, in the language of the renormalization group the most relevant tunnelling operator correspond to quasiparticles with the lowest algebraic decay. In the case of the Anti-Pfaffian the lowest algebraic decay is given by $\frac{1}{2}$ and it corresponds to two quasiparticles, the $e/4$ and $e/$ anyon. In this case we also need to take into account that quasiparticles with a smaller charge have a larger magnetic length, and therefore a larger bare tunnelling amplitude. So also in the case of the Anti-Pfaffian it is expected that the interference current is due to tunnelling of the $e/4$ quasiparticle. 

\subsection{\texorpdfstring{$\nu = 7/3$}{nu = 7/3} state}
\begin{table}[!htb]
\begin{center}
\begin{tabular}{l||c|c|c|c|c}
Model ($\nu= 7/3$)& $Qe$ & $g_c$ & $g_n$ & $g$ & $\eta$ \\
\hline\hline
$\overline{\text{BS}}_{2/3}$ & e/3 & 1/3 & 5/8 & 23/24 & --\\
 \hline
 $\text{BS}_{1/3}^\psi$ & e/3 & 1/3 & 3/8 & 17/24 & +\\
\hline
$\overline{\text{RR}}_{k=4}$ & e/6 & 1/12 & 1/4 & 1/3 & --\\
\hline
Laughlin${}_{1/3}$ & e/3 & 1/3 & 0 & 1/3 & 
\end{tabular}
\caption{Properties of quasiparticles for different edge models of the $\nu=\frac{7}{3}$ state. The ${e/3}$ Laughlin quasiparticle is present in the Laughlin $\nu=2+1/3$ state and all other non-Abelian states. Data obtained from Ref.~\onlinecite{bishara2009b}.}\label{tab:quasiparticles_7_3}
\end{center}
\vspace{-0.4cm}
\end{table}

\begin{figure}[!htb]
\includegraphics[width=.5\textwidth]{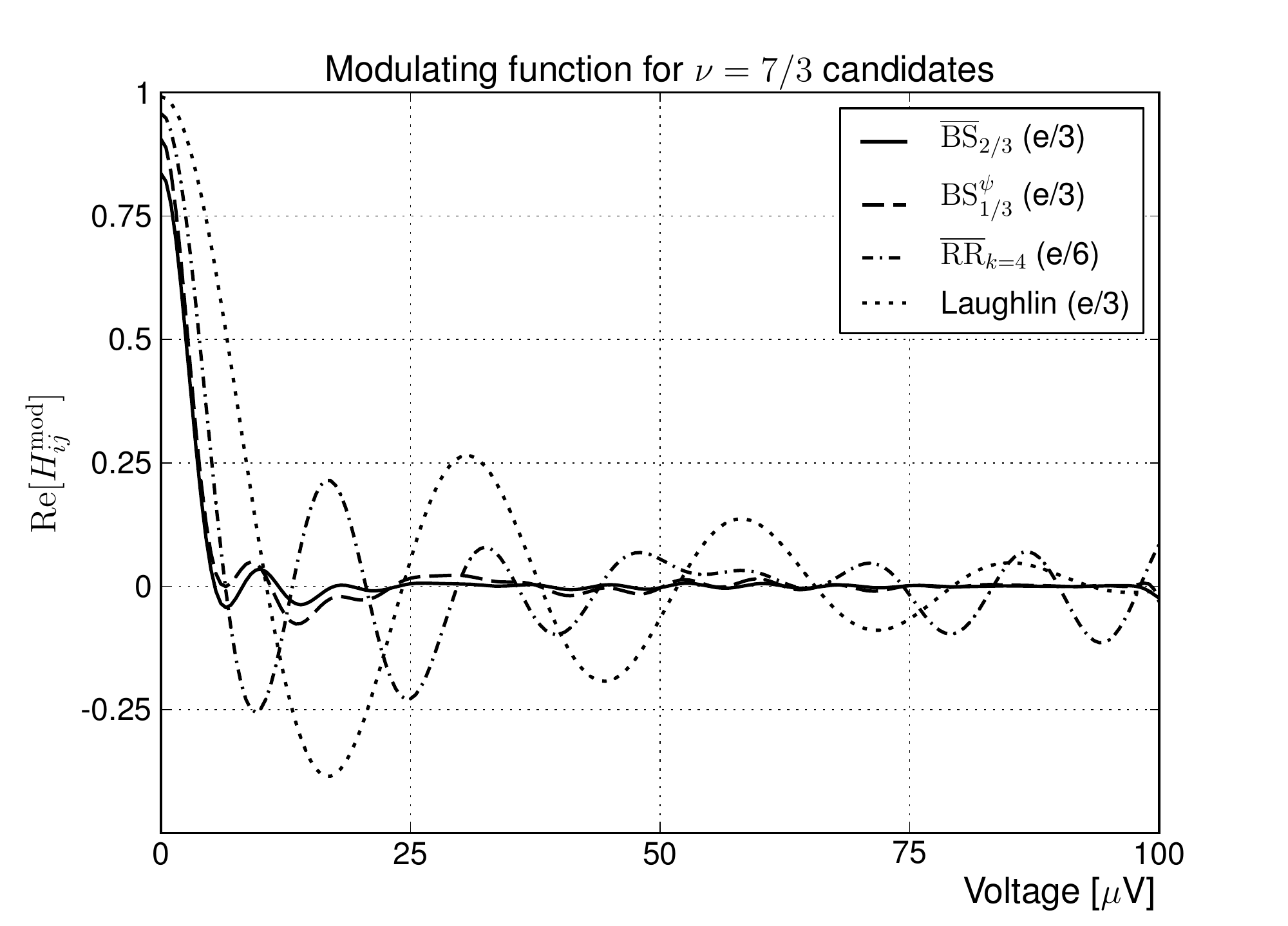}
\caption{The modulating function of four candidate states for the $\nu = 7/3$ state. The proposed states and corresponding quasiparticles are listed in table \ref{tab:quasiparticles_7_3}. The parameters used for this plot are $v_c = 5\cdot 10^3$ [m/s], $v_n = 1.4\cdot 10^3$ [m/s], $T = 1$ [mK], $a = 2.4$ [$\mu$m] and $b = 2.1$ [$\mu$m].}\label{fig:particles_7_3}
\end{figure}

The next state we look at is the $\nu = 7/3$ plateau\citep{willett1987,pan1999}. The trial states and the corresponding quasiparticles with lowest algebraic decay are listed in table \ref{tab:quasiparticles_7_3}. These trial states are the Abelian Laughlin\citep{laughlin1983} state at $\nu = 2 + 1/3$, the particle-hole conjugate of the Read-Rezayi\citep{read1999} state at $k = 4$, and two Bonderson-Slingerland states\citep{bonderson2008}. The BS states are formed through a hierarchical construction of a non-Abelian candidate state, in this case the Pfaffian and Anti-Pfaffian state. Figure \ref{fig:particles_7_3} shows the modulating function for the proposed states.

In addition to plotting the tunnelling current for a number trial states, figures \ref{fig:particles_5_2} and \ref{fig:particles_7_3} show the effect of different values of $g_n$ and $g_c$ on the $R$ function. The general rule is that a larger value of $g_i$ corresponds to a larger damping on the contributing frequency. In particular, a larger sum of $g_n+g_c$ corresponds to an $R$ function which decays more rapidly for increasing $V$. 

\subsection{\texorpdfstring{$\nu = 12/5$}{nu = 12/5} state}
\begin{table}[!htb]
\begin{center}
\begin{tabular}{l||c|c|c|c|c}
Model ($\nu= 12/5$)& $Qe$ & $g_c$ & $g_n$ & $g$ & $\eta$ \\
\hline\hline
$\text{HH}_{2/5}$ & e/5 & 1/5 &  2/5 & 3/5 & + \\
\hline
$\overline{\text{RR}}_{k=3}$ & e/5 & 1/10 & 3/10 & 2/5 & --\\
\hline
 $\text{BS}_{2/5}$ & e/5 & 1/10 & 1/8 & 9/40 & +\\
\hline
Laughlin${}_{2/5}$ & 2e/5 & 2/5 & 0 & 2/5 & 
\end{tabular}
\caption{Properties of quasiparticles for different edge models of the $\nu=\frac{12}{5}$ state. The ${2e/5}$ Laughlin quasiparticle is present in all the listed states. Data obtained from Ref.~\onlinecite{bishara2009b}. }\label{tab:quasiparticles_12_5}
\end{center}
\vspace{-0.4cm}
\end{table} 

\begin{figure}[!htb]
\includegraphics[width=.5\textwidth]{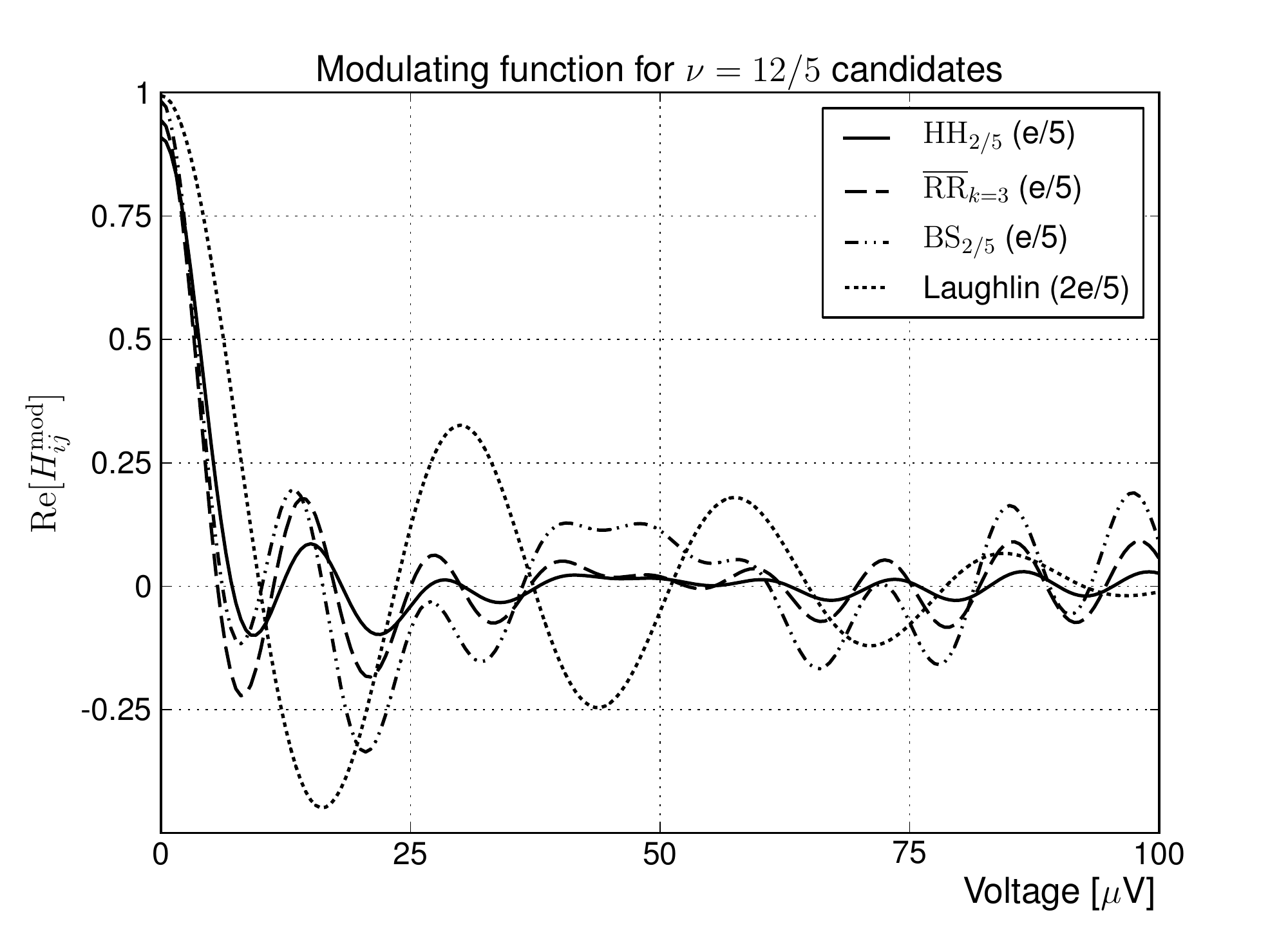}
\caption{The modulating function of four candidate states for the $\nu = 7/3$ state. The proposed states and corresponding quasiparticles are listed in table \ref{tab:quasiparticles_12_5}. The parameters used for this plot are $v_c = 5\cdot 10^3$ [m/s], $v_n = 1.4\cdot 10^3$ [m/s], $T = 1$ [mK], $a = 2.4$ [$\mu$m] and $b = 2.1$ [$\mu$m].}\label{fig:particles_12_5}
\end{figure}

The last plateau we discuss is\citep{xia2004} at $\nu = 12/5$. There are numerical studies\citep{rezayi2009,sreejith2011,bonderson2013} each of which suggest a different quantum Hall trial state for the $\nu = 12/5$ plateau. The edge states we discuss here are the particle-hole conjugate of the Read-Rezayi state\citep{read1999,rezayi2009} at $k=3$, a Haldane-Halperin edge\citep{haldane1983b,halperin1984,sreejith2011}, and a Bonderson-Slingerland state\citep{bonderson2008,bonderson2013}. The corresponding quasiparticles with lowest algebraic decay are listed in table \ref{tab:quasiparticles_12_5}. The modulating functions for these states are plotted in \ref{fig:particles_12_5}.

From the plots on the $\nu =5/2$, $\nu = 7/3$ and $\nu = 12/5$ we find empirically that the parameters $g_i$ control the amplitudes of the different oscillations present in the $R$ function. These are the oscillations discussed in Section~\ref{sec:frequencyanalysis}. We find that a larger $g_i$ causes a relatively smaller amplitude of the corresponding oscillation. This empirical rule is supported by the discussion on dephasing in Section~\ref{sec:hightemperature}. Here it was found that for a typical length or temperature scale the $R$ function is exponentially suppressed as a function of increasing temperature or increasing circumference of the interferometer. These scales are partially determined by $\bigl[\sum_i \frac{g_i}{v_i}\bigr]^{-1}$. Here we find empirically that also the relative amplitude of each oscillation is inversely related to the corresponding algebraic decay.

\section{The Aharonov-Bohm effect and the interference current}\label{sec:abphase}
\subsection{Weak tunnelling and the AB phase}
The AB-phase is determined by the magnetic field strength $B$, the area of the interferometer and the quasiparticle charge. It is given by the number of unit flux quanta for a quasiparticle with charge $Q$ piercing through the interferometer
\begin{align}
e^{i\Phi/\Phi_Q},\qquad\text{where }~ 
\begin{cases}
\Phi = 2\pi B \times \text{Area}\\
\Phi_Q = \frac{h}{Qe}
\end{cases}
~.\label{abphase}
\end{align}
Here $\Phi$ is $2\pi$ times the total number of flux quanta through the interferometer and $\Phi_Q$ is a unit flux quantum for a quasiparticle with charge $Qe$. This expression only applies in the weak tunnelling limit, where quasiparticles with the smallest algebraic decay are the most relevant operators in the language of the Renormalization Group. In this limit the interferometer is said to be in the Aharonov-Bohm regime and throughout this work we assume this always applies. 

In contrast, in the strong tunnelling limit the tunnelling current effectively pinches off the area within the interferometer, thereby forming a quantum dot. This is called Coulomb blockade\citep{rosenow2007,halperin2011}. In this limit electrons tunnelling between the quantum dot and the fluid outside the interferometer form the most relevant operators.  The AB phase is no longer determined by expression \eqref{abphase}, see e.g. Ref.~\onlinecite{zhang2009} for the case of the integer QHE.

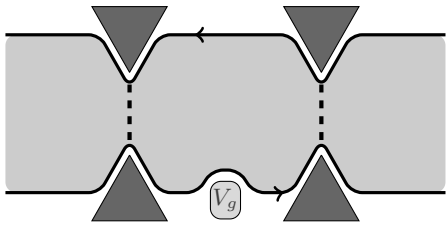
\begin{figure}[tb]
\begin{tikzpicture}
\path [fill = gray!40, rounded corners] 
(.5,.1) -- ++ (1.25,0) -- ++ (.4,.7) -- ++ (.4,-.7) -- ++ (.5,0) -- ++(.2,.3) -- ++ (.35,0) 
-- ++ (.2,-.3) -- ++ (.5,0)-- ++ 
(.4,.7) -- ++ (.4,-.7) -- ++ (1.25,0) -- ++ (0,2.1) -- ++ (-1.25,0) -- ++ 
(-.4,-.7) -- ++ (-.4,.7) -- ++ (-1.75,0) -- ++ (-.4,-.7) -- ++ 
(-.4,.7) -- ++ (-1.25,0) -- cycle; 
 
\draw [rounded corners, very thick, black,style={,postaction={decorate},
        decoration={markings,mark=at position 0.62 with {\arrow[draw=black]{>}}}}]
(0.5,.1)    -- ++ (1.25,0) -- ++ (.4,.7) -- ++ (.4,-.7) -- ++
(.5,0) -- ++(.2,.3) -- ++ (.35,0) -- ++ (.2,-.3) -- ++ (.5,0) -- ++ 
(.4,.7) -- ++ (.4,-.7) -- ++ (1.25,0);

\draw [rounded corners, very thick, black,style={,postaction={decorate},
        decoration={markings,mark=at position 0.46 with {\arrow{<}}}}]
(0.5,2.2) -- ++ (1.25,0) -- ++ (.4,-.7) -- ++ (.4,.7) -- ++ (1.75,0) -- ++ 
(.4,-.7) -- ++ (.4,.7)  -- ++ (1.25,0);

\draw [dashed, ultra thick] (2.15,.8) -- ++(0,.8);
\draw [dashed, ultra thick] (4.7,.8) -- ++(0,.8);

\draw [fill = darkgray!80] (2.15,.6) -- ++(.5,-.875) -- ++(-1,0)-- ++ (.5,.875);
\draw [fill = darkgray!80] (4.7 ,.6) -- ++(.5,-.875) -- ++(-1,0)-- ++ (.5,.875);
\draw [fill = darkgray!80] (2.15,1.7) -- ++(.5,.875) -- ++(-1,0)-- ++ (.5,-.875);
\draw [fill = darkgray!80] (4.7 ,1.7) -- ++(.5,.875) -- ++(-1,0)-- ++ (.5,-.875);

\draw [rounded corners, fill = darkgray!20] (3.225 ,-.25) -- ++(.4,0) -- ++(0,.5)-- ++ (-.4,0) -- cycle;
\node [darkgray] at (3.425 ,0) {$V_{g}$};
\end{tikzpicture}
\caption{Idea of the setup of an interferometer with a side gate. By applying a voltage on the side gate the electrons are repelled thereby deforming the edge of the quantum Hall liquid. As a function of the side-gate voltage the effective area of the interferometer and the length of the lower edge grow or shrink. This changes both the AB phase and the $R$ function.}\label{fig:setupsidegate}
\end{figure}

\subsection{Manipulating the AB phase through a side gate}

The AB phase is manipulated by either varying the magnetic field strength or deforming the effective area of the interferometer. We are interested in the latter case. In practice\citep{an2011,willett2009} the area is changed through a side-gate voltage. This setup is depicted in figure \ref{fig:setupsidegate} with the side-gate voltage given by $V_g$ (not to be confused with the voltage bias between the two edges, $\omega_Q$). By charging the side gate the Coulomb interaction repels electrons inside the interferometer, effectively deforming the area of the quantum Hall fluid. If we ignore the interference effects due to the $R$ function or quasiparticle braiding, then the current shows the following oscillating behaviour due to the AB phase
\begin{align}
I_B &= I_0 + I_{\text{osc}}\times \cos\left(\Phi_{AB}(V_g)/\Phi_Q + \delta\right)\label{pureab}\\
\Phi_{AB}(V_g) &= 2\pi\frac{B}{h/Qe} \times \text{Area}(V_g) ~. \nonumber
\end{align}
This oscillating signal arises in the weak tunnelling limit. One typically assumes the change in area is linear with respect to the side-gate voltage, meaning $\text{Area}(V_g) \propto V_g$. The Coulomb interaction and localization effects can alter this behaviour and cause small, non-linear fluctuations as a function of the side-gate voltage\citep{halperin2011}. This is called the Coulomb dominated regime (not to be confused with Coulomb blockade). In this regime the edge and the area inside the interferometer readjust to keep the dot neutral. Quasiparticles still tunnel along the point contacts and the interference current is still visible, but the corresponding AB phase does not follow expression \eqref{pureab}. We assume the interferometer is not Coulomb dominated and the change in area is linear with respect to the side-gate voltage.

Recent experiments\citep{willett2013a,willett2013b,an2011} observe on the order of $\lesssim 5$ full oscillations when the side-gate voltage is varied. This applies to the $\nu =5/2$ state, with a magnetic field strength of \mbox{$B \sim 5.5$ [T]}. The area of the interferometer is estimated in the range of\mbox{0.1 - 0.4 [$\mu $m${}^2$]}, depending on the exact geometry of the device. For an interferometry area of \mbox{0.15 [$\mu$m${}^2$]} and a quasiparticle with charge \mbox{$Qe = e/4$} this corresponds to a total of roughly 50~unit flux quanta. If we assume the interferometer is in the AB dominated regime, then a generous estimate of the change in area is about \mbox{$\leq$ 10\%} when five full oscillations are observed.

\subsection{The interference current: combining the AB phase and the \texorpdfstring{$H^{\text{mod}}$}{H-mod} function}
When the side-gate is used to change the area of the interferometer, then almost inevitably the length of the edge between the two point contacts changes as well. This change in length causes interference effects through the modulating function $H_{ij}^{\text{mod}}$. Including this in the expression for the interference current gives
\begin{multline}
I_B = I_0 \\+ I_{\text{osc}}\times \text{Re}\left[e^{i\Phi_{AB}(V_g)/\Phi_Q  +i\delta}H_{ij}^{\text{mod}}(\omega_Q; V_g)\right]~.\label{nonpureab}
\end{multline}
The function $H_{ij}^{\text{mod}}$ implicitly depends on the side-gate voltage $V_g$ through the length of the lower edge, $a(V_g)$. Whether the change in $H_{ij}^{\text{mod}}$ as a function of $V_g$ is significant is determined by the change in the length of the edge, the velocity of the edge modes and the voltage bias between the two edges $\omega_Q$. 

For instance, in the experiment of Ref.~\onlinecite{willett2013a} the quantum Hall fluid inside the interferometer is cigar-shaped with the ends of the cigar corresponding to the point contacts. We can picture the scenario in which the side-gate voltage deforms the lower edge uniformly, such that a 5\% change in the area of the interferometer is accompanied with relatively negligible change in the length of the edge. In this scenario the function $H_{ij}^{\text{mod}}(\omega_Q; V_g)$ is approximately constant as a function of $V_g$. 

\begin{figure}[tb]
\includegraphics[width=.5\textwidth]{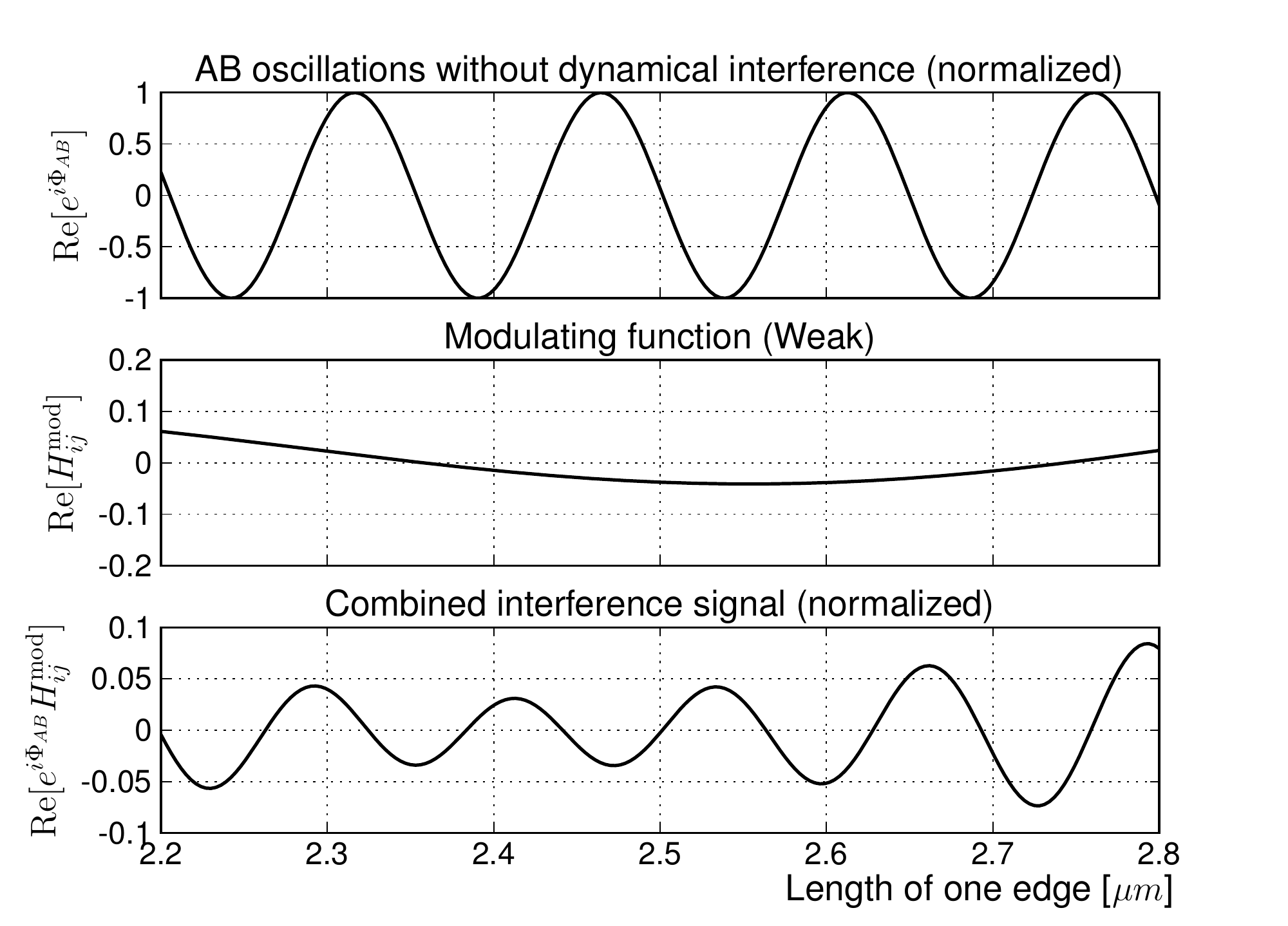}
\caption{Plot of estimated AB oscillations as a function of the varying edge length. This variation is caused by the side-gate voltage and we assume a linear relation between the area of the interferometer, the side-gate voltage and the length of the edge. The parameters used are $Q=e/4$, $g_c= \frac{1}{10}$, $g_n = \frac{1}{8}$, $v_c = 8\cdot 10^3$ [m/s], $v_n = 3\cdot 10^3$ [m/s], $b=2.5$ [$\mu$m], $a=2.25$--$2.75$ [$\mu$m], $V=50$ [$\mu$V], $T=0$ [K]. The plot is ``weak'' in the sense that the modulating function does not change much over the plotted range.}\label{fig:ab_signal_weak}
\end{figure}

\begin{figure}[tb]
\includegraphics[width=.5\textwidth]{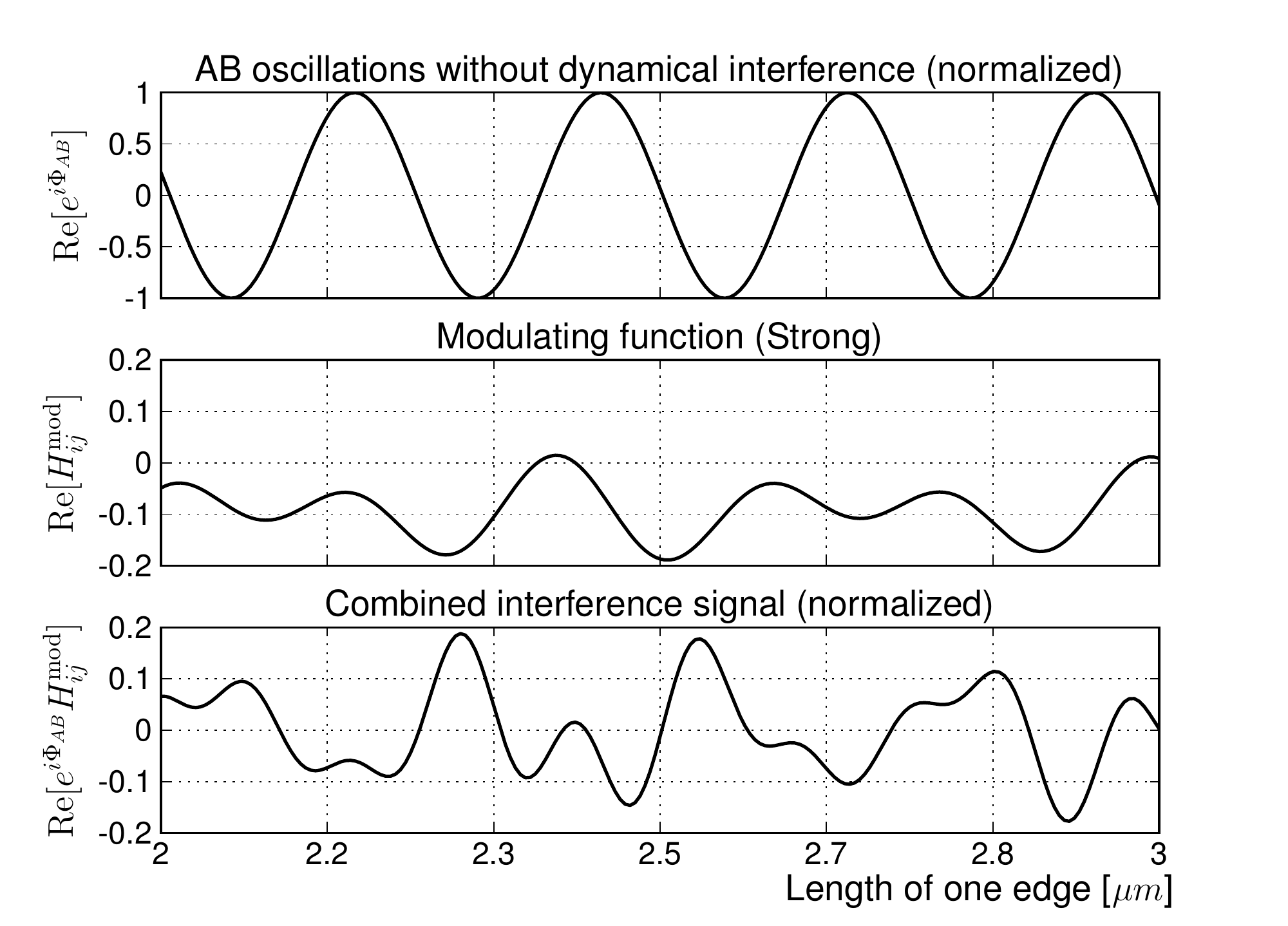}
\caption{The same plot as in figure \ref{fig:ab_signal_weak}, but with different velocities, namely  $v_c = 9\cdot 10^2$ [m/s], $v_n = 6\cdot 10^2$ [m/s] and the range over which the edge length is varied is larger, $a=2.0$--$3.0$ [$\mu$m]. These slower velocities and larger range lead to a $H_{ij}^{\text{mod}}$ function which varies significantly more than that of figure \ref{fig:ab_signal_weak}.}\label{fig:ab_signal_strong}
\end{figure}

The other possibility is that the change in $a(V_g)$ is not small. The device used in the experiment of Ref.~\onlinecite{an2011} has a circular shape, and it is possible that the change in edge length is relatively larger than that of Ref.~\onlinecite{willett2013a}. It then depends on the remaining parameters, the velocity and voltage bias, if the change in $H_{ij}^{\text{mod}}(\omega_Q; V_g)$ is large enough to be observable. 

In figure \ref{fig:ab_signal_weak} and \ref{fig:ab_signal_strong} we have plotted these two scenarios. Figure \ref{fig:ab_signal_weak} is the ``weak'' case in which the function $H_{ij}^{\text{mod}}$ remains largely constant while $V_g$ is varied. In the lower panel of this figure the function $H_{ij}^{\text{mod}}$ causes a small modulation of the total interference signal. The interference due to a varying edge length is difficult to observe through measurement of this signal. Figure \ref{fig:ab_signal_strong} shows the ``strong'' case where the change in $H_{ij}^{\text{mod}}$ is much larger. These plots differ in the values used for the velocities and edge lengths, keeping all other parameters fixed. The frequencies of the oscillations at which $H_{ij}^{\text{mod}}$ varies are given by Eq.~\eqref{frequenciesforvaryingedgelength}.

\subsection{Frequency analysis of interference current}
In plotting the figures \ref{fig:ab_signal_weak} and \ref{fig:ab_signal_strong} we assume a linear relation between the area and the side gate voltage, $\text{Area} \propto V_g$, and the length of the edge and side-gate voltage, $a(V_g)\propto V_g$. Under this assumption the interference due to the AB effect oscillates at some frequency with respect to the varying edge length $a(V_g)$. We denote this frequency by $\phi_{AB}$, 
\begin{align}
e^{i\Phi/\Phi_Q} = e^{2\pi i \phi_{AB} \cdot a(V_g)}
\end{align}
In other words, $\phi_{AB}$ corresponds to the frequency of the oscillations appearing in the upper panels of figures \ref{fig:ab_signal_weak} and \ref{fig:ab_signal_strong}. Fixing the proportionality constant between the change in area and the change in edge length equal to $C_1$, i.e. $\Delta\text{Area}(V_g) = C_1\times \Delta a(V_g)$, then $\phi_{AB} = C_1\times\frac{B}{h/Qe}$. The proportionality constant depends on the exact details of the interferometric device, and the change of both area and edge length is performed through the side-gate voltage $V_g$. The charge of the quasiparticle in the fractional regime can then obtained by looking at the ratio of this frequency compared to that in the integer regime where $Q=1$,
\begin{align}
\frac{\phi_{AB}({\nu = \text{fractional}})}{\phi_{AB}({\nu = \text{integer}})} = Q\label{eq:chargeratio}
\end{align}

\begin{figure}[!tb]
\includegraphics[width=.5\textwidth]{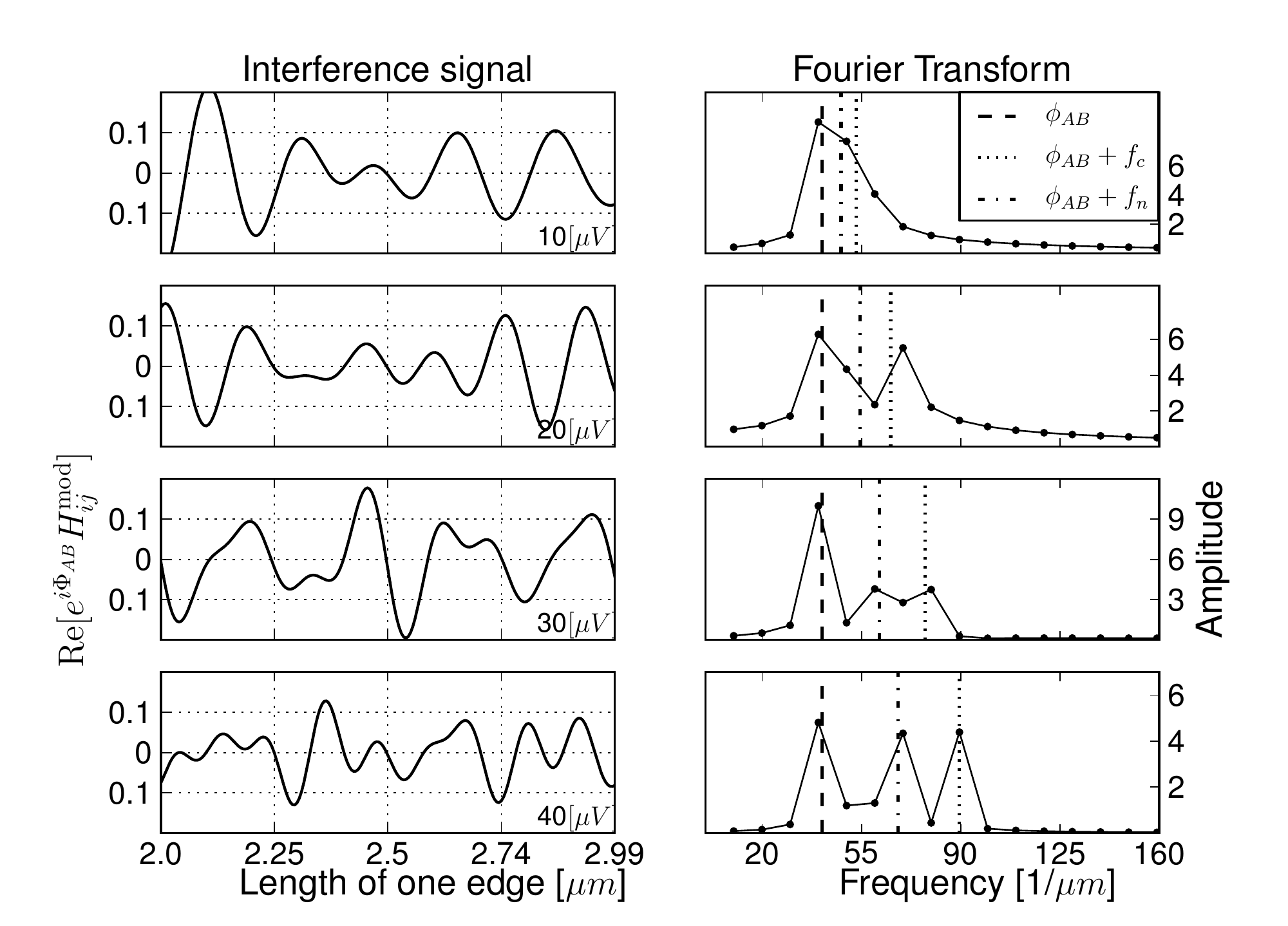}
\caption{The four figures on the left plot the interference current $\text{Re}[e^{i\phi_{AB}\cdot a}H_{ij}^{\text{mod}}]$ as a function of varying the edge length $a = [2.0 - 3.0]$ [$\mu$m] for an applied voltage bias of $V_{\text{bias}} = 10$, $20$, $30$ and $40$ [$\mu$V]. The voltage bias $V_{\text{bias}}$ should not be confused with the side-gate voltage. The figures on the right are the corresponding Fourier transforms. The peaks correspond to Eq.~\eqref{freqpeaks}. The remaining parameters are $g_c = g_n = 1/8$, $T = 0$ [K], $Qe= e/$, $b=2.5$ [$\mu$m], $v_n=5\cdot 10^2$ [m/s], $v_c=9\cdot 10^2$ [m/s]. Finally, $\phi_{AB} = C_1\times \frac{B}{h/Qe} = 4.11$ [$\mu$m$^{-1}$] with $C_1 = 1.0\cdot 10^{-2}$ [$\mu$m] and $B = 6.8$ [T].}\label{fig:int_ab_fourier}
\end{figure}

In figure \ref{fig:ft_analysis_edge} we showed that the as a function of a varying edge length the modulating function $\hmod$ oscillates with frequencies $\frac{QeV}{v_i h}$. A similar analysis shows that the combined signal of the AB oscillations and the modulating function oscillates at three frequencies, given by
\begin{multline}
\text{frequency peaks} = \\ \left\lbrace \phi_{AB}, \phi_{AB} +  \frac{QeV}{v_n h}, \phi_{AB} +  \frac{QeV}{v_c h} \right\rbrace~. \label{freqpeaks}
\end{multline}
These frequencies correspond to the signals appearing in the lower panels of figures \ref{fig:ab_signal_weak} and \ref{fig:ab_signal_strong}. In particular the ``pure" AB oscillations corresponding to $\phi_{AB}$ remain present and the quasiparticle charge can be measured through formula \eqref{eq:chargeratio} even if the oscillations in $H_{ij}^{\text{mod}}$ are strong. The shifted peaks $\phi_{AB} +  \frac{QeV}{v_i h}$ provide an experimental probe of the velocity of the edge modes. Numerical estimates\citep{hu2009} indicate that $v_c > v_n$, meaning the largest frequency corresponds to the velocity of the neutral mode.

If the terms $\frac{QeV}{v_i h}$ in \eqref{freqpeaks} are small compared to $\phi_{AB}$, then the frequency peaks overlap in a Fourier analysis and become indiscernible. To enhance the visibility of the different peaks we can either reduce $\phi_{AB}$, increase the range over which the edge length is changed or increase $\frac{QeV}{v_i h}$. The frequency $\phi_{AB}$ and the variation in edge length are both determined by the geometric properties of the interferometric device and the side-gate voltage. 

Increasing $\frac{QeV}{v_i h}$ can be accomplished by injecting a larger current into the system which is equivalent to increasing the voltage bias $\omega_Q$. The effect of a larger current on the frequency spectrum is demonstrated in figure \eqref{fig:int_ab_fourier}, where the applied voltage bias is increased from 10 [$\mu$V] to 40 [$\mu$V] in steps of 10 [$\mu$V], while keeping all other parameters fixed. For a voltage bias of 10 [$\mu$V] the frequency peaks merge and are indistinguishable. This is due to the relative magnitude of $\frac{QeV}{v_i h}$ and $\phi_{AB}$. At a voltage bias of 40 [$\mu$V] three frequency peaks emerge corresponding to the frequencies \eqref{freqpeaks}.

\section{Discussion}
We have calculated the tunnelling current through a Fabry-P\'{e}rot fractional quantum Hall interferometer in linear response theory for a broad class of edge theories. Our main result is an expression for the tunnelling current in terms of Carlson's $R$ function at finite temperatures and in terms of the confluent Lauricella hypergeometric function at zero temperature. This expression arises as the tunnelling current is related to the Fourier transform of the quasiparticle propagators. In the conformal limit these propagators have a universal form, which is the reason behind the generality of our result.

Our result applies to both Abelian and non-Abelian edge theories with an arbitrary number of edge modes -- the neutral and charged degrees of freedom -- each of which is characterized by its own edge velocity and chirality. In addition our result is applicable to interferometers with different edge lengths between the point contacts and our result can be straightforwardly extended to include more than two point contacts as explained in Section~\ref{sec:linearresponse}.

We have implemented a numerical scheme to calculate Carlson's $R$ function and the confluent Lauricella hypergeometric function, and the corresponding interference current using a series representation. This numerical scheme is written in NumPy and Fortran and publicly available\citep{github}. We are also making available all the code that reproduce the plots in this work.

The interference in the tunnelling current is attributed to the Aharonov-Bohm phase, the dynamical interference induced by the voltage bias between the edges and the statistical properties of the quasiparticles. Recent experiments\citep{willett2013a,an2011} measure the Aharonov-Bohm phase by deforming the area inside the interferometer through a plunger gate. This setup also changes the edge length between the point contacts which induces interference effects through the dynamical interference. We show that the total interference results in oscillations in the tunnelling current as a function of the edge length. We have determined the frequency of these oscillations in terms of the edge velocities and the source-drain voltage, i.e. Eq. \eqref{freqpeaks}. These frequencies can be used to measure the edge velocities.

The visibility of the frequency peaks depends among other things on the geometry of the interferometer and the range over which the length of the edge is varied. If there are many AB oscillations, while the change in edge length is small then the dynamical interference effects are hardly discernible from the AB oscillations. It is possible that the change in edge length of current interferometric devices is negligible and the interference effects we describe are indeed not measurable. In this case our proposed experiment requires an alteration of the interferometer, using for instance a different geometry.

Another way to increase the visibility of the frequency peaks is to increase the strength of the injected current. The frequencies are directly proportional to the source-drain voltage bias. The frequency of the AB oscillations is independent of the source-drain voltage, while the frequencies of the oscillations due to dynamical interference increase with larger source-drain voltages. 

Alternatively, this dependency can be used to check if the dynamical oscillations play a role in experiments which focus on the AB interference. If the effect of dynamical oscillations play a role in experiments which measure AB oscillations, then our results predict that this becomes apparent by running the experiment multiple times at different source-drain voltages. 

\section*{Acknowledgements}
JKS and OS were supported by Science Foundation Ireland Principal Investigator award 08/IN.1/I1961. S.H.S. acknowledges support from EPSRC grants EP/I031014/1 and EP/I032487/1. 
\appendix
\section{Fourier transform of propagators}\label{sec:integrals}
We require the Fourier transform of the two-point propagator. We start with the finite-temperature case and no spatial dependence,
\begin{align}
P_{g}(t) = \left(\pi T\right)^{g} \sin\left[\pi T(\delta + it)\right]^{-g}~.\label{app:pg}
\end{align}
Here $\delta>0$ is an infinitesimal integral regulator which is taken to zero in the end. To compute its Fourier transform \mbox{$P_{g}(\omega) = \int dt e^{i\omega t} P_{g}(t)$} we follow Ref.~\onlinecite{martin2005} and substitute $\delta +it = i\frac{x}{2\pi T} + \frac{1}{2T}$. This leads to
\begin{align}
P_{g}(t) &=  \left(\pi T\right)^{g} \cosh\left(\frac{x}{2}\right)^{-g}\nonumber\\
& = \left(2\pi T\right)^{g} e^{- g x/2} \left(1+e^{-x}\right)^{-g}~.
\end{align}
With this substitution the limits of the contour are $\pm\infty+i(\pi-\delta)$. The contour is deformed so that it runs over the real line of $x$, which can be done provided there are no singularities that prevent this deformation. The function $\cosh(x/2)$ is zero at the points $x_n = (2n+1)\pi i$, for $n$ integer. It is therefore the presence of the integral regulator $\delta$, which allows for the deformation.

After substitution the resulting integral is an integral representation of the Euler beta function \citep{gradshteyn2007}. We have
\begin{align}
P_{g}(\omega) &= \frac{e^{\frac{\omega}{2T}}}{ \left(2\pi T\right)^{1-g}} \int_{-\infty}^{\infty}  e^{-\left(\frac{g}{2}-i\frac{\omega}{2\pi T}\right) x}  \left(1 + e^{-x}\right)^{-g}dx \nonumber\\
&  =  \frac{e^{\frac{\omega}{2T}}}{ \left(2\pi T\right)^{1-g}} B\left(\frac{g}{2}+i\frac{\omega}{2\pi T},\frac{g}{2}-i\frac{\omega}{2\pi T}\right).\label{app:fourierprop}
\end{align}
Here we have taken $\delta\rightarrow 0$ in the final result.

Through a similar manipulation the Fourier transforms of products of two-point propagators with unequal arguments can be obtained. This results in the Fourier transform for $G^>_{ij}$, Eq.~\eqref{correlator}. We first note that with the substitution $\delta +it = i\frac{x}{2\pi T} + \frac{1}{2T}$ we have
\begin{align}
 P_{g}\left(t+\xi\right)
&= \left(2\pi T\right)^{g} e^{-\pi T\xi g}
e^{- \frac{g}{2}x} A_g(\xi)\\
\text{where}\quad A_g(\xi) &= \left(1+ e^{-2\pi T \xi}e^{-x}\right)^{-g}~.\nonumber
\end{align}
When applying this substitution to the Fourier transform of the product of four propagators (setting $g= g_n+g_c$) we obtain
\begin{multline}
\int_{-\infty}^{\infty}dt \big[e^{i\omega t} 
P_{g_c}(t+\tfrac{a}{v_c}) P_{g_c}(t-\tfrac{b}{v_c})\\\times P_{g_n}(t+\eta\tfrac{a}{v_n}) 
 P_{g_n}(t-\eta\tfrac{b}{v_n})\big]
 = \\
 e^{\frac{\omega}{2T}}\left(2\pi T\right)^{2g-1}   
e^{\pi T(b-a)(\tfrac{g_c}{v_c}+\eta\tfrac{g_n}{v_n})} \\
\times \int_{-\infty}^{\infty} dx\big[ e^{-(g-i\frac{\omega}{2\pi T}) x} 
A_{g_c}(\tfrac{a}{v_c})
A_{g_c}(-\tfrac{b}{v_c})\\ \times
A_{g_n}(\eta\tfrac{a}{v_n})
A_{g_n}(-\eta\tfrac{b}{v_n})\big] ~.\label{app:integral1}
\end{multline}
The resulting integral is an integral definition of Carlson's $R$ function \citep{carlson1963}, see Eq.~\eqref{app:carlsonr}. This function is a scaling function and is closely related to the Lauricella hypergeometric function\citep{lauricella1893,mathai2009} $F_D^{(n)}$ . This Lauricella function is a multivariable generalization of the Gauss hypergeometric function of one variable and the Appell hypergeometric function of two variables \citep{gradshteyn2007}. For our purposes it is convenient to use the $R$ function to represent our main result, although the two representations are interchangeable, see Eq.~\eqref{app:r-lau}.

Applying the integral representation \eqref{app:carlsonr} gives for the integral
\begin{align}
%&\int_{-\infty}^{\infty}dt e^{i\omega t} P_{2h}(t-a/v) = e^{i\omega a/v}P_{2h}(\omega)\\
%&\int_{-\infty}^{\infty}dt e^{i\omega t} P_{2h}(t-a/v)P_{2h}(t+b/v) = e^{2\frac{\pi T h}{v}(b-a)} R\left(2h+i\frac{\omega}{2\pi T}; \{ 2h,2h\}, 4h;\right) P_{2h}(\omega)\\
%\int_{-\infty}^{\infty}dt e^{i\omega t} P^c_{h_c}(t-a/v_c)P^c_{h_c}(t+b/v_c)P^n_{h_n}(t-\eta a/v_n)P^n_{h_n}(t+\eta b/v_n)
&\text{\eqref{app:integral1}}
 = e^{\pi T (b-a)\left(\frac{g_c}{v_c} + \eta \frac{g_n}{v_n}\right)} P_{2g}(\omega) \nonumber\\
  &\times R\Big(g-i\frac{\omega}{2\pi T}; \{ g_c,g_c,g_n,g_n\}; \nonumber\\
  &\qquad\quad e^{-2\pi T\frac{a}{v_c}},e^{2\pi T\frac{b}{v_c}},e^{-\eta 2\pi T\frac{a}{v_n}},e^{\eta 2\pi T\frac{b}{v_n}} \Big)\label{app:fouriercarlson}
\end{align}
where $P_g(\omega)$ is given by \eqref{app:fourierprop}. The resulting $R$ function is Carlson's $R$ function. Note that the order in which the parameters appear is important.

\section{Carlson's \texorpdfstring{$R$}{R} function}\label{app:rfunction}
\subsection{Main properties of the \texorpdfstring{$R$}{R} function}
We first introduce a notation. We define $\mathcal{G}_n$ as the ordered set with $n$ elements given by
\begin{align}
\mathcal{G}_n = \lbrace g_1,\ldots, g_n\rbrace
\end{align}
and we set
\begin{align}
\gamma = \sum_{i=1}^n g_i~.
\end{align}
In the main text we usually work with the case where $n=4$ and the ordered set corresponds to $\mathcal{G}_n = \lbrace g_c, g_c, g_n, g_n\rbrace$, and $\gamma = 2(g_h+g_n) = 2g$. Carlson's $R$ function is treated in Ref.~\onlinecite{carlson1963} and is defined through the integral representation
\begin{multline}
R\left(\alpha; \mathcal{G}_n; \lbrace z_i\rbrace \right)=\frac{1}{B\left(\alpha,\gamma - \alpha\right)} \\  \times\int_{-\infty}^\infty e^{-\alpha x}\Bigl[\prod_{i=1}^n(1+z_i e^{-x})^{-g_i}\Bigr] dx~.
 \label{app:carlsonr}
\end{multline}
Here, $B(x,y)$ is the Euler beta function, and $\lbrace z_i\rbrace$ is the ordered set $\lbrace z_i\rbrace_{i=1}^n = \lbrace z_1,\ldots z_n\rbrace$. 

We require $\text{Re}\left[\alpha\right] > 0$ and $\text{Re}\left[\gamma-\alpha\right]>0$ for convergence of the integral. Furthermore, we take the $z_i$'s to be real and positive. The $R$ function is symmetric under the simultaneous interchange of $g_i\leftrightarrow g_j$ and $z_i\leftrightarrow z_j$. In the text the $z_i$ correspond to the exponentials~$e^{\pm 2\pi T\eta_i \frac{ a}{v_i}}$. 

The $R$ function is a scaling function, i.e. it is homogeneous. This follows directly from the integral definition \eqref{app:carlsonr}
\begin{multline}
R\left(\alpha;  \mathcal{G}_n; \{ z_1,\ldots z_n\}\right) = \\  \lambda^{\alpha} R\left(\alpha;  \mathcal{G}_n; \{\lambda z_1,\ldots, \lambda z_n\} \right)~.\label{app:homog}
\end{multline}
We also have the Euler-type transformation
\begin{multline}
R\left(\alpha; \mathcal{G}_n ; z_1,\ldots z_n\right) = \\
\Bigl[\prod_{i=1}^n z_i^{-g_i}\Bigr] R\left(\gamma-\alpha; \mathcal{G}_n ; z_1^{-1},\ldots z_n^{-1}\right)~.\label{app:intrep}
\end{multline}
For some special values the $R$ function with $n$ arguments reduces to one with $m<n$ arguments. For instance
\begin{multline}
R\left(\alpha; \mathcal{G}_n; \{z_1,\ldots z_{k}, z,\ldots, z \}\right) =\\ R\left(\alpha;\{g_1, \ldots, g_{k}, \tilde{g}\} ; \{ z_1,\ldots z_{k},z \} \right)\label{app:reduction1} 
\end{multline}
where $\tilde{g} = g_{k+1} + \cdots + g_n$.  We also have the case 
\begin{multline}
B(\alpha,\gamma-\alpha)R\left(\alpha ;\mathcal{G}_n;\{ z_1,\ldots z_k,0,\ldots, 0\}\right) = \\
B(\alpha,\gamma-\alpha - \tilde{g})R\left(\alpha; \mathcal{G}_k;\{ z_1,\ldots z_k\}\right)\label{app:reduction2}
\end{multline}
The $R$ function is closely related to the Lauricella hypergeometric function \citep{lauricella1893,carlson1963,mathai2009}. We define the Lauricella function through its series representation
\begin{multline}
F_D^{(n)}\left(\alpha;  \mathcal{G}_n; \gamma ;\{ 1- w_1,\ldots 1 - w_n \} \right)= \\
 \sum_{m_1=0}^\infty\!\! \cdots\!\! \sum_{m_{n}=0}^\infty  \frac{(\alpha)_{\sum_i m_i }}{(\gamma)_{\sum_i m_i }} 
\left[\prod_{i=1}^n \frac{(g_i)_{m_i}}{m_i!} \left(1-w_i\right )^{m_i}\right]
\label{app:expansion}
\end{multline}
where $(\alpha)_m = \Gamma[\alpha+m]/\Gamma[\alpha]$ is the Pochhammer symbol and we require $|1-w_i| < 1$ and $\text{arg}(1-w_i)>0$ for convergence of the series.

To demonstrate the relation between the two functions we define $z_n \equiv \text{max}(z_1,\ldots,z_n)$ as the largest parameter of the $z_i$'s. Because of the identity \eqref{app:carlsonr} we can always set this parameter to be the last argument of the $R$ function. Furthermore, we will demand $z_i \neq z_j$ for $i\neq j$, which can always be accomplished through the reduction property \eqref{app:reduction1}. The relation between $R$ and $F_D$ is given by
\begin{multline}
R\left(\alpha; \mathcal{G}_n; \{z_1,\ldots z_n\}\right) = \\
 z_n^{-\alpha}F_D^{(n-1)}\bigl(\alpha;  \mathcal{G}_{n-1}; \gamma ; \{ 1- \frac{z_1}{z_n},\ldots 1 - \frac{z_{n-1}}{z_n}\}\bigr)\label{app:r-lau}
\end{multline}
The arguments of the Lauricella function all satisfy \mbox{$|1-z_i/z_n|<1$} and \mbox{$\text{arg}(1-z_i/z_n)=0$} meaning we have convergence of the series \eqref{app:expansion}.
 
The Lauricella hypergeometric function is a generalization of the single-variable Gauss hypergeometric function, denoted by ${}_2F_1$, and the two-variable Appell hypergeometric function, $F_1$. We have
\begin{align}
&R\left(\alpha;\{g_1\};\{ z_1\}\right) =  z_1^{-\alpha}\\
&R\left(\alpha;\{g_1,g_2\};\{ z_1,z_2\}\right) =\nonumber\\
&\qquad z_2^{-\alpha}~ {}_2F_1(\alpha; \{ g_1\};g_1+g_2; \{ 1-\tfrac{z_1}{z_2}\})\nonumber\\
&R\left(\alpha;\{g_1,g_2,g_3\};\{ z_1,z_2,z_3\}\right) = \nonumber\\
&\qquad z_3^{-\alpha} ~F_1(\alpha; \{ g_1,g_2\};g_1+g_2+g_3; \{ 1-\tfrac{z_1}{z_3},1-\tfrac{z_2}{z_3}\}).\nonumber
\end{align}
Here we assume $z_3 > z_2 > z_1$.

\subsection{High temperature behaviour}\label{app:asymptotic}
Consider again the expression for $G_{ij}^>(\omega)$, Eq.~\eqref{correlator}. This is proportional to the integral
\begin{align}
\mathcal{I} \equiv \int_{-\infty}^\infty dt e^{i\omega t} \prod_{i=1}^m P_{g_i}(t + \xi_i)
\end{align}
where the $\xi_i$ correspond to the energy scales set by the velocity and edge lengths, $\xi \sim \pm \eta_i\frac{a}{v_i}$ and $P_g(t)$ is given by Eq.~\eqref{app:pg}. We are interested in the behavior for this function when $T$ grows large. For this we substitute \mbox{$\delta + it \rightarrow ix + \frac{1}{2T}$}, which gives
\begin{align}
P_{g_i}\left(t+\xi_i\right) &= \left(\pi T\right)^{g_i} \cosh \left(\pi T\left(x + \xi_i\right)\right)^{-g_i}
\end{align}
and the integral becomes
\begin{align*}
\mathcal{I}  = \left(\pi T\right)^{2g} e^{\frac{\omega}{2T}}\int_{-\infty}^\infty dx e^{i\omega x} \prod_i \cosh \left(\pi T\left(x + \xi_i\right)\right)^{-g_i}
\end{align*}
To be consistent with the main text we set $\sum_i g_i = 2g$. We split the integral into two domains, and pull out an exponential from the $\cosh$ function. This gives
\begin{align}
&e^{-\frac{\omega}{2T}}\frac{\mathcal{I}}{\left(2\pi T\right)^{2g}}  =\nonumber\\  &\int_{0}^{\infty}dx e^{-(2\pi T g- i\omega) x} \prod_i (e^{\pi T \xi_i}+e^{-2\pi T(x + \frac{\xi_i}{2})})^{-g_i}+ \nonumber\\
&\int_{-\infty}^{0}dx e^{(2\pi T g + i\omega) x} \prod_i (e^{2\pi T\left(x + \frac{\xi_i}{2}\right)}+e^{-\pi T \xi_i})^{-g_i} \label{app:integralweapproximate}
\end{align}
Consider the first integral. We perform an integration by parts, and obtain a boundary term and a remainder,
\begin{align}
\int_{0}^{\infty}dx  e^{-(2\pi T g- i\omega) x} &\prod_i (e^{\pi T \xi_i}+e^{-2\pi T(x + \frac{\xi_i}{2})})^{-g_i}  \nonumber\\
=\frac{1}{2\pi T g -i\omega}\Big[&\prod_i (e^{\pi T \xi_i}+e^{-\pi T\xi_i})^{-g_i}
     \nonumber\\
&+\int_{0}^{\infty}dx e^{-(2\pi T g- i\omega) x} f(x)\Big]\label{app:partialintegration}
\end{align}
where
\begin{multline}
f(x) =2\pi T\prod_i (e^{\pi T \xi_i}+e^{-2\pi T(x + \frac{\xi_i}{2})})^{-g_i} \nonumber\\
\times\sum_{j} g_j\frac{e^{-2\pi T(x + \frac{\xi_i}{2})}}{ e^{\pi T \xi_i}+e^{-2\pi T(x + \frac{\xi_i}{2})} }
\end{multline}
We can estimate an upper bound for the remainder term. For this we note that $f(x)$ is positive on the integration domain and bounded by
\begin{align*}
f(x) %& \leq 2\pi T\prod_i (e^{\pi T \xi_i}+e^{-\pi T \xi_i})^{-g_i} 
%\sum_{j} g_j\frac{e^{-\pi T\xi_i}}{ e^{\pi T \xi_i}+e^{-\pi T \xi_i} } \\
&\leq 4\pi Tg \prod_i (e^{\pi T \xi_i}+e^{-\pi T \xi_i})^{-g_i}, ~~x\in [0,\infty)
\end{align*}
This gives an upper bound on the remainder given by
\begin{multline}
\frac{1}{2\pi T g -i\omega}\left|\int_{0}^{\infty}dx e^{-(2\pi T g- i\omega) x} f(x)\Big] \right|\leq\\
 \frac{4\pi Tg}{2\pi T g -i\omega}  \prod_i (e^{\pi T \xi_i}+e^{-\pi T \xi_i})^{-g_i} \int_{0}^{\infty}dx e^{-(2\pi T g- i\omega) x} \\
= \frac{ 4\pi Tg}{(2\pi T g- i\omega)^2} \prod_i (e^{\pi T \xi_i}+e^{-\pi T \xi_i})^{-g_i}
\end{multline}
The product also appears in the expression for the boundary term in Eq.~\eqref{app:partialintegration}. This product therefore determines the asymptotic behavior of the boundary term in the high temperature limit, and also acts as an upper bound on the remainder term. A similar analysis can be applied to the second integral in Eq.~\eqref{app:integralweapproximate}. It follows that the asymptotic behavior of the integral $\mathcal{I}$ in the high temperature limit is given by
\begin{align}
\mathcal{I} \sim (2\pi T)^{2g-2} e^{-\pi T \sum_i |\xi_i| g_i}
\end{align}
The factor $(2\pi T)^{2g-2}$ is the high temperature behavior of expression~\eqref{app:fourierprop}. This shows that the high temperature behavior of the modulating function is given by the exponential~$\exp({-\pi T \sum_i |\xi_i| g_i})$.

\subsection{Computing the \texorpdfstring{$R$}{R} function}\label{app:computingr}
For $n=1$ and $n=2$ the $R$ function reduces to the Gauss and Appell hypergeometric functions respectively for which various efficient numerical implementations exist. For $n>3$ no numerical implementation is available and we can either perform numerical integration or compute the expansion \eqref{app:expansion} to some finite order. Numerical integration of the integral \eqref{app:carlsonr} takes into account the Beta function as well, which is why we use the series expansion instead. We will follow Ref.~\onlinecite{laarhoven1988} to cast this series expansion into a more tractable form suitable for a numerical implementation.

The main result of Ref.~\onlinecite{laarhoven1988} is that the multivariate Taylor expansion \eqref{app:expansion} can be written as the single summation
\begin{multline}
F_D^{(n)}\left(\alpha;\mathcal{G}_n; \gamma ;\{ 1- w_1,\ldots 1 - w_n\}\right) \\
= 1+ \sum_{m=1}^\infty  \frac{(\alpha)_{m}}{(\gamma)_{m}} \Lambda_m(t_1,\ldots,t_m). \label{app:explam}
\end{multline}
Here $(\alpha)_n = \Gamma[\alpha+n]/\Gamma[\alpha]$ is the Pochhammer symbol and $\Lambda_m$ is the \emph{cycle index} (of the symmetric group $S_m$) of the variables $t_j$. Defining the variables $t_j$ ($j= 1,\ldots, m$) 
\begin{align}
t_j = \sum_{i=1}^n g_i(1-w_i)^j\label{app:t-j}
\end{align} 
then the cycle index $\Lambda_m$ of this set $\lbrace t_j\rbrace_{j=1}^m$ is given by
\begin{multline}
\Lambda_m (t_1,\ldots, t_m) =\\ \sum_{\substack{k_1,\ldots, k_m \\ k_1+ 2k_2\cdots +mk_m = m}}\left[\prod_{j=1}^m \frac{1}{k_j!}\left(\frac{t_j}{j}\right)^{k_j}\right]~. \label{app:summing}
\end{multline}
The summation over the $k_i$'s \eqref{app:summing} is constrained by $\sum_{j=1}^m jk_j = m$, which makes its computation for large $m$ rather involved. It's more efficient to use an iterative approach, as $\Lambda_m$ can be expressed in terms of $\lbrace\Lambda_n\rbrace_{n<m}$. Defining $\Lambda_0=1$ we have for $m\geq 1$ 
\begin{align}
\Lambda_m(t_1,\ldots, t_m) = \frac{1}{m}\sum_{j=1}^m t_j \Lambda_{m-j}(t_1,\ldots, t_{m-j})\label{lambdam}
\end{align}
Let us also give the corresponding expansion for the $R$-function. For that we again assume $z_n$ is the largest argument of the function. Then
\begin{multline}
R\left(\alpha;\{ g_1,\ldots, g_n\}; \lbrace z_1,\ldots, z_n\rbrace \right) 
%&=  z_n^{-a} F_D^{(n-1)}\left(a;\lbrace b_1,\ldots, b_{n-1}\rbrace; b ; 1- z_1/z_n,\ldots 1 - z_{n-1}/z_n\right) \\
=\\ z_n^{-\alpha}\sum_{m=0}^\infty  \frac{(\alpha)_{m}}{(\gamma)_{m}} \Lambda_m (\tau_1,\ldots , \tau_m)\label{app:rseriesexpansion}
\end{multline}
where $\gamma = \sum_{i=1}^n g_i$ and $z_n = \text{max}(z_1,\ldots,z_n)$ as before. The $\tau_j$ ($i=1,\ldots, n-1$) are given by
\begin{align}
\tau_j = \sum_{i=1}^{n-1} g_i(1-z_i/z_n)^j ~.\label{app:tau-j}
\end{align}
This algorithm is due to Laarhoven and Kalker\citep{laarhoven1988}. 

In the main text the $R$ function which enters the expression for the interference term is a multivariate expansion in terms of the scales $1-\tfrac{z_i}{z_n} =$ \mbox{$1-\exp({-2\pi T(\tfrac{a_i}{v_i} - \tfrac{a_j}{v_j}}))<1$}. For large temperature scales ($\geq 15$ mK) the arguments approach the radius of convergence, $(1-z_i/z_n)\lesssim 1$, and the rate of convergence of the series becomes extremely slow, especially when the frequency $\omega_Q$ becomes large as well. This requires a very large number of terms in the expansion, which becomes problematic since the algorithm for $\Lambda_m$ scales as order $\mathcal{O}(N^2)$ with $N$ the number of terms in the series. In this regime numerical integration does not seem to be an alternative, as the standard integration schemes suffer from slow convergence as well. 

The situation is somewhat improved by using a series acceleration. We have chosen a series acceleration via the \emph{Combined} \emph{Nonlinear}-\emph{Condensation} \emph{Transformation} (CNCT) as outlined in Ref.~\onlinecite{ulrich1999}. The algorithm works in two steps. First, the (largely monotone) series \eqref{app:explam} is transformed into an alternating series via a Van Wijngaarden transformation. Alternating series are known to converge better using a series acceleration. Second, this alternating series is accelerated via a nonlinear sequence transformation. For our purposes we have chosen Levin's $u$ transformation \citep{ulrich1999}, although other choices yield similar results.

The advantage of the CNCT method is that only a handful of terms of the original series are needed to obtain a high precision estimate of the series. This method significantly improves the rate of convergence of many series \cite{ulrich1999}. However, the method requires the capability to compute ``random'' terms in the series \eqref{app:rseriesexpansion}. To be specific, to perform the Van Wijngaarden transformation we require the terms 
$\frac{(a)_M}{(c)_M}\Lambda_{M}$ with $M={2^k(j+1)-1}$ and $j$ and $k$ integers, see Ref.~\onlinecite{ulrich1999}. Typically we need all terms with $j,k<30$ for a decent precision in the final answer. But note that the index $M$ grows exponentially. This is problematic, because our algorithm is designed to determine $\Lambda_m$ iteratively and this iteration process grows as $\mathcal{O}(N^2)$. The CNCT method and similar acceleration methods therefore do not fully resolve the issue of slow convergence. To avoid this problem our plots are performed at low temperature ($T=1$ [mK] or $T=0$ [mK]).

A second problem that arises is a lack of precision in the terms computed. We found that the typical double floating point accuracy can lead to problems when evaluating the series for large $\omega_Q$ ($> 100$ [mK]) and values of the velocities and distance scales as mentioned in Section~\ref{sec:plots}. This issue is resolved by making use of high-precision floating point accuracy\citep{mpmath}. The downside to this is that the computation of a large number of terms is extremely slow. In particular, we cannot simultaneously make use of the CNCT algorithm and high-precision floating point accuracy.

We have implemented this algorithm through a combination of {\tt NumPy}\citep{numpy} and {\tt Fortran}, making use of {\tt F2PY}\citep{f2py}. In some cases we also made use of the high-precision floating-point arithmetic package {\tt mpmath}\citep{mpmath}. All plots are generated using {\tt matplotlib}\citep{matplotlib}.
\section{Zero temperature case}\label{app:zeroT}
In the zero temperature case the KMS relation of the $G$ correlators, see Eq.~\eqref{eq:kmsrelation}, no longer applies. Within our approximation we do have the relation \mbox{$G^>_{ij}(t) = G^<_{ji}(-t)$}. The expression of the tunnelling current at zero temperature is therefore given by
\begin{multline}
I_B(\omega_Q) = 
 Qe \Big(\sum_{i=1}^N |\Gamma_i|^2 \left[G^{>}_{ii}(\omega_Q) - G^{<}_{ii}(\omega_Q)\right] ~ + \\
2 \sum_{\substack{i<j}}^N 
|\Gamma_i\Gamma_j^*|  \text{Re}\left[ e^{i\Phi_{ij}+i\alpha_{ij}}\left[G^{>}_{ij}(\omega_Q) - G^{<}_{ij}(\omega_Q)\right]\right]\Big)  ~.\label{tunnellingexpression-zeroT}
\end{multline}
The analysis of the $G^>$ correlator is the same as in the finite temperature case, with the exception that we use the zero temperature expression of the propagator $P_g(t)$. In particular \eqref{correlator} still applies, but with the propagator given by
\begin{align}
P_g(t) &= \frac{1}{(\delta +it)^g}~.
\end{align}
The expression for $G^>$ and $G^<$ then boils down to
\begin{multline}
G^{>}_{ij}(\omega) - G^{<}_{ji}(\omega) = a_{\text{vac}}\Aij v_c^{-2g_c}v_n^{-2g_n}\int_{-\infty}^\infty dt e^{i\omega t} \\
\Big[P_{g_c}(t+\frac{a}{v_c})  P_{g_c}(t-\frac{b}{v_c}) P_{g_n}(t+\eta \frac{a}{v_n}) P_{g_n}(t-\eta \frac{b}{v_n}) -\\
P_{g_c}(-t+\frac{a}{v_c})  P_{g_c}(-t-\frac{b}{v_c}) P_{g_n}(-t+\eta \frac{a}{v_n}) P_{g_n}(-t-\eta \frac{b}{v_n})
\Big]\label{generalzerotintegral}
\end{multline}
We have not found a reference or method to treat this Fourier transform directly. It can be treated for the special case of a symmetric interferometer and a single edge mode, where \mbox{$v_c=v_n$} and \mbox{$a=b$}. This special case is treated in Appendix~\ref{zerot-singlemode}. Alternatively, we can start with the expression for the tunnelling current of the finite temperature case and take the zero temperature limit. This approach allows for more general values of the physical parameters and is performed in Section~\ref{zerot-directlimit}. Finally, we suggest in Section~\ref{multimodesintegral} a solution to the integral \eqref{generalzerotintegral}, obtained by taking the zero temperature limit from the finite temperature expression.

As in the finite temperature case we find for the tunnelling current 
\begin{align}
I_B(\omega_Q) = \frac{Qe}{v_c^{2g_c}v_n^{2g_n}}  a_{\text{vac}} |\Gamma(\omega_Q)|^2 I_{2g}(\omega_Q)\text{sgn}(\omega_Q)
\end{align}
with $I_{2g}$ given by \eqref{app:ig} and the effective tunnelling amplitude equals
\begin{multline}
|\Gamma|^2 = 
\sum_{i=1}^N |\Gamma_i|^2 + \\2 \sum_{\substack{i<j}}^N 
|\Gamma_i\Gamma_j^*|  \text{Re}\left[\Aij e^{i\Phi_{ij}+i\alpha_{ij}}H_{ij}^{\text{mod}}(\omega_Q) \right]
\end{multline}
The modulating function $H_{ij}^{\text{mod}}(\omega_Q)$ is given by \eqref{hmodatzerot} in the symmetric interferometer case with a single mode, and by \eqref{hmodatzerotgeneralcase} in the more general case.

\subsection{Zero temperature: tunnelling current for a single mode by computation of the Fourier transform}\label{zerot-singlemode}
We start with the Fourier transform of the correlator $G_{ii}(\omega)$. This corresponds to the tunnelling current through a single point contact, see Eq.~\eqref{tunnellingexpression-zeroT}. We require the Fourier transform of the propagator $P_g(t)$, which is given by\citep{gradshteyn2007}
\begin{align}
P_g(\omega) &= I_g(\omega) \Theta(\omega) \\
\text{where}\quad I_g(\omega) &\equiv \frac{2\pi}{\Gamma[g]} |\omega|^{g-1} \label{app:ig}
\end{align}
and $\Theta(\omega)$ is the step function. Then
\begin{align}
G_{ii}^>(\omega) - G_{ii}^<(\omega) = a_{\text{vac}}v_c^{-2g_c}v_n^{-2g_n} I_{2g}(\omega)\text{sgn}(\omega)~.
\end{align}
For the expression of the interference term we set \mbox{$v=v_c=v_n$} and $a=b$. The required integral is (see Eq.~\eqref{generalzerotintegral})
\begin{multline}
K_g(\omega) = \int_{-\infty}^\infty dt e^{i\omega t} [ P_g(t+\frac{a}{v}) P_g(t-\frac{a}{v})\\
 -P_g(-t-\frac{a}{v})) P_g(-t+\frac{a}{v})]~.
\label{app:kintegral}
\end{multline}
We consider the separate cases where \mbox{$g<1$} and \mbox{$g>\frac{1}{2}$}. The two cases overlap, and we find a single expression applicable for all values of $g$. For \mbox{$g<1$} the integral regulator is not required, so we set \mbox{$\delta = 0$}. With some careful manipulations of the fractional powers of $i$ we obtain
\begin{align}
K_{g<\frac{1}{2}}(\omega) 
={}&  4\sin(\pi g)\text{sgn}(\omega)\int_{\frac{a}{v}}^\infty dt \frac{\sin(|\omega| t)}{
(t^2-(\tfrac{a}{v})^2)^{g}} \nonumber\\
={}& \Gamma\bigl[g+\frac{1}{2}\bigr]\Bigl(\frac{|\omega|a}{2v}\Bigr)^{\frac{1}{2}-g}\nonumber\\
&\times J_{g-\frac{1}{2}}
\bigl(\frac{a|\omega|}{v}\bigr)I_{2g}(\omega)\text{sgn}(\omega)~.\label{app:glessthanone}
\end{align}
The function $J_{g}(x)$ is the Bessel function of the first kind. The integral is found in Ref.~\onlinecite{gradshteyn2007}. For the case of $g>\frac{1}{2}$ we need an integral representation of the confluent hypergeometric function ${}_1F_1$
\begin{multline}
\int_{-\infty}^\infty (\beta + it)^{-g}(\gamma + it)^{-g} e^{i\omega t} dt 
=\\  e^{-\gamma\omega}  {}_1F_1(g;2g;(\gamma-\beta)\omega) I_{2g}(\omega) \Theta(\omega)~.
\end{multline}
This applies when \mbox{$\text{Re}[\beta],\text{Re}[\gamma]>0$} and \mbox{$\text{Re}[g]> \frac{1}{2}$}. With this integral representation we find
\begin{multline}
K_{g>\frac{1}{2}}(\omega) 
=e^{-\delta|\omega|}  e^{i\frac{ |\omega| a}{v}} \\\times {}_1F_1\Bigl(g;2g;-2i\frac{|\omega| a}{v}\Bigr) I_{2g}(\omega) \text{sgn}(\omega)~.
\end{multline}
For these specific parameters the confluent hypergeometric function ${}_1F_1$ reduces to the Bessel function of the first kind \citep{gradshteyn2007}
\begin{multline}
e^{i\frac{|\omega|a}{v}} {}_1F_1\Bigl(g;2g;-2i\frac{|\omega|a}{v}\Bigr) = \\ \Gamma\Bigl[g+\frac{1}{2}\Bigr]\Bigl(\frac{|\omega|a}{2v}\Bigr)^{\frac{1}{2}-g} J_{g-\frac{1}{2}}\Bigl(\frac{|\omega|a}{v}\Bigr)~.\nonumber
\end{multline}
Therefore both cases (\mbox{$g>\frac{1}{2}$} and \mbox{$g<1$}) match in the limit of \mbox{$\delta\rightarrow 0$} and expression \eqref{app:glessthanone} extends to all values of $g>0$. Finally, we have for the zero temperature expression of the modulating function of a symmetric interferometer
\begin{align}
H_{ij}^{\text{mod}}(\omega) &= \Gamma\bigl[g+\frac{1}{2}\bigr]\Bigl(\frac{|\omega|a}{2v}\Bigr)^{\frac{1}{2}-g} J_{g-\frac{1}{2}}\Bigl(\frac{|\omega|a}{v}\Bigr)~.\label{hmodatzerot}
\end{align}

\subsection{Obtaining the zero temperature limit from the finite temperature expression}\label{zerot-directlimit}
The more general case in which we consider multiple modes with different edge velocities involves a more complicated Fourier transform which we are not able to determine directly. Instead, we use the result for finite temperatures and take the limit of \mbox{$T\downarrow 0$}. 

We require the zero temperature limit of the modulating function, $H_{ij}^{\text{mod}}$ see Eq~\eqref{modulatingfunction}. To perform this limit we make use of the series representation of the $R$ function, Eq.~\eqref{app:expansion}. This gives
\begin{multline}
\lim_{T\downarrow 0} R\left(\alpha-i\tfrac{\omega}{2\pi T}; \mathcal{G}_n; e^{2\pi T x_1},\ldots , e^{2\pi T x_{n}}\right) =\\
\lim_{T\downarrow 0}  e^{-(2\pi T \alpha-i\omega) x_{n}}\sum_{m_1=0}^\infty\!\! \cdots\!\!\!\! \sum_{m_{n-1}=0}^\infty 
 \frac{(\alpha-i\tfrac{\omega}{2\pi T})_{\sum_{i=1}^{n-1}m_i }}{(\gamma)_{\sum_{i=1}^{n-1}m_i }}\\
\times\left[\prod_{i=1}^{n-1} \frac{(g_i)_{m_i}}{m_i!} \left(1-e^{-2\pi T x_{n,i}}\right )^{m_i}\right]
 ~.\label{zeroTlimitRfunction}
\end{multline}
The $x_i$'s correspond to the (real valued) energy scales associated with the edge modes, i.e. \mbox{$\pm\frac{a}{v_c}$} and so on. We assume \mbox{$x_{n} \geq x_i$} for all $i$ and we write \mbox{$x_n-x_i = x_{n,i}\geq 0$}. The limit is determined term-by-term. We first note the approximation
\begin{align}
\left(1-e^{-2\pi Tx_{n,i}}\right )^{m_i} = (2\pi T)^{m_i}x_{n,i}^{m_i} + \ldots~. \label{app:tempexpansion}
\end{align}
The dots are of higher order in $T$. Combining this with the \mbox{$(a-i\tfrac{\omega}{2\pi T})_{m}$} term, where \mbox{$m=\sum_{i=1}^{n-1} m_i$}, we obtain for the zero temperature limit
\begin{multline}
\lim_{T\downarrow 0} \prod_{j=1}^{n-1}\bigl( 2\pi Tx_{n,j}\bigr)^{m_j}\prod_{k=0}^{m-1} \left(\alpha-i\frac{\omega}{2\pi T} + k\right) =\\
\prod_{j=1}^{n-1} (-i\omega x_{n,j})^{m_j}+\ldots~.
\end{multline}
The higher order corrections of \eqref{app:tempexpansion} vanish in this limit. Plugging this back into \eqref{zeroTlimitRfunction} gives
\begin{multline}
\lim_{T\downarrow 0} R\left(\alpha-i\tfrac{\omega}{2\pi T}; \mathcal{G}_n; \{ e^{2\pi T x_1},\ldots , e^{2\pi T x_{n}} \}\right) = \\
e^{i\omega x_{n}}\sum_{m_1=0}^\infty\!\! \cdots\!\!\!\! \sum_{m_{n-1}=0}^\infty 
 \frac{1}{(\gamma)_{\tilde{m} }}
\left[\prod_{i=1}^{n-1} \frac{(g_i)_{m_i}}{m_i!} (-i\omega x_{n,j})^{m_i}\right]= \\
e^{i\omega x_n}\Phi_2^{(n-1)}(\mathcal{G}_{n-1}; \gamma;  \{-i\omega x_{n,1},\ldots, -i\omega x_{n,n-1}\} )
\end{multline}
with $\tilde{m}=\sum_{i=1}^{n-1}m_i$. The resulting series is called the \emph{confluent Lauricella hypergeometric function}\citep{mathai2009}
\begin{multline}
\Phi_2^{(n)}(\mathcal{G}_n; \gamma; \{w_1,\ldots, w_n\}) = \\
\sum_{m_{1}=0}^\infty\!\!\cdots\!\!\sum_{m_{n}=0}^\infty  \frac{1}{(\gamma)_{\sum_i m_i}} 
\left[\prod_{i=1}^n \frac{(g_i)_{m_i}}{m_i!} w_i^{m_i}\right]~. \label{app:confllaur}
\end{multline}
This series is a multivariable generalization of the confluent hypergeometric $\Phi_2^{(n=2)}$ function \citep{gradshteyn2007}. The expression for the modulating function is
\begin{multline}
H^{\text{mod}}_{ij}(\omega) =  e^{i\omega x_n} \times\\
\Phi_2^{(n-1)}(\mathcal{G}_{n-1}; \gamma;  \{-i\omega x_{n,1},\ldots, -i\omega x_{n,n-1}\} )~.
\label{hmodatzerotgeneralcase}
\end{multline}
Here we recall that the $x_i$ correspond to all combinations of \mbox{$\eta_i\frac{a}{v_i}$} and \mbox{$-\eta_i\frac{b}{v_i}$}, the parameter $x_n$ satisfies $x_n > x_i$ for $i < n$ and $x_{n,i}\equiv x_n - x_i > 0$. As a sanity check we look at the case treated in Appendix~\ref{zerot-singlemode}, which corresponds to the symmetric interferometer and a single channel. The confluent Lauricella function reduces to the confluent hypergeometric function, \mbox{$\Phi_2^{(1)}(b,c;x)={}_1F_1(b,c;x)$}, which follows from the series representation. And so
\begin{multline}
\lim_{T\downarrow 0} R\left(g-i\frac{\omega}{2\pi T};\lbrace g,g\rbrace; \{ e^{2\pi T \frac{a}{v}}, e^{-T \frac{a}{v}}\}\right)
= \\ e^{i\omega \frac{a}{v}}{}_1F_1\left(g;2g; -2i\omega \frac{a}{v}\right)~.
\end{multline}
This matches with the result \eqref{app:glessthanone}.

The series expansion of the confluent Lauricella function \eqref{app:confllaur} is of the same form as the non-confluent Lauricella function, \eqref{app:expansion}. The same combinatoric trick as explained in Appendix~\ref{app:computingr} can be used to rewrite this multivariable series as a single expansion in terms of cycle indices, see Section~\ref{app:computingr}. This expansion is given by
\begin{align}
\Phi_2^{(n)}(\mathcal{G}_n; \gamma; \{w_1,\ldots, w_n\}) =
\end{align}

We find that the convergence of the confluent series is much better than the non-confluent (finite temperature) case. In general, we do not require as many terms in the series. However, for the physical values of the velocity, distance and voltage used in the main text we find that double floating precision is still not sufficient and we require high-precision floating point numbers\citep{mpmath}. 

\subsection{Zero temperature: multiple modes}\label{multimodesintegral}
We have obtained the general expression for the zero temperature case by taking the zero temperature limit of the finite temperature expression. The same result can also be obtained by taking the Fourier transform of the zero-temperature expression for the $G^>$ correlators. Since these calculations must produce the same answer we obtain the following integral representation of the confluent Lauricella hypergeometric function. With \mbox{$P_g(t) = (\delta + it)^{-g}$} we have
\begin{multline}
\int_{-\infty}^\infty dt e^{i\omega t} \big[\prod_j P_{g_j}(t+x_j) - \prod_k P_{g_k}(-t+x_k)\big] = \\
I_{2g}(\omega)\text{sgn}(\omega) \times\\
e^{i\omega z_{n}}\Phi_2^{(n-1)}\big(\mathcal{G}_{n-1}; \gamma ;
 -i\omega x_{n,1}, \cdots, -i\omega x_{n,{n-1}})~.
\end{multline}
Here $x_{n,i} = x_n - x_i > 0 $ for all $i<n$, all $g_i > 0$ and $\delta$ is taken to zero in the end. The function $I_{g}$ is given by \eqref{app:ig} and $\Phi_2^{(n-1)}$ is the confluent Lauricella hypergeometric function, which has the series representation \eqref{app:confllaur}. 
\bibliography{./main-all-refs} 
\addcontentsline{toc}{chapter}{References}
\end{document}